\documentclass[twocolumn,tighten]{aastex631}

\usepackage{multirow}
\usepackage{bm}
\received{May 24, 2021}
\revised{June 26, 2021}
\accepted{June 29, 2021}

\shortauthors{D\'ek\'any et al.}
\graphicspath{{./}{figures/}}
\begin{document}

\title{Metallicity estimation of RR~Lyrae stars from their $I$-band light curves}

\author[0000-0001-7696-8331]{Istv\'an D\'ek\'any}
\affiliation{Astronomisches Rechen-Institut, 
	Zentrum f\"ur Astronomie der Universit\"at Heidelberg,
	M\"onchhofstr. 12-14, 69120 Heidelberg, Germany}

\author[0000-0002-1891-3794]{Eva K. Grebel}
\affiliation{Astronomisches Rechen-Institut, 
	Zentrum f\"ur Astronomie der Universit\"at Heidelberg,
	M\"onchhofstr. 12-14, 69120 Heidelberg, Germany}

\author[0000-0002-6495-0676]{Grzegorz Pojma\'nski}
\affiliation{Astronomical Observatory, University of Warsaw, Al. Ujazdowskie 4, 00-478 Warszawa, Poland}

%\collaboration{2}{(et al.)}

%% Note that the \and command from previous versions of AASTeX is now
%% depreciated in this version as it is no longer necessary. AASTeX 
%% automatically takes care of all commas and "and"s between authors names.

%% AASTeX 6.31 has the new \collaboration and \nocollaboration commands to
%% provide the collaboration status of a group of authors. These commands 
%% can be used either before or after the list of corresponding authors. The
%% argument for \collaboration is the collaboration identifier. Authors are
%% encouraged to surround collaboration identifiers with ()s. The 
%% \nocollaboration command takes no argument and exists to indicate that
%% the nearby authors are not part of surrounding collaborations.

%% Mark off the abstract in the ``abstract'' environment. 
\begin{abstract}

We have revisited the problem of metallicity prediction of RR~Lyrae stars from their near-infrared light curves in the Cousins $I$ waveband. Our study is based on high-quality time-series photometry and state-of-the-art high-resolution spectroscopic abundance measurements of 80 fundamental-mode (RRab) and 24 first-overtone (RRc) stars, spanning $\sim$[$-2.7$,$+0.18$]~dex and $\sim$[$-3$,$-0.5$]~dex ranges, respectively. Employing machine-learning methods, we investigated various light-curve representations and regression models to identify their optimal form for our objective. Accurate new empirical relations between the [Fe/H] iron abundance and the light-curve parameters have been obtained using Bayesian regression for both RRab and RRc stars with mean absolute prediction errors of 0.16 and 0.18~dex, respectively. We found that earlier $I$-band [Fe/H] estimates had a systematic positive bias of up to $\sim$0.4~dex in the metal-poor regime. Our new predictive models were deployed on large ensembles of RR~Lyrae stars to obtain photometric metallicity distribution functions (MDFs) for various old stellar populations in and around the Milky Way. We find that the mode of the old bulge component's MDF is approximately $-1.4$~dex, in remarkable agreement with the latest spectroscopic result. Furthermore, we derive MDF modes of $-1.83$, $-2.13$, and $-1.77$~dex for the Large and Small Magellanic Clouds, and the Sagittarius dwarf galaxy, respectively.

\end{abstract}

%% Keywords should appear after the \end{abstract} command. 
%% The AAS Journals now uses Unified Astronomy Thesaurus concepts:
%% https://astrothesaurus.org
%% You will be asked to selected these concepts during the submission process
%% but this old "keyword" functionality is maintained in case authors want
%% to include these concepts in their preprints.
\keywords{RR Lyrae variable stars(1410) -- Metallicity(1031) -- Light curves(918)}

\section{Introduction} \label{sec:intro}

RR~Lyrae stars are keystone objects of Galactic archeology. These pulsating, low-mass, horizontal-branch stars belong to the oldest stellar populations of the Universe \citep[see][for a review]{catelan_pulsating_2015}. Thanks to their relatively high luminosities and characteristic large-amplitude light curves caused by radial pulsation, they are straightforward to identify, making them excellent population tracers. Large time-domain photometric surveys have discovered them in vast numbers in the Milky Way and other members of the Local Group \citep[e.g.,][to mention only a few]{clementini_gaia_2019,soszynski_over_2019,dekany_near-infrared_2020,stringer_identifying_2020}. They follow tight period--luminosity--metallicity relations at infrared wavelengths \citep[e.g.,][]{marconi_new_2015,muraveva_rr_2018}, enabling us to employ them as high-precision standard candles and reddening estimators. Moreover, the existence of a precise relationship between their light-curve shapes and metal abundances discovered by \citet{jurcsik_determination_1996} does not only make it feasible to determine their distances solely from photometric observations, but also gives us simple means to obtain metallicity distributions of old stellar populations, opening the possibility to constrain their early formation histories \citep[see][and references therein]{savino_age_2020}.

In order to unlock the full potential of RR~Lyrae stars as all-in-one photometric tracer objects, both precise and accurate methods for the prediction of their metallicities are required for all common photometric wavebands used by current and future surveys. The number of RR~Lyrae stars with low-resolution spectroscopic metal abundance measurements based on hydrogen and calcium line widths as metallicity proxies (i.e., the ``$\Delta$S-method'', see, e.g., \citealt{layden_metallicities_1994}) have recently proliferated \citep[e.g.,][]{liu_probing_2020,muhie_kinematics_2021}, mainly thanks to large surveys such as SEGUE \citet{yanny_segue_2009} and LEGUE \citet{deng_lamost_2012}. However, such observations are resource-intensive and currently unfeasible for faint stars in distant and/or highly reddened populations. In addition to being cheaper and much more straightforward to obtain, photometric RR~Lyrae metallicities have also been shown to have a similar precision as [Fe/H] estimates obtained with the $\Delta$S-method. Following up on the pioneering work by \citet[hereafter JK96]{jurcsik_determination_1996}, who calibrated a simple linear formula for the estimation of the [Fe/H] from $V$-band light curves, several studies have extended their work to various other photometric wavebands \citep[e.g.,][]{smolec_metallicity_2005,nemec_metal_2013,ngeow_palomar_2016,skowron_ogle-ing_2016,hajdu_data-driven_2018,iorio_chemo-kinematics_2020}.

A common shortcoming of all these formulae is that they were derived from either sub-optimal data sets or from one another, either via intermediate data sets or by simply using transformation formulae between photometric systems. This is due to the long-prevailing shortage of objects with both high-quality light curves and accurate and high-precision spectroscopic metallicity measurements. The various prediction models had to be trained on relatively few objects offering a limited coverage of the underlying parameter space, which in itself can induce biased predictions. Furthermore, the underlying metallicity measurements were dominantly obtained by low-resolution spectroscopy or spectro-photometry, often forming heterogeneous samples, which in turn had to be calibrated to even fewer high-resolution (HR) spectroscopic [Fe/H] abundance measurements, leading to the coexistence of various ``metallicity scales''. Finally, adapting a prediction model from one waveband to another by the transformation of the light-curve parameters has the additional pitfall that such transformations themselves can be metallicity-dependent \citep[see, e.g.,][]{skowron_ogle-ing_2016}.

In the course of the past decade, HR spectroscopic analyses of RR~Lyrae stars have been slowly accumulating thanks to the devotion of tremendous amounts of telescope time and the painstaking effort of several authors \citep[see][and references therein]{crestani_deltaS}. Recently, \citet{mullen_metallicity_2021} capitalized on this progress to establish new prediction formulae of the metallicity from optical ($V$) and mid-infrared ($W1$ and $W2$) light curves based on a sizable data set of low-resolution spectroscopic [Fe/H] estimates, which, in turn, were tied to recent HR spectroscopic data. Their work revealed large systematics in earlier estimators, which were resulting from the various aforementioned problems.

In this work, we revisit the problem of metallicity estimation from Cousins $I$-band light curves of fundamental-mode RR~Lyrae (RRab) stars, and, for the first time, attempt to establish a similar estimator for the first-overtone RR~Lyrae (RRc) stars as well. This particular photometric waveband is of particular interest in contemporary research due to vast amounts of $I$-band photometric time series of RR~Lyrae stars acquired by the Optical Gravitational Lensing Experiment \citep[OGLE,][]{udalski_ogle-iv_2015}. What makes OGLE unique among the large number of recent and ongoing photometric surveys is that it provides homogeneous, high-quality light curves for billions of objects over large areas of the observationally most challenging regions of the Local Group, namely the Galactic bulge and mid-plane, and the Small and Large Magellanic Clouds. These data were utilized by a plethora of studies, employing the RR~Lyrae stars not only as distance indicators \citep[e.g.,][]{jacyszyn-dobrzeniecka_ogle-ing_2017}, but also as metallicity tracers \citep[e.g.,][]{skowron_ogle-ing_2016,pietrukowicz_properties_2020}.

Such studies either relied on the metallicity estimator from the early work of \citet[][hereafter S05]{smolec_metallicity_2005}, or predictive formulae established for optical wavebands and transformed into the $I$ band. The S05 $I$-band formula was established on the basis of a very small data set available at the time, comprising metallicity estimates from low-resolution spectra and spectro-photometry on the JK96 scale. As an alternative, \citet{skowron_ogle-ing_2016} employed the \citet[][hereafter N13]{nemec_metal_2013} estimator to explore the metallicity distribution of the Magellanic Clouds using OGLE data, by transforming it from the Kepler photometric band to the $I$-band; and their method was further applied by \citet{jacyszyn-dobrzeniecka_ogle-ing_2017}. The disadvantage of this approach is that it adds noise and potential systematics to the predictions due to the uncertainties and the metallicity-dependence of the transformations. 

$I$-band photometric metallicities were also used as intermediate calibration data. \citet{hajdu_data-driven_2018} trained a deep-learned [Fe/H] predictive model for the near-infrared $K_s$-band using $I$-band predictions of the [Fe/H] from the S05 formula. Their estimator was deployed on data from the VVV survey \citep{minniti_vista_2010} to explore the southern thick disk \citep{dekany_near-infrared_2018}. In view of the number of currently available HR spectroscopic [Fe/H] measurements, the recalibration of the $I$-band photometric estimation of RR~Lyrae stars became timely. To this end, in order to minimize the possible sources of noise and systematics from intermediate data and calibrations, we take the strategy of relying solely on homogeneized HR spectroscopic measurements.

This paper is structured as follows. First, we describe the photometric and spectroscopic data that form the basis of our analysis in Sect.~\ref{sec:data}. In addition, we present the parametric representations of the photometric time series that we use throughout the study, together with the procedures to obtain them. In Sect.~\ref{sec:prediction}, we search for the optimal form for the predictive models of the metallicity for RRab and RRc stars and estimate their performances using machine-learning techniques; then obtain their final parameter distributions via Bayesian regression analysis and compare the resulting predictive models to the S05 estimator and discuss the nature and possible sources of the observed differences. In Sect.~\ref{sec:MDFs}, we deploy our new [Fe/H] estimators on large data sets from various Local Group environments and derive the metallicity distribution functions (MDFs) of their old stellar populations. We also compare the resulting distributions with those obtained by the widely employed S05 formula. Finally, we summarize our findings and conclusions in Sect.\ref{sec:conclusions}.

\newpage

\section{Observational Data} \label{sec:data}

We surveyed the literature for RR~Lyrae stars with both high-resolution spectroscopic metallicity determinations and well-sampled, accurate light curves in the Cousins $I$ photometric passband. The largest single catalog of contemporary spectroscopic RR~Lyrae metallicities was published recently by \citet{crestani_deltaS}, and is based mostly on echelle spectra from the du Pont telescope at Las Campanas Observatory, complemented by data from various other instruments. According to \citet{crestani_deltaS}, their measurements, together with the additional results by \citet{for_chemical_2011}, \citet{chadid_spectroscopic_2017}, and \citet{sneden_rrc_2017} form a homogeneous sample (from here on, referred to as the CFCS sample), covering $-3.1 \lesssim [{\rm Fe}/{\rm H}] \lesssim 0.2$, i.e.,  virtually the full range of possible metallicities of RR~Lyrae stars. We used this data set as our reference metallicity scale, and complemented it with additional data sets from the literature that have a sufficient number of common objects with the CFCS sample in order to account for possible systematic offsets. The CFCS sample was thus complemented by additional metallicity measurements from \citet[][C95]{clementini_composition_1995}, \citet[][F96]{fernley_metal_1996}, \citet[][L96]{lambert_chemical_1996}, \citet[][L13]{liu_abundances_2013}, \citet[][N13]{nemec_metal_2013}, \citet[][G14]{govea_chemical_2014}, \citet[][P15]{pancino_chemical_2015}, and \citet[][A18]{andrievsky_relationship_2018}.
The resulting spectroscopic data set contains metallicities for 183 RRab and 49 RRc stars in the aforementioned metallicity range, but improving its coverage.

We searched the literature for existing $I$-band photometric time series of the RR~Lyrae stars in the combined spectroscopic data set from various surveys and individual studies. Although the OGLE Collection of Variable Stars\footnote{\url{http://ogledb.astrouw.edu.pl/~ogle/OCVS/}} (OCVS) includes tens of thousands of Galactic RR~Lyrae stars, its footprint covers low Galactic latitudes toward the bulge and the southern mid-plane that spectroscopic studies tend to purposely avoid due to large interstellar extinction and point-source density. This mutual avoidance resulted in merely 2 objects in common with the spectroscopic sample. For several field RR~Lyrae stars from our closer Galactic neighborhood, highly accurate $I$-band light curves were published by \citet{monson_standard_2017}. The All Sky Automated Survey \citep[ASAS,][]{pojmanski_all_1997}, although monitoring the sky in mostly the $V$ band, also acquired well-sampled light curves in the $I$ band for hundreds of field RR~Lyrae stars; a large part of this sample was previously analyzed by \citet{szczygiel_galactic_2009}. We found that 94 RRab and 34 RRc stars from the spectroscopic data set had $I$-band light curves from the aforementioned photometric studies. Together with an additional star (RR~Gem) with photometric data from \citet{jurcsik_blazhko_2005}, these 129 objects with combined HR spectroscopic and $I$-band time-series photometric data form the basis of our study.

\subsection{Light curves and their representation} \label{subsec:lightcurves}

The optimal periods of the ASAS light curves were determined with the procedure discussed by  \citet{dekany_into_2019}; consisting of a robust, non-linear regression of a truncated Fourier series in conjunction with iterative outlier rejection, and the optimal number of terms (i.e., Fourier order) determined by cross-validation. We found that the light curves from other sources already had highly accurately  determined periods, therefore we directly adopted those values from the respective studies.

\subsubsection{Gaussian Process Regression} \label{subsubsec:gpr}

Although the traditional direct Fourier fitting (DFF) provides a good generic model representation of periodic light curves, it shows considerable volatility in case of uneven phase-sampling and/or noisy photometry, in which cases it lacks a good bias-variance trade-off at any choice of Fourier order. To obtain a more robust model representation of the light-curve shape, we employed Gaussian process regression (GPR) on the phase-folded data by the following procedure. First, we considered the $[-0.5, 1.5]$ range of the pulsation phase by repeating the measurements beyond the $[0,1]$ phase range in order to obtain natural periodic boundary conditions for the GPR fit. We applied GPR on the resulting data set using the product of an exponential sine-squared and a constant kernel for modeling the inherent shape of the light curves, and we added a white kernel component to include the observational noise in the model. The optimum solutions were found by maximizing the log-marginal likelihood using the {\tt scikit-learn} software library.

\begin{figure} %[ht!]
	\plotone{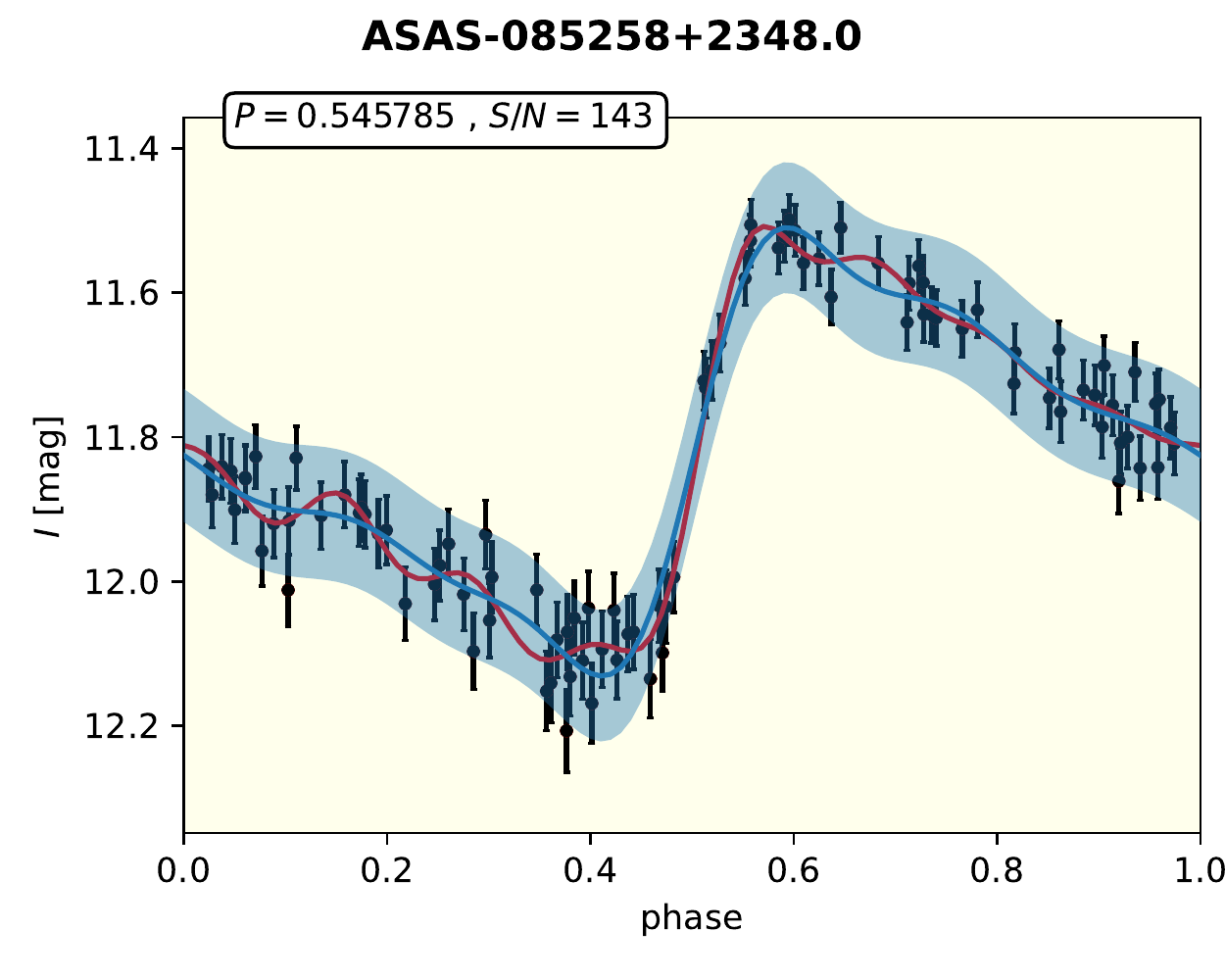}
	\caption{Phase-folded ASAS $I$-band light curve of the RRab star ASAS-085258+2348.0. The objects period (in days) and the $S/N$ are shown in the inset. The individual photometric measurements and their errors are shown with black points and error bars. The red and blue curves show the best-fitting models obtained by DFF (8-th order Fourier sum) and GPR, respectively. The blue shaded area shows the $95\%$ confidence interval of the GPR model. \label{fig:dff_gpr}}
\end{figure}

Figure~\ref{fig:dff_gpr} shows the difference between the best-fitting DFF and GPR models of an RRab star's ASAS light curve of modest quality. The GPR model can provide the complexity of a high-order Fourier sum without suffering from high variance in the absence of a large amount of high-quality data. For the latter, the difference between the results from the two methods diminishes. In addition to its robustness, another advantage of the GPR is that it yields a probabilistic model, from which an arbitrary number of samples can be drawn, and can then be used for computing empirical confidence intervals and error distributions of the light-curve parameters.

We transformed the fitted GPR models into the Fourier parameter space by evaluating the mean of the predictive GPR distribution over a dense equidistant phase grid, and fit a 20-order Fourier sum to the resulting points. We estimated the uncertainties in the Fourier parameters by drawing 500 random samples from the GPR model, repeating the entire regression procedure for each realization, and subsequently computing the standard deviations of the parameter distributions obtained in this way.

As the result of the above procedure, the light curves are represented by the parameters

\begin{equation}
	\{P, A_i, \phi_{i1}=\phi_i-i\phi_1~|~i=1,2,3\}.
\end{equation}\label{eq:lcparam}

\noindent Here, $P$ is the period, and $A_i$ and $\phi_i$ are the amplitude and phase in the $i$-th Fourier term of the following form:

\begin{equation}
	m(t) = \sum^{20}_{i=0} A_i \sin \left( 2i\pi \frac{t-t_0}{P} + \phi_i \right).
\end{equation}\label{eq:lcmodel}

\noindent The resulting light-curve parameters and their uncertainties for our entire sample of RR~Lyrae stars are listed in Table~\ref{tab:rrlyraes}.

While the restriction of the above representation to the first few Fourier terms causes some potential information loss by disregarding the finest details of some RRab light curves described by higher-order parameters, in practice, such features become rather subtle for longer periods and are easily washed out by photometric noise. In addition, the use of noisy high-order parameters as regressors for metallicity prediction could easily lead to overfitting due to our modest sample size.

\vfill\eject

\subsubsection{Principal component analysis} \label{subsubsec:pca}

The traditional Fourier decomposition discussed above provides a simple and intuitive parametric representation of the RR~Lyrae light curves; however, the amplitudes and phases of different orders show strong correlations, which might complicate feature selection and affect predictive performance when they are considered as descriptive variables in a regression problem such as metallicity prediction. In order to eliminate such potential effects, we performed a principal component analysis (PCA) on the Fourier parameters.

PCA is a linear transformation that decomposes a multivariate data set into a set of orthogonal components, resulting in a transformed feature space in which the individual dimensions of the data set are linearly uncorrelated and form a sequence according to the amount of explained variance, with the transformed data set showing the highest variance along the first dimension. To attain a robust transformation, the training set of the PCA should ideally fully encompass and densely populate the entire parameter space in question. We achieved this by performing PCA on the entire OGLE collection of RR~Lyrae stars, i.e., the catalogs of the Galactic bulge and disk, and the Large and Small Magellanic Clouds (LMC and SMC). First, all light curves in the aforementioned catalogs were fitted following the procedure in Sect.~\ref{subsubsec:gpr}. We selected the training sets form the PCA by applying the following quality criteria to the data:

\vskip5mm
\noindent RRab:
\begin{equation}
C>0.93;~S/N>200;~N_{\rm ep.}>100;~0.1<A_t<1.1 \label{eq:crit_rrab}
\end{equation}

\noindent RRc:
\begin{equation}
C>0.93;~S/N>100;~N_{\rm ep.}>100;~0.1<A_t<0.4 \label{eq:crit_rrc}
\end{equation}

\noindent Here, $C$ denotes the phase coverage\footnote{$1-$maximum phase lag}, $S/N = A_t\sqrt{N_{\rm ep.}} / \sigma$, $A_t$ is the peak-to-valley amplitude of the light-curve model, $N_{\rm ep.}$ is the number of epochs, and $\sigma$ is the residual standard deviation.

We performed PCA on the resulting $\sim$39000 RRab and $\sim$21000 RRc stars by first standardizing the features in Eq.~\ref{eq:lcparam} to zero mean and unit variance, and then performing singular value decomposition on the data matrix using the {\tt scikit-learn} software library. Table~\ref{tab:pca} lists the ratios of the explained variance ($\hat\sigma_k$), singular values ($s_k$), and the components of the principal axes in the original feature space (i.e., right singular vectors).

As an additional, alternative representation of the light curves, we transformed the Fourier parameters into the linearly uncorrelated $u_k;~k\in\{1...6\}$ feature space established by the PCA. Errors in the $u_k$ parameters were estimated from the GPR models in the same way as for the Fourier parameters themselves (see Sect.~\ref{subsubsec:gpr}).
The $u_k$ parameters of our RR~Lyrae calibration data set, together with their uncertainties, are listed in Table~\ref{tab:pcapars}.

%\begin{longrotatetable}
\begin{splitdeluxetable*}{llllllBcccccccccc}
	%\tablenum{1}
	%\movetabledown=1in
	\tablecaption{Names and photometric parameters of the RR~Lyrae stars in the calibration data set \label{tab:rrlyraes}}
	%	\tablewidth{0pt}
	%\tablewidth{700pt}
	%\tabletypesize{\tiny}
	\tablehead{
		\colhead{Name} & \colhead{Gaia DR2 source\_id} & \colhead{ASAS ID} & \colhead{type} & \colhead{Blazhko} & \colhead{ref.} & \colhead{$N_{\rm ep}$} & 
			\colhead{$P$} & \colhead{$\langle I \rangle$} & \colhead{$A_{t}$} & \colhead{$A_1$} & \colhead{$A_2$} & \colhead{$A_3$} &
			 \colhead{$\phi_{21}$} & \colhead{$\phi_{31}$} & \colhead{$S/N$}
	}
	%\decimalcolnumbers
	\startdata
	AA Aql &  4224859720193721856 & 203815-0253.5 & RRab & 0 & ASAS & 151 & 
		0.361788 & 11.38 & 0.932 & 0.265(.005) & 0.164(.005) & 0.110(.005) & 
			8.865(.052) & 5.499(.075) & 333 \\
	AA Aqr &  2608819623000543744 & 223604-1000.9 & RRab & 0 & ASAS & 123 & 
		0.608883 & 12.36 & 0.736 & 0.224(.011) & 0.107(.011) & 0.106(.011) & 
			8.900(.145) & 5.600(.190) & 121 \\
	AE Boo &  1234729400256865664 & 144735+1650.7 & RRc  & 0 & M17 & 288 & 
		0.314887 & 10.26 & 0.234 & 0.120(.001) & 0.009(.001) & 0.008(.001) & 
			9.641(.048) & 6.597(.055) & 350 \\
	AM Vir &  3604450388616968576 & 132333-1639.9 & RRab & 1 & ASAS & 235 & 
		0.615087 & 10.93 & 0.449 & 0.157(.003) & 0.078(.003) & 0.047(.003) & 
			9.336(.062) & 6.234(.092) & 257 \\
	AO Peg &  1786827307055763968 & 212703+1836.0 & RRab & 0 & ASAS & 133 & 
		0.547243 & 12.37 & 0.616 & 0.207(.012) & 0.117(.012) & 0.063(.012) & 
			8.887(.171) & 5.935(.250) & 98 \\
	\enddata
	\tablecomments{This table is available in its entirety in machine-readable form.}
\end{splitdeluxetable*}
%\end{longrotatetable}

\begin{deluxetable*}{ccrcccccc}
	%\tablenum{1}
	\tablecaption{Results of the Principal Component Analysis\label{tab:pca}}
	\tablewidth{0pt}
	\tablehead{
		\colhead{PCA comp.($k$)} & \colhead{$\hat{\sigma^2_k}$} & \colhead{$s_{(k)}$} & \colhead{$P$} &
		\colhead{$A_1$} & \colhead{$A_2$} & \colhead{$A_3$} & \colhead{$\phi_{21}$} & \colhead{$\phi_{31}$}
	}
	%\decimalcolnumbers
	\startdata
	\multicolumn9c{RRab} \\
	1 & 0.820 & 438.86 &  0.337 & -0.420 & -0.417 & -0.433 &  0.417 &  0.419   \\
	2 & 0.109 & 159.78 & -0.683 & -0.392 & -0.399 & -0.280 & -0.288 & -0.243   \\
	3 & 0.055 & 113.64 &  0.644 & -0.143 & -0.275 & -0.047 & -0.486 & -0.501   \\
	4 & 0.008 & 43.41  & -0.027 &  0.597 & -0.077 & -0.623 &  0.301 & -0.398   \\
	5 & 0.004 & 31.69  &  0.055 & -0.445 &  0.764 & -0.438 & -0.028 & -0.154   \\
	6 & 0.004 & 30.04  & -0.022 & -0.309 & -0.047 &  0.391 &  0.646 & -0.577   \\
	\hline
	\multicolumn9c{RRc} \\
	1 & 0.449 & 239.26 & -0.153 &  0.474 &  0.558 &  0.445 & -0.253 & -0.423   \\
	2 & 0.281 & 189.16 & -0.603 & -0.388 & -0.032 & -0.351 & -0.506 & -0.325   \\
	3 & 0.098 & 111.58 &  0.412 &  0.042 & -0.118 &  0.114 & -0.822 &  0.354   \\
	4 & 0.088 & 105.73 &  0.647 & -0.083 &  0.082 & -0.407 & -0.022 & -0.634   \\
	5 & 0.056 & 84.34  & -0.076 &  0.423 &  0.445 & -0.693 & -0.018 &  0.369   \\
	6 & 0.029 & 61.25  &  0.140 & -0.661 &  0.685 &  0.139 &  0.055 &  0.226   \\
	\enddata
	%\tablecomments{}
\end{deluxetable*}

\begin{deluxetable*}{lrrrrrr}
	%\tablenum{1}
	%\movetabledown=1in
	\tablecaption{Light-curve parameters in the principal component space of the RR~Lyrae stars in the calibration data set \label{tab:pcapars}}
	%	\tablewidth{0pt}
	%\tablewidth{700pt}
	%\tabletypesize{\tiny}
	\tablehead{
		\colhead{Name} & \colhead{$u_1$} & \colhead{$u_2$} & \colhead{$u_3$} & \colhead{$u_4$} & \colhead{$u_5$} & \colhead{$u_6$}
	}
	%\decimalcolnumbers
	\startdata
	AA Aql  &  $-3.877(.180)$  &  $ 0.211(.113)$ & $-1.525(.158)$ & $-0.402(.155)$ & $ 0.335(.174)$ & $ 0.024(.114)$ \\
	AA Aqr  &  $-1.569(.410)$  &  $-0.704(.302) $& $ 0.751(.429)$ & $-0.739(.339)$ & $-0.530(.397)$ & $ 0.198(.361)$ \\
	AE Boo  &  $-0.742(.264)$  &  $ 0.289(.251) $& $-0.153(.397)$ & $ 0.265(.169)$ & $-0.513(.260)$ & $-0.432(.111)$ \\
	AM Vir  &   $1.680(.177)$  &  $-0.027(.123) $& $-0.136(.190)$ & $ 0.026(.137)$ & $ 0.032(.103)$ & $ 0.145(.155)$ \\
	AO Peg  &  $-0.802(.554)$  &  $ 0.085(.320) $& $-0.002(.539)$ & $-0.207(.400)$ & $ 0.454(.391)$ & $-0.772(.408)$ \\
	\enddata
	\tablecomments{This table is available in its entirety in machine-readable form.}
\end{deluxetable*}

\subsection{Spectroscopic data} \label{subsec:spectroscopy}

The metallicities in our compiled spectroscopic data set are based on direct equivalent width measurements of iron absorption lines in HR spectra. For several studies, separate abundance estimates were given from both neutral (FeI) and ionized (FeII) iron lines, in which cases we averaged them for each individual spectrum. We made one exception from this by using the FeII values in the case of the study by \citet[][]{fernley_metal_1996}, whose FeI measurements are systematically underabundant by $\sim$ 0.1---0.15~dex, which they suspected to be the result of a bias in their temperature estimates or non-LTE effects. In case of the CFCS sample, the mean absolute difference between the FeI- and FeII-based metallicity measurements is only 0.013~dex, an order of magnitude smaller than the mean of their individual [Fe/H] error estimates provided separately for their FeI and FeII measurements. A similar, although generally smaller discrepancy between the individual uncertainties for at least one of the FeI or FeII measurements and the actual spread between them can be observed in case of the other studies as well, making the quoted [Fe/H] errors hard to interpret. We note that \citet[][L96]{lambert_chemical_1996}, \citet[][L13]{liu_abundances_2013} and \citet[][A18]{andrievsky_relationship_2018} did not provide separate [Fe/H] measurements for FeI and FeII lines, but only combined estimates.

In addition to homogeneous [Fe/H] measurements, robust estimates of their uncertainties are also indispensable for an accurate predictive modeling, due to the strong heteroskedasticity of the data. Although most of the authors provided individual error estimates, they were computed with different methods and under different assumptions. Upon close inspection of the error estimates in the various studies, we concluded that in many cases, the individual error estimates are in tension not only with the actual differences between the FeI- and FeII-based estimates, but also with the standard deviations obtained from multiple measurements of the same object with the same instrument, from the same authors. These are possibly due to the implicit inclusion of systematic error estimates in the quoted uncertainties. Furthermore, some authors did not provide error estimates along with their [Fe/H] measurements at all. 

Due to these complications in interpreting the heterogeneous [Fe/H] uncertainties found in the literature, we opted to compute our own error estimates from the available data set itself. For this purpose, we took advantage of the fact that with only a few exceptions, the authors acquired multiple spectra for several objects using the same instrument, each yielding individual metallicity estimates. Assuming that a given author attained similar precision in the [Fe/H] determination for different objects using the same instrument, we estimated the statistical errors in the metallicities by pooled variances of the published [Fe/H] values for each instrument-author combination, using repeated [Fe/H] measurements for the same objects as individual pools. For those few cases where this approach was not possible, we adopted the uncertainties provided by the authors, or in their total absence, we assumed an uncertainty of 0.1~dex.

According to \citet{crestani_deltaS}, the CFCS sample is comprised of homogeneous metallicities, which means that they were obtained by the same method using the same input physics. However, this might not be the case for the rest of the data in our combined spectroscopic sample, therefore it was necessary to bring all the other subsamples to the CFCS scale. Assuming a simple metallicity-independent transformation between the CFCS sample and all other results from the literature, we estimated the systematic shifts between the data sets by taking the mean of the [Fe/H] differences between the individual measurements for common objects in both compared samples, weighted by the adopted uncertainties discussed earlier. The resulting shifts range from $-0.26$~dex to $+0.16$~dex, and presumably arise from the combination of systematic differences between the adopted Solar metallicities, as well as slightly different input physics and spectral fitting algorithms. 

Table~\ref{tab:specdata} lists the individual [Fe/H] values, their uncertainties, and the computed shifts ($\Delta$) with respect to the CFCS data, and the corresponding spectroscopic references for the objects that also have $I$-band photometric data available from the literature.

\begin{deluxetable*}{ccccl}
	%\tablenum{1}
	%\movetabledown=1in
	\tablecaption{Spectroscopic metallicities of RR~Lyrae stars compiled from the literature \label{tab:specdata}}
	%	\tablewidth{0pt}
	%\tablewidth{700pt}
	\tablehead{
	\colhead{Name} & \colhead{[Fe/H]} & \colhead{$\Delta$} & \colhead{ref.} & \colhead{instrument}   
	}
	%\decimalcolnumbers
	\startdata
	AA Aql & $-0.49 (.10)$ &  0 &  C21  &  Subaru  \\
	AA Aql & $-0.34 (.06)$ &  0 &  C21  &  SALT    \\
	AA Aql & $-0.32 (.04)$ &  0 &  L13  &  Subaru  \\   
	AA Aqr & $-2.31 (.10)$ &  0 &  C21  &  UVES    \\
	AE Boo & $-1.62 (.17)$ &  0 &  C21  &  HARPS   \\
	\enddata
	\tablecomments{This table is available in its entirety in machine-readable form.}
\end{deluxetable*}

\section{Predictive modeling of the metallicity} \label{sec:prediction}

The combined photometric--spectroscopic data set discussed in Sect~\ref{sec:data} was further narrowed down by selecting objects with phase coverage of $C>0.8$, along with $S/N>100$ and $S/N>70$ in the $I$-band light curves for RRab and RRc stars, respectively, in order to constrain the sample to stars with unbiased light-curve parameters. The remaining data were visually inspected, and a few additional objects were culled where undersampled photometry was suspected to bias their light curves. Some of the manually rejected objects had phase gaps overlapping their light-curve minima or maxima, others showed strong systematics in their light curves due to undersampled (semi-)periodic modulation known as the Blazhko effect \citep[see, e.g., ][for a review]{2016pas..conf...22S}. In case of strong amplitude/phase modulation, an incomplete coverage of the Blazhko cycle can result in a light-curve model that is significantly biased with respect to the mean light variation of the star, justifying its omission.

The resulting {\it development data set} comprises 379 individual spectroscopic [Fe/H] measurements of 82 RRab stars ranging from $-2.6$ to $+0.2$ dex, 22 of which have single, and 60 have multiple independent metallicity determinations; as well as 147 individual [Fe/H] values of 24 RRc stars ranging from $-3$ to $-0.5$ dex, among which 20 have multiple measurements. Importantly, there are only 2 RRc [Fe/H] measurements above $-0.9$ dex, currently posing a serious limitation to any photometric predictive model of the metallicity for these objects.  We also note that according to both the respective authors of the photometric databases that form the basis of our study, as well as our own assessment, our development set contains 24 RRab and 6 RRc stars with (adequately sampled) Blazhko modulation. They are marked in the fifth column of Table~\ref{tab:rrlyraes}. For the large majority of these objects, the size of the modulation is rather small. Their possible influence on the metallicity prediction will be investigated in the following sections.

We divide the regression problem of predicting the [Fe/H] from the $I$-band light curves into two sub-problems: feature and model selection, and final estimation of the model parameters. In the latter, given a regression model type and an optimal set of features (i.e., descriptive variables), we estimate the final model parameters by Bayesian regression, which has two main advantages. Firstly, it allows us to estimate credible intervals and covariances from the posterior distributions instead of mere point estimates of the regression model's parameters; and secondly, it provides a means to properly take the heteroskedastic uncertainties in the descriptive variables into account by a probabilistic modeling of latent regressors. However, due to the costly nature of Bayesian inference, we use a more conventional, frequentist's approach for the preceding feature and model selection procedure.

We opted to use the individual metallicity measurements in the predictive modeling instead of binning the spectroscopic data into one [Fe/H] value per object, in order to avoid introducing biases in the error distributions by small-number statistics for objects with only a few measurements. In addition, using single, unbinned metallicity measurements provides natural sample weights for the target variable.

\subsection{Feature and model selection} \label{subsec:fmsel}

For selecting the optimal set of features, we considered both the Fourier and the PCA representations of the GPR light-curve models discussed in Section~\ref{subsec:lightcurves}. In both scenarios, we considered linear least-squares models with either linear or second-order polynomial basis functions. We searched for the best-performing regression model using the Sequential Feature Selection (SFS) algorithm implemented in the {\tt scikit-learn} software library. The SFS is a greedy algorithm, attempting to find the global optimum of the model performance, measured by some metric, through a series of local optima. From a pool of $N$ features, it selects a subset of $M$ features by sequentially adding (forward-SFS) or removing (backward-SFS) the feature to/from the previous subset that leads to the next local optimum of the chosen performance metric measured by cross-validation (CV). For example, a subset of $M=3$ features from a pool of $N=6$ is selected in 3 iterations by forward-SFS. First, the algorithm selects one feature by evaluating the metric for all 6 possible univariate models by CV, and selects the best-performing one. In the second iteration, the second feature is selected by evaluating all 5 possible bivariate models that include the feature selected in the first iteration, and choosing the best performing one. The final model is selected in the last iteration by evaluating the 4 possible multivariate models with 3 features that include the previously selected two. Likewise, backward-SFS also performs 3 iterations in the $N=6$, $M=3$ case, but performing more computations overall, and it may lead to different feature subsets. The best feature set is searched by running the algorithm for different values of $M$, evaluating each model by CV, and selecting the best-performing one.

The forward-SFS algorithm was applied to both the RRab and RRc data using the mean absolute error (MAE) as performance metric, estimated by 20-fold CV repeated 10 times, selecting up to 6 features in each case. We found that the RRab stars SV~Eri and AN~Ser were strong outliers in all cases, with deviations of $\sim$0.6~dex from all trial models, and were consequently removed from the development data set. Their peculiar behavior cannot be related to their photometry, since both stars have accurate, well-sampled, unmodulated $I$-band light curves, and their photometric properties in the Gaia EDR3 database do not indicate any hint of binarity or blend. The metal-poor SV~Eri ($[{\rm Fe/H}]=-2.22$, \citealt{crestani_deltaS}), however, is known to stand out in terms of its unusually large period increase, and is suspected to be rapidly crossing the instability strip in the final stages of its horizontal-branch evolution \citep{2007A&A...476..307L}. The metal-rich AN~Ser on the other hand, has unusually low alpha-element abundance ($[{\rm Fe/H}]=+0.05$ and $[\alpha/{\rm Fe}]=-0.2$ according to \citealt{chadid_spectroscopic_2017}). AN~Ser has 3 independent metallicity determinations, while only one [Fe/H] measurement was reported by \citet{crestani_deltaS} for SV~Eri.

\begin{figure*}
	\gridline{
		\fig{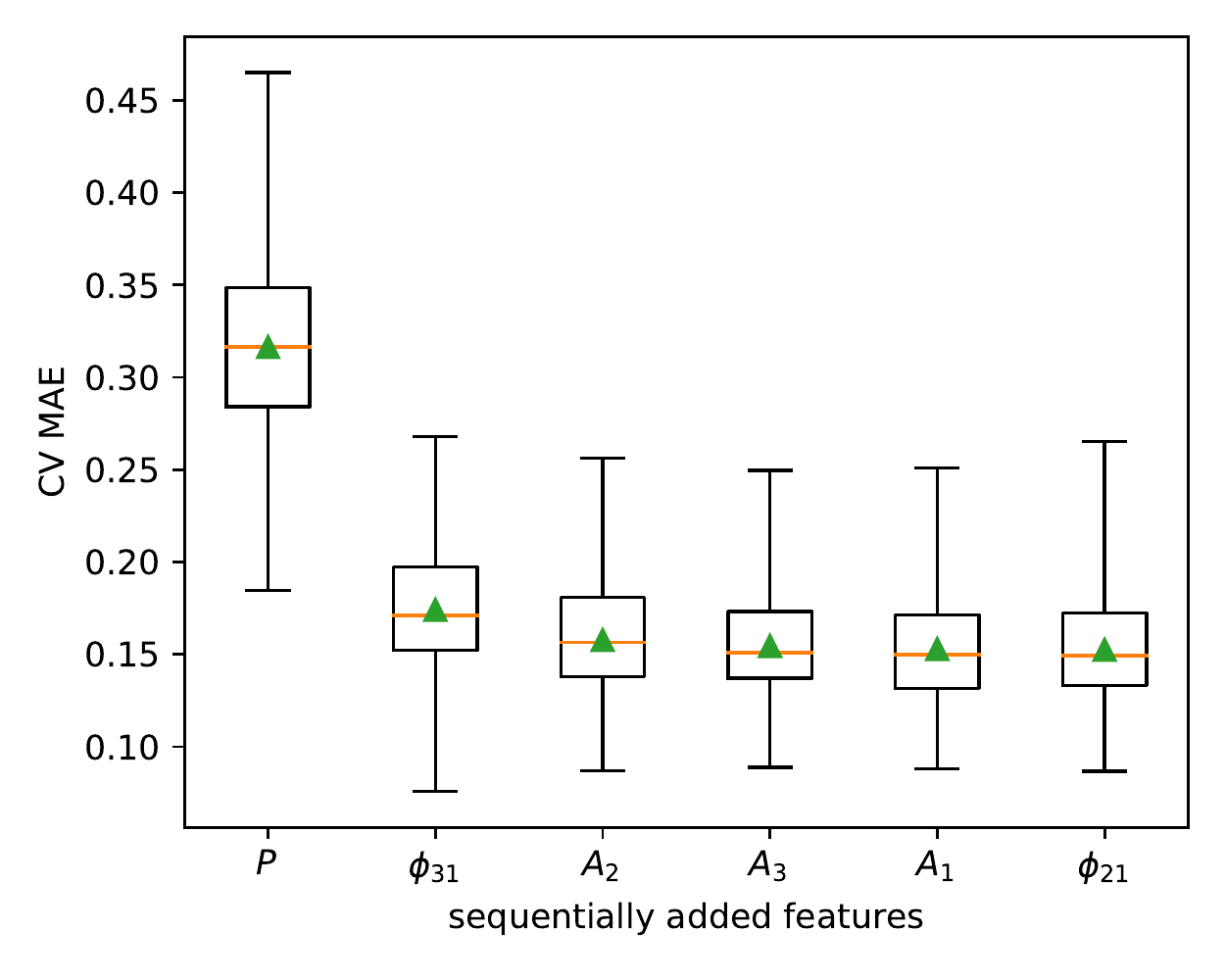}{0.3\textwidth}{(a)}
		\fig{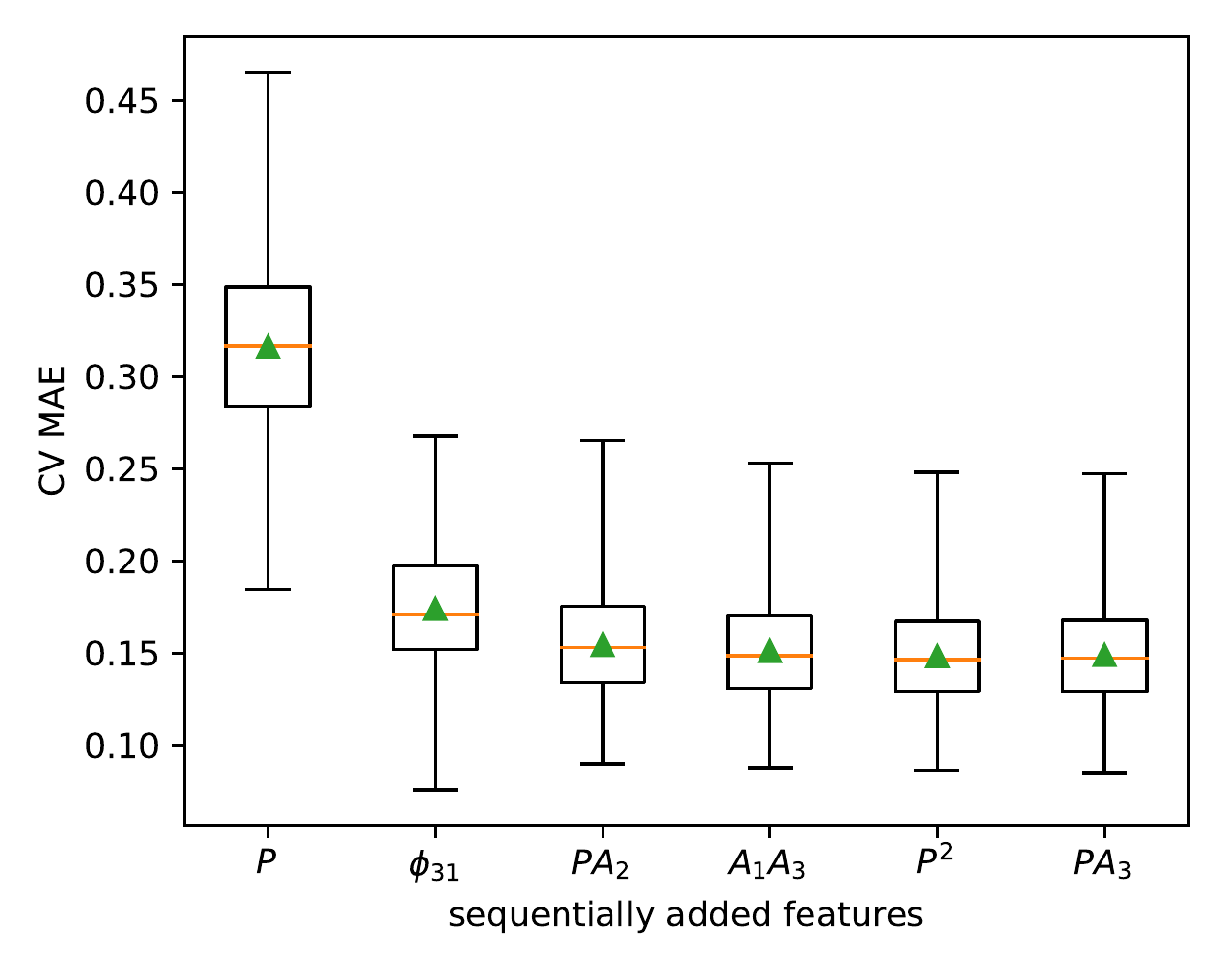}{0.3\textwidth}{(b)}
	}
	\gridline{
		\fig{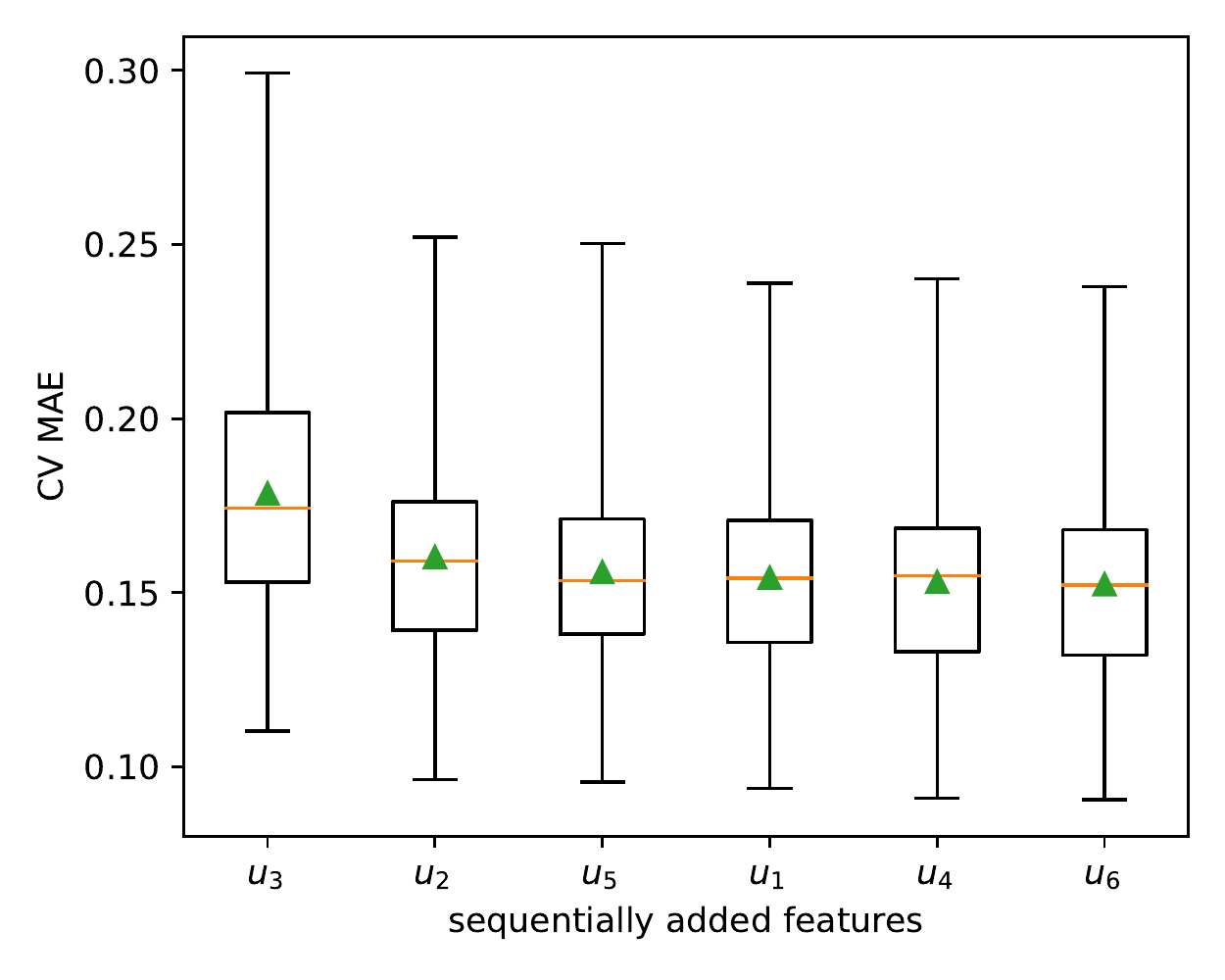}{0.3\textwidth}{(c)}
		\fig{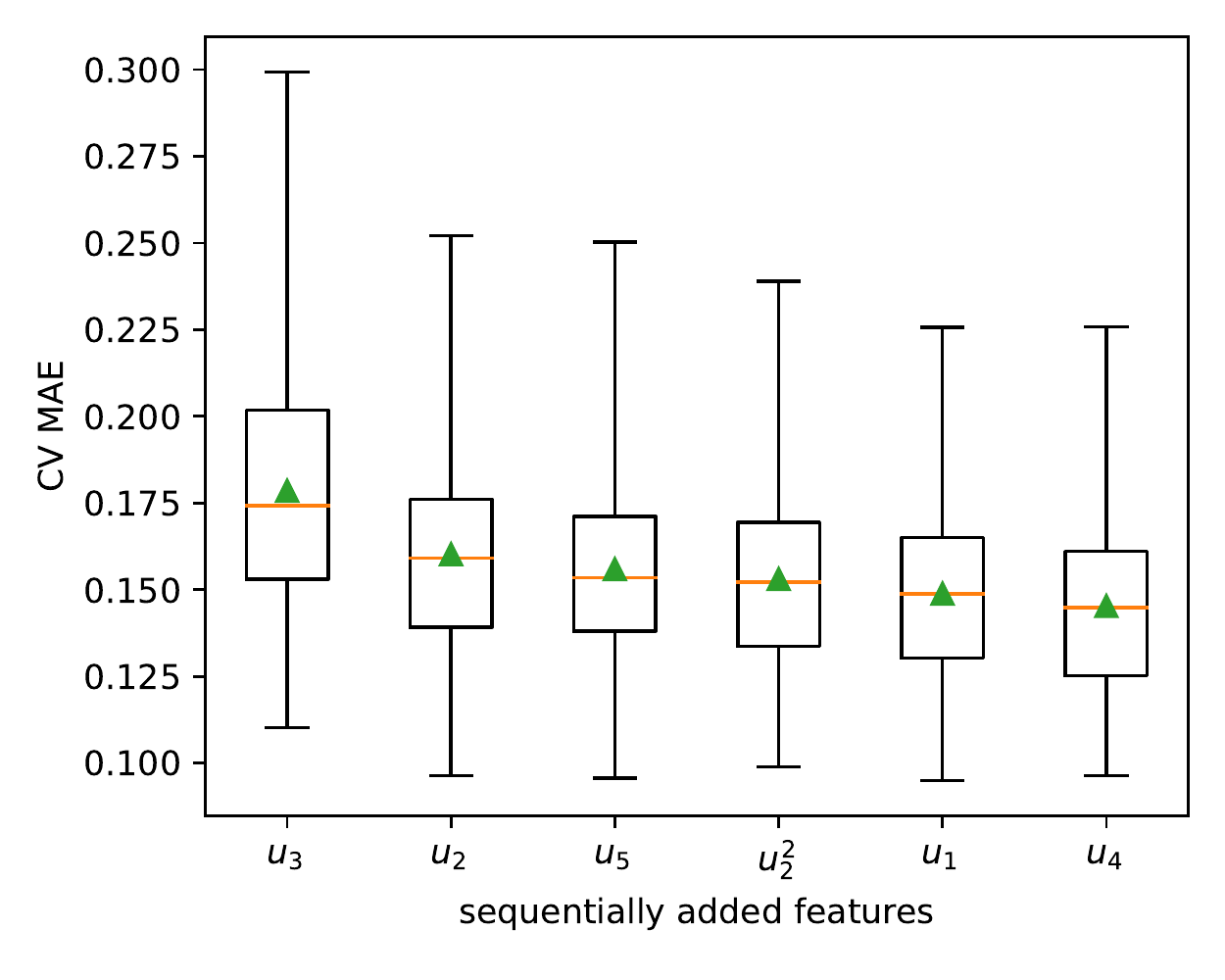}{0.3\textwidth}{(d)}
	}
	\caption{Boxplots of the distributions of the MAE measured by CV for different feature subsets for the RRab stars. The labels on the horizontal axes show the sequentially added features selected to be optimal by the SFS algorithm, while increasing the number of used features from 1 to 6 (left to right in each panel). The rectangular boxes encompass the interquartile ranges of the distributions with lines at the medians, with horizontal orange lines at the medians and green triangles at the means. The whiskers indicate the full ranges of the distributions. Panels (a) and (b) show the results from linear least-squares models using linear and quadratic Fourier parameters as their feature space, respectively. Panels (c) and (d) show the same but for the feature spaces transformed by PCA (see Sect.~\ref{subsubsec:pca}).
		\label{fig:sfs_perf_rrab}}
\end{figure*}

\begin{figure*}
	\gridline{
		\fig{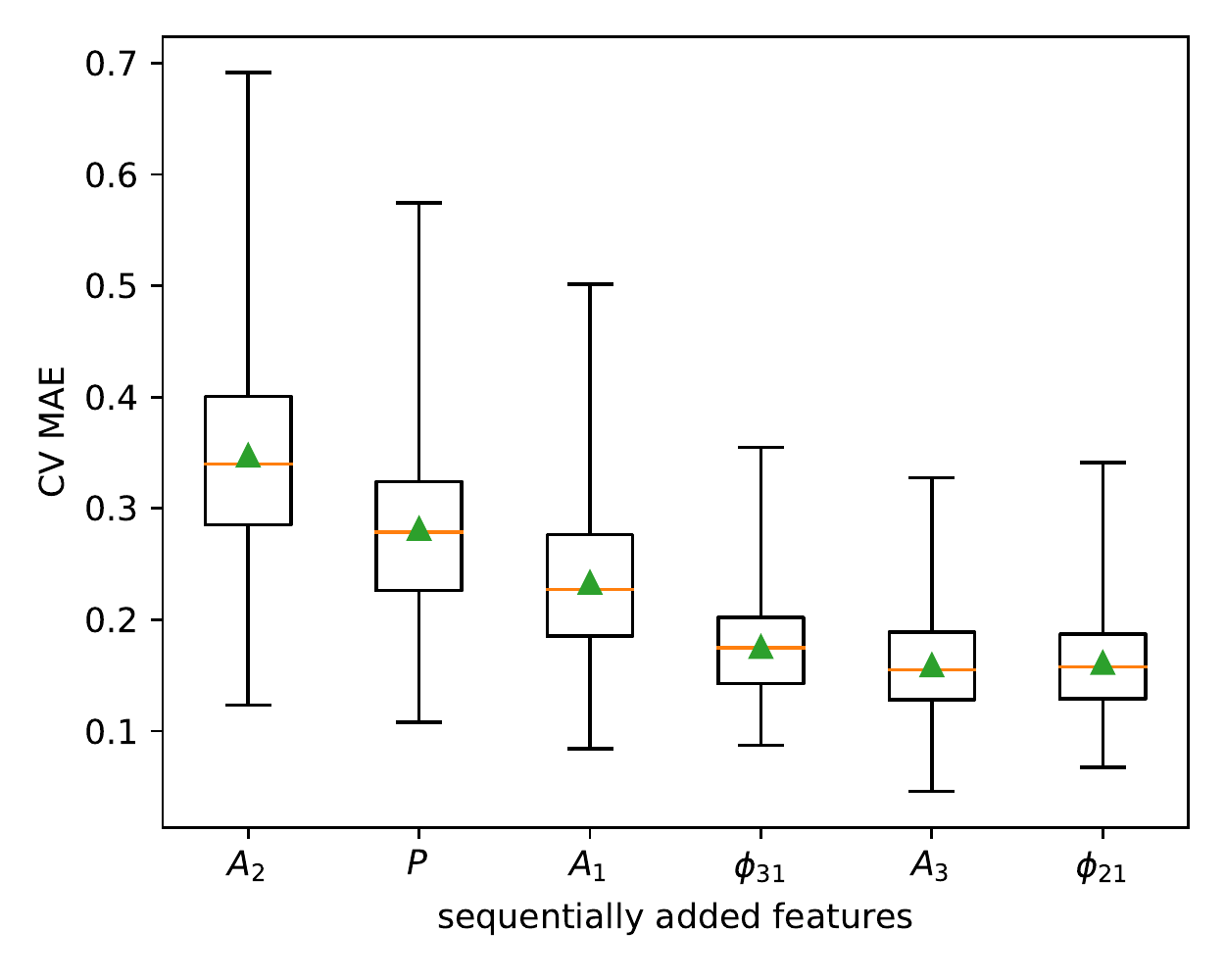}{0.3\textwidth}{(a)}
		\fig{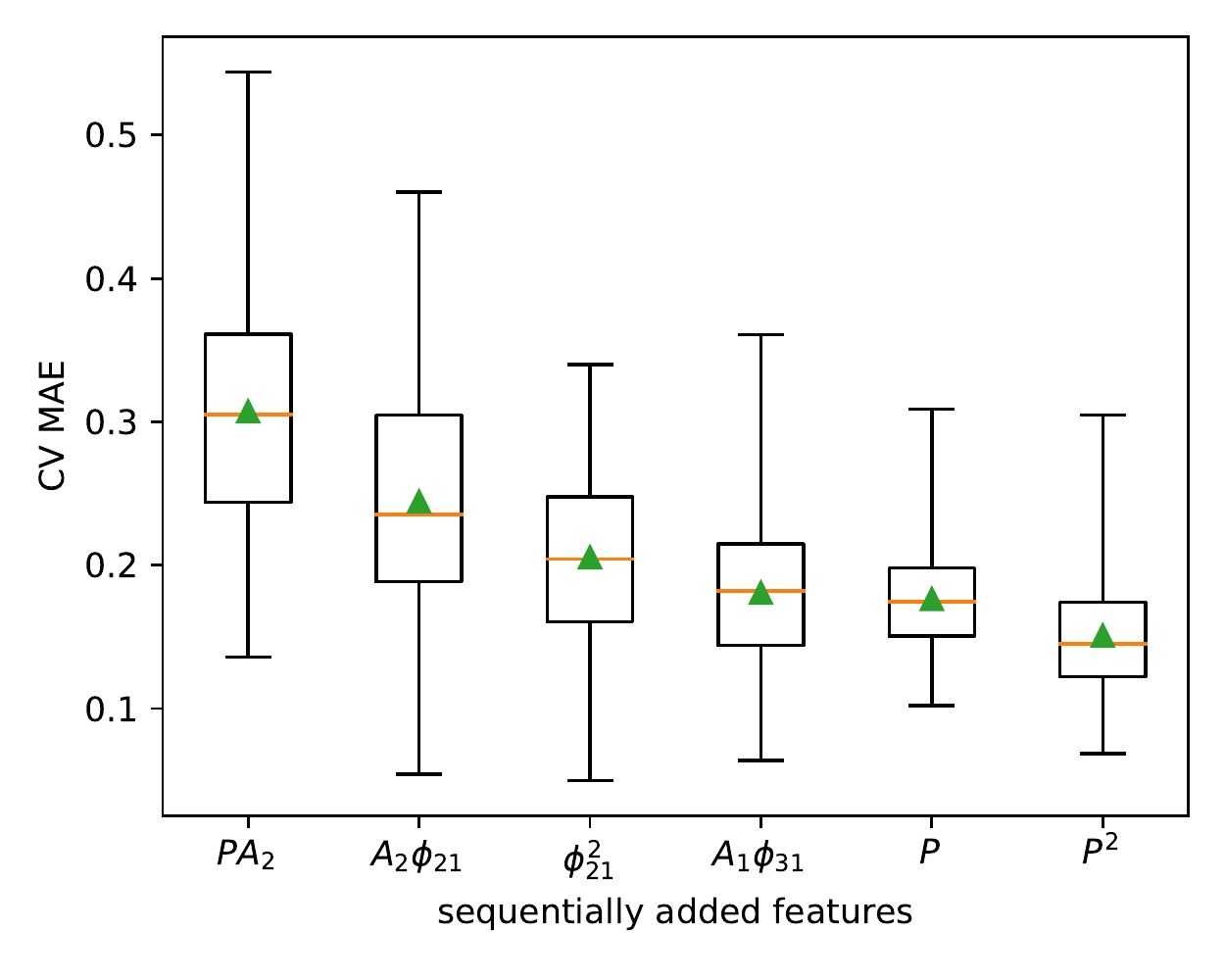}{0.3\textwidth}{(b)}
	}
	\gridline{
		\fig{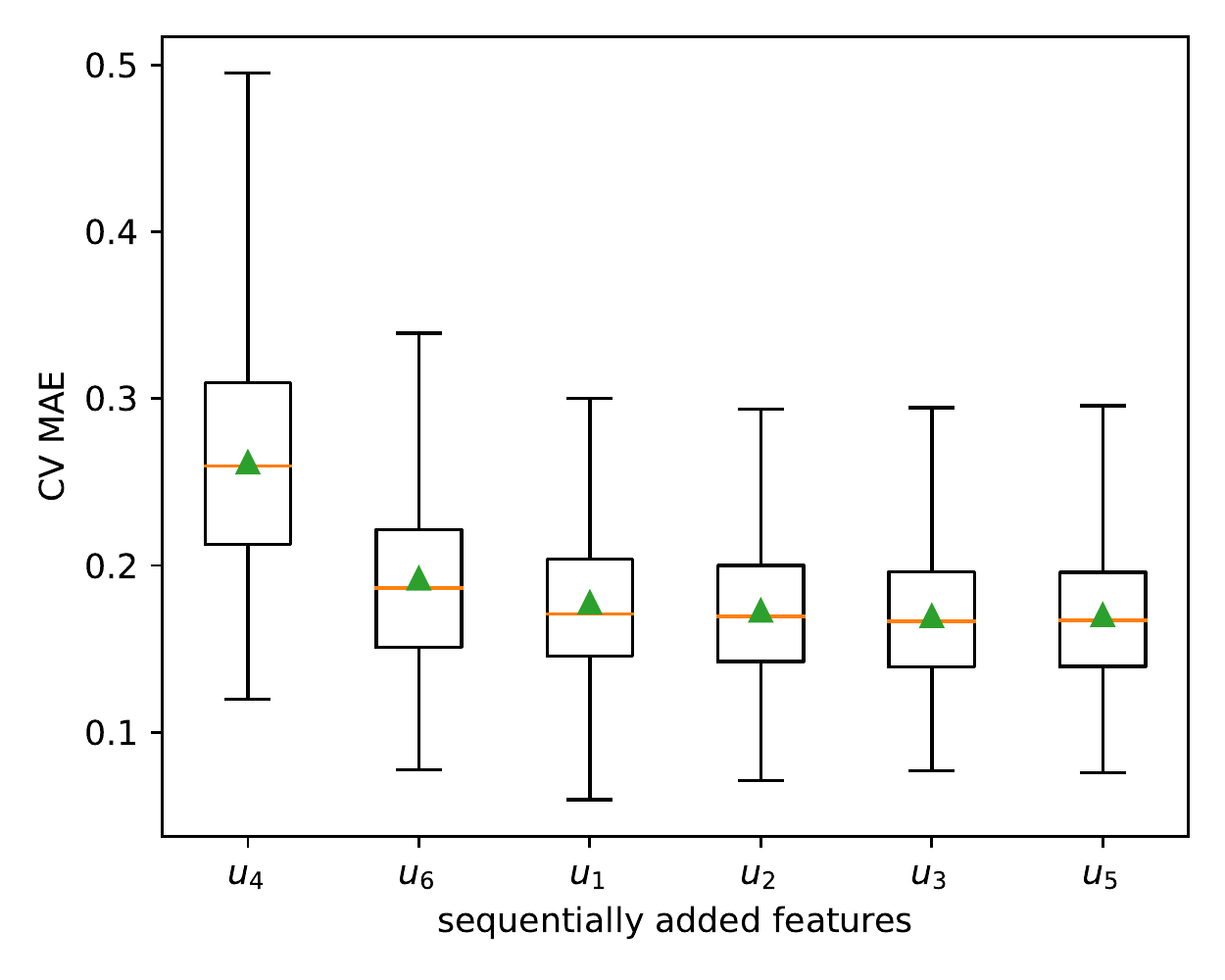}{0.3\textwidth}{(c)}
		\fig{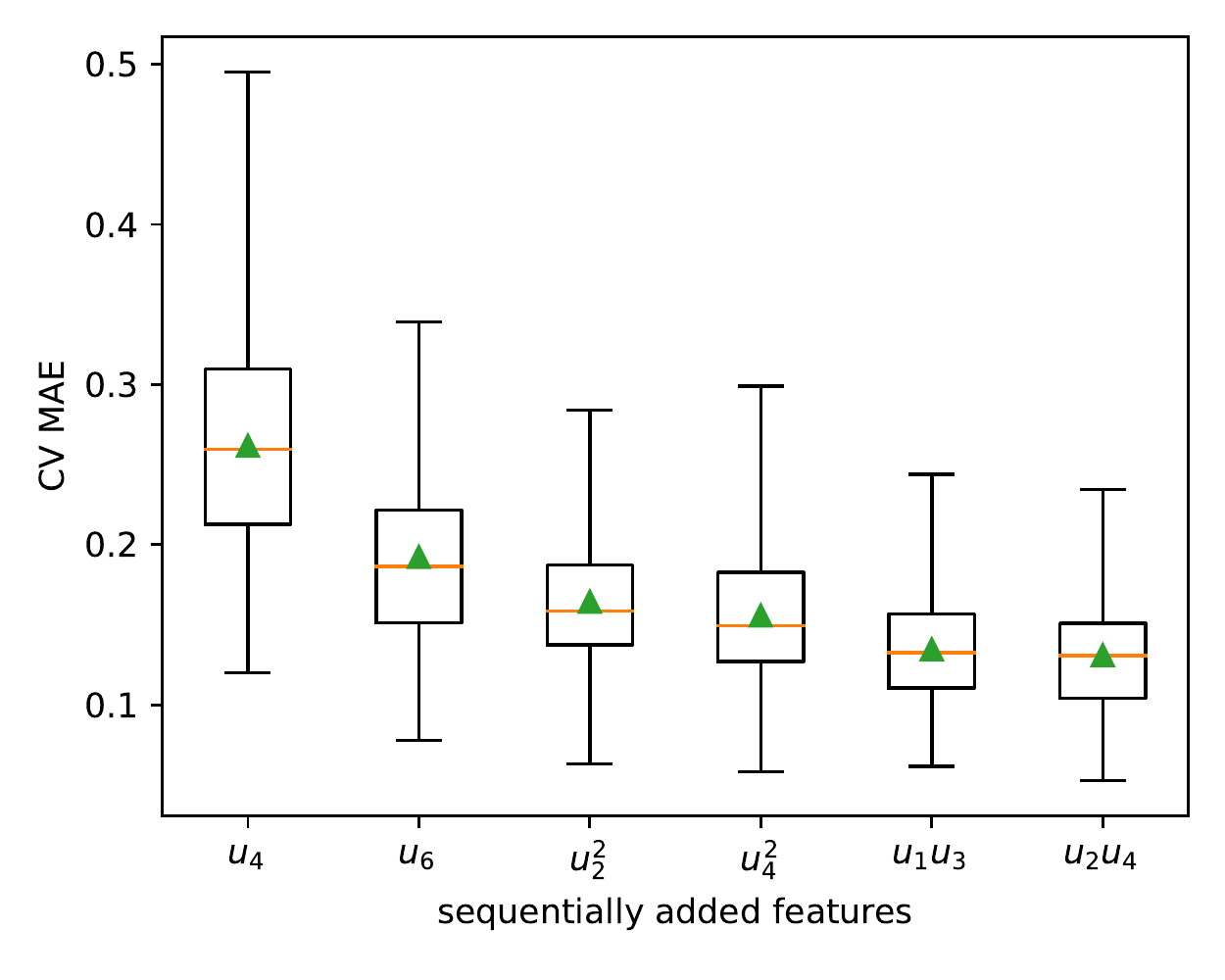}{0.3\textwidth}{(d)}
	}
	\caption{The same as Fig~\ref{fig:sfs_perf_rrab} but for RRc stars.
		\label{fig:sfs_perf_rrc}}
\end{figure*}

Figures~\ref{fig:sfs_perf_rrab} and \ref{fig:sfs_perf_rrc} show the distributions of the MAE from CV for the different feature subsets selected by SFS, while Table~\ref{tab:performance} summarizes the results. For RRab stars, the two most important Fourier parameters for predicting the [Fe/H] are the period and the $\phi_{31}$, while adding $A_2$ as a third feature consistently improves the model's performance, independently from the random realization of the CV splitting. However, adding more than 3 features by SFS, the predictive performance measured by MAE does not improve beyond its uncertainty. These findings are fully consistent with the earlier analysis by \citet{smolec_metallicity_2005}. Using quadratic basis functions, a virtually identical performance can be achieved by using the $\{P,\phi_{31}, PA_2\}$ feature subset, with no significant further improvement by adding more features. Using the linearly uncorrelated $u_i$ parameters of the principal component feature space leads to similar results. The combination of $u_3$ and $u_2$ in a simple linear model provides good predictive performance, which is marginally improved by adding $u_5$, while the rest of the features proved unimportant in predicting the metallicity. However, there is no clear benefit in using linearly uncorrelated features instead of the standard Fourier parameters and no improvement can be attained by using a second-order polynomial basis, either. We concluded that for the RRab stars, based on the currently available data, a three-parameter linear model using the classical $\{P,\phi_{31}, A_2\}$ feature set is optimal for the prediction of the metallicity.

\begin{deluxetable*}{cc|cll|cll}
	%\tablenum{1}
	%\movetabledown=1in
	\tablecaption{Performances of various regression models and feature sets for predicting the [Fe/H]  of RR~Lyrae stars\label{tab:performance}}
	%	\tablewidth{0pt}
	%\tablewidth{700pt}
	%\tabletypesize{\tiny}
	\tablehead{
		\multicolumn2c{} \vline & \multicolumn3c{RRab} \vline & \multicolumn3c{RRc} \\
		\colhead{model} & \colhead{$M$} \vline & \colhead{$\langle$MAE$\rangle$} & \colhead{$\sigma$(MAE)} & \colhead{features} \vline  & \colhead{$\langle$MAE$\rangle$} & \colhead{$\sigma$(MAE)} & \colhead{features}
	}
	%\decimalcolnumbers
	\startdata
	\multirow{6}{*}{Fourier(1)} 
	& 1  & 0.316 & 0.050 & $\{P\}$ & 
								0.348 & 0.088 & $\{A_2\}$ \\
	& 2  & 0.174 & 0.032 & $\{P,\phi_{31}\}$ &
								0.282 & 0.086 & $\{A_2,P\}$ \\
	& 3  & 0.158 & 0.031 & $\{P,\phi_{31},A_2\}$ & 
								0.226 & 0.068 & $\{A_2,P,A_1\}$ \\
	& 4  & 0.154 & 0.030 & $\{P,\phi_{31},A_2,A_3\}$ & 
								0.182 & 0.039 & $\{A_2,P,A_1,\phi_{31}\}$ \\
	& 5  & 0.153 & 0.030 & $\{P,\phi_{31},A_2,A_3,A_1\}$ & 
								0.170 & 0.041 & $\{A_2,P,A_1,\phi_{31},A_3\}$ \\
	& 6  & 0.152 & 0.031 & $\{P,\phi_{31},A_2,A_3,A_1,\phi_{21}\}$ & 
								0.171 & 0.041 & $\{A_2,P,A_1,\phi_{31},A_3,\phi_{21}\}$ \\
	\hline                                                                                                        
	\multirow{6}{*}{Fourier(2)} 
	& 1  & 0.316 & 0.050 & $\{P\}$ & 
			0.307 & 0.080 & $\{PA_2\}$ \\    
	& 2  & 0.174 & 0.032 & $\{P,\phi_{31}\}$ & 
			0.244 & 0.083 & $\{PA_2,A_2\phi_{21}\}$ \\
	& 3  & 0.154 & 0.030 & $\{P,\phi_{31},PA_2,\}$ & 
			0.205 & 0.060 & $\{PA_2,A_2\phi_{21},\phi_{21}^2\}$ \\
	& 4  & 0.151 & 0.029 & $\{P,\phi_{31},PA_2,A_1A_3,\}$ & 
			0.181 & 0.051 & $\{PA_2,A_2\phi_{21},\phi_{21}^2,A_1\phi_{31}\}$ \\
	& 5  & 0.148 & 0.029 & $\{P,\phi_{31},PA_2,A_1A_3,P^2\}$ & 
			0.176 & 0.037 & $\{PA_2,A_2\phi_{21},\phi_{21}^2,A_1\phi_{31},P\}$ \\
	& 6  & 0.149 & 0.029 & $\{P,\phi_{31},PA_2,A_1A_3,P^2,PA_3\}$ & 
			0.151 & 0.041 & $\{PA_2,A_2\phi_{21},\phi_{21}^2,A_1\phi_{31},P,P^2\}$ \\
	\hline                                                                                                       
	\multirow{6}{*}{PCA(1)}     
	& 1  & 0.179 & 0.036 & $\{u_3\}$ & 
			0.262 & 0.072 & $\{u_4\}$ \\
	& 2  & 0.160 & 0.029 & $\{u_3,u_2\}$ & 
			0.192 & 0.053 & $\{u_4,u_6\}$ \\
	& 3  & 0.156 & 0.029 & $\{u_3,u_2,u_5\}$ & 
			0.178 & 0.044 & $\{u_4,u_6,u_1\}$ \\
	& 4  & 0.154 & 0.027 & $\{u_3,u_2,u_5,u_1\}$ & 
			0.173 & 0.041 & $\{u_4,u_6,u_1,u_2\}$ \\
	& 5  & 0.153 & 0.027 & $\{u_3,u_2,u_5,u_1,u_4\}$ & 
			0.170 & 0.043 & $\{u_4,u_6,u_1,u_2,u_3\}$ \\
	& 6  & 0.152 & 0.027 & $\{u_3,u_2,u_5,u_1,u_4,u_6\}$ & 
			0.170 & 0.043 & $\{u_4,u_6,u_1,u_2,u_3,u_5\}$ \\
	\hline                                                                                                    
	\multirow{6}{*}{PCA(2)}     
	& 1  & 0.179 & 0.036 & $\{u_3\}$ & 
			0.262 & 0.072 & $\{u_4\}$ \\
	& 2  & 0.160 & 0.029 & $\{u_3,u_2\}$ & 
			0.192 & 0.053 & $\{u_4,u_6\}$ \\
	& 3  & 0.156 & 0.029 & $\{u_3,u_2,u_5\}$ & 
			0.164 & 0.041 & $\{u_4,u_6,u_2^2\}$ \\
	& 4  & 0.153 & 0.028 & $\{u_3,u_2,u_5,u_2^2\}$ & 
			0.156 & 0.042 & $\{u_4,u_6,u_2^2,u_4^2\}$ \\
	& 5  & 0.149 & 0.026 & $\{u_3,u_2,u_5,u_2^2,u_1\}$ & 
			0.134 & 0.035 & $\{u_4,u_6,u_2^2,u_4^2,u_1u_3\}$ \\
	& 6  & 0.145 & 0.026 & $\{u_3,u_2,u_5,u_2^2,u_1,u_4\}$ & 
			0.131 & 0.035 & $\{u_4,u_6,u_2^2,u_4^2,u_1u_3,u_2u_4\}$ \\
	\enddata
	%\tablecomments{}
\end{deluxetable*}

For RRc stars, there is a gradually diminishing performance increase by adding up to 5 Fourier parameters as features, including the amplitudes of the first three Fourier terms. However, the amplitude of the third term can become hard to accurately measure for some RRc stars in the absence of precision photometry. For this reason, and due to the modest improvement of the MAE by including $A_3$, we restrict our linear regression model to the $\{P, A_1, A_2, \phi_{31}\}$ feature set. In case of including second-order polynomial features, SFS leads to a similar optimum by selecting $\{PA_2, A_2\phi_{21}, \phi_{21}^2, A_1\phi_{31}\}$, and adding more terms slightly further improves the performance. Similar results can be attained with the $\{u_4,u_6,u_1\}$ feature subset, without significant further improvement by increasing the number of features. Using a quadratic formula however, the mean MAE shows a gradual decrease down to $\sim$0.13~dex with 5 features. However, since such a complicated formula is highly error-prone in case of noisy light curves, and because the selection of features 3--5 is fairly sensitive to the random initialization of the CV in the SFS, we constrain our further analysis to the linear Fourier features $\{P, A_1, A_2, \phi_{31}\}$.

%
%Finally, there is no advantage of using the $u_i$ orthogonal feature space with respect to the standard Fourier parameters, as can be seen in the bottom panels of Fig.~\ref{fig:sfs_perf_rrc}.

In order to probe whether the Blazhko effect systematically affects the feature selection, we also carried out the same analysis after the exclusion of all objects  that showed signs of modulation in their light curves, resulting in modified development sets of 245 and 116 individual metallicity measurements of 56 RRab and 18 RRc stars, respectively. In spite of the significant decrease in the amount of data, this modification did not cause any significant change with respect to the original analysis, as far as the selection of linear features are concerned. In case of quadratic features however, SFS converges to different local optima for $M>2$, while becoming more volatile to the random initialization of the algorithm. This however does not cause any significant change in the relative performances of the models and is  more likely a direct effect of the reduced data set sizes (particularly for the RRc stars), rather than due to the exclusion of the Blazhko stars.

From the above analysis, we draw as conclusion that based on the currently available data, linear  models using the Fourier parameters as regressors yield satisfactory performance in predicting the metallicity for both RRab and RRc stars. The possibility of improvement by using more subtle features and/or more complicated models is not ruled out, but is not supported by the data, and should be further explored, once significantly more observational data become available. We select $\{P,\phi_{31}, PA_2\}$ and $\{P, A_1, A_2, \phi_{31}\}$ as optimal feature subsets for RRab and RRc stars, respectively. In the following Section, we determine the parameters of the corresponding linear predictive models of the [Fe/H] by Bayesian regression.

\subsection{Bayesian Regression of the Metallicity}\label{subsec:bayesian}

In the Bayesian approach to regression problems, parameters of a predictive model are considered to be random variables, and their optimal (i.e., most probable) values are inferred from posterior densities conditional on observed data. An important advantage of this approach over frequentist methods is that uncertainties in the model parameters can be expressed as credible intervals, and their posterior distributions can be used for making further inferences, e.g., to analyze the propagation of error in the model's predictions. Another key benefit of Bayesian regression lies in the possibility of taking the errors in the covariate features explicitly into account. In frequentist approaches, the descriptive variables are considered to be fixed values, which may lead to a biased regression model, e.g., in case they stem from diverse measurements or other processes that lead to heteroskedastic uncertainties. The latter problem clearly affects the regression of the metallicity to light-curve parameters, since the uncertainties in the latter have a large spread as a result of the heterogeneous quality of the photometric time-series acquired by different instruments under diverse conditions for objects with different apparent magnitudes.

The uncertainties in the covariates are introduced into our probabilistic model via latent regressors $\xi_i$:
\begin{eqnarray}
	\nonumber y_i &=& f(\bm{\xi}_i,\bm{\theta}) + \epsilon_i \\	
	        \bm{\xi}_i &=& \bm{x}_i + \bm{\nu}_i
\end{eqnarray}
\noindent where $(x_i,y_i)$ are the observed covariates and response data of the $i$-th example respectively; $f$ is the functional form of the regression model with parameters $\bm{\theta}$; $\bm{\xi}_i$ are the latent (i.e., the {\em actual}, unobservable) covariate features, and the $\epsilon_i$ and $\bm{\nu}_i$ terms are random variables representing the stochasticity in the model. In our specific regression problem, $\bm{x}_i$ is the vector of light-curve features (i.e., the $\{P,\phi_j,A_j\}$ or $\{u_j\}$ parameters), while $y_i$ is (are) the spectroscopic [Fe/H] measurement(s) of the $i$-th RR~Lyrae star in the development data set.

Inferences about the $\bm{\theta}$ and $\bm{\xi}_i$ random variables, conditional on $(x_i,y_i)$, are drawn using Bayes' theorem:
\begin{equation}
	P(\bm{\theta}, \bm{\xi} | \bm{y}, \bm{x}) \propto P(\bm{y},\bm{x}|\bm{\xi},\bm{\theta}) \cdot P(\bm{\theta},\bm{\xi} | \bm{x})
\end{equation}
\noindent Here, the first term on the right side of the equation is the joint likelihood of the observed data given the model parameters and the latent regressors, and the second term is the prior probability distribution of the latter, given the observed covariates. Their product is proportional to the joint posterior density of $\bm{\theta}$ and $\bm{\xi}_i$. We assume that $\epsilon_i$ and $\bm{\nu_i}$ follow zero-mean normal distributions with variances of the former made equal to the squared error estimates of the [Fe/H] measurements (see Sect~\ref{subsec:spectroscopy}). Likewise, we use informative priors on $\bm{\xi}_i$ by setting the variances of $\bm{\nu_i}$ to be equal to the error estimates of the light-curve parameters obtained from the GPR models, as discussed in Sect.~\ref{subsubsec:gpr}. For the elements of $\bm{\theta}$, we pose weakly informative priors assuming normal distributions with means obtained by ordinary linear least-squares regression. 

In view of our conclusions from Sect.~\ref{subsec:fmsel}, we implemented the above probabilistic regression model using the {\tt pymc3} software library for the two types of RR~Lyrae stars in the following forms:

\vskip5mm
\noindent RRab:
\begin{equation}
[{\rm Fe}/\rm{H}] = \theta_P P + \theta_{\phi_{31}}\phi_{31} + \theta_{A_2}A_2 + \theta_0
\end{equation}

\noindent RRc:
\begin{equation}
[{\rm Fe}/\rm{H}] = \theta_P P + \theta_{\phi_{31}}\phi_{31} + \theta_{A_1}A_1 + \theta_{A_2}A_2 + \theta_0
\end{equation}

\noindent Posterior distributions were computed numerically via Hamiltonian Markov chain Monte Carlo simulations using the No-U-Turn Sampler (NUTS) of \citet{hoffmann_nuts}. We performed the sampling in multiple parallel Markov chains and assessed convergence numerically through the $\hat R$ statistic and visually using trace- and posterior density plots using the {\tt ArviZ} software package.

\begin{figure*}
	\plotone{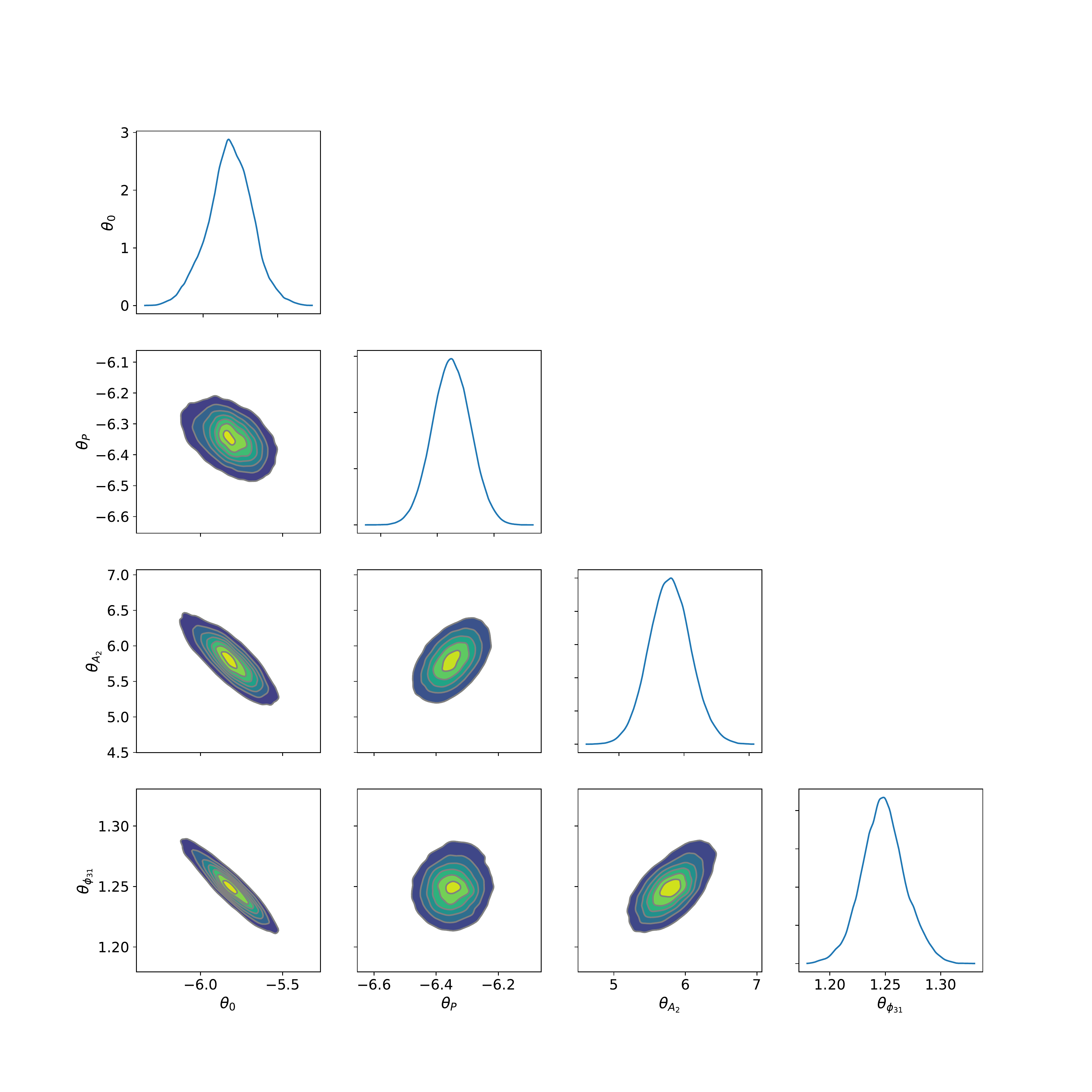}
	\caption{Posterior distributions of the parameters of the metallicity prediction model for RRab stars. Kernel density estimates of marginal posterior MCMC samples are shown for single parameters and all possible parameter pairs.
		\label{fig:posterior_rrab}}
\end{figure*}

\begin{figure*}
	\plotone{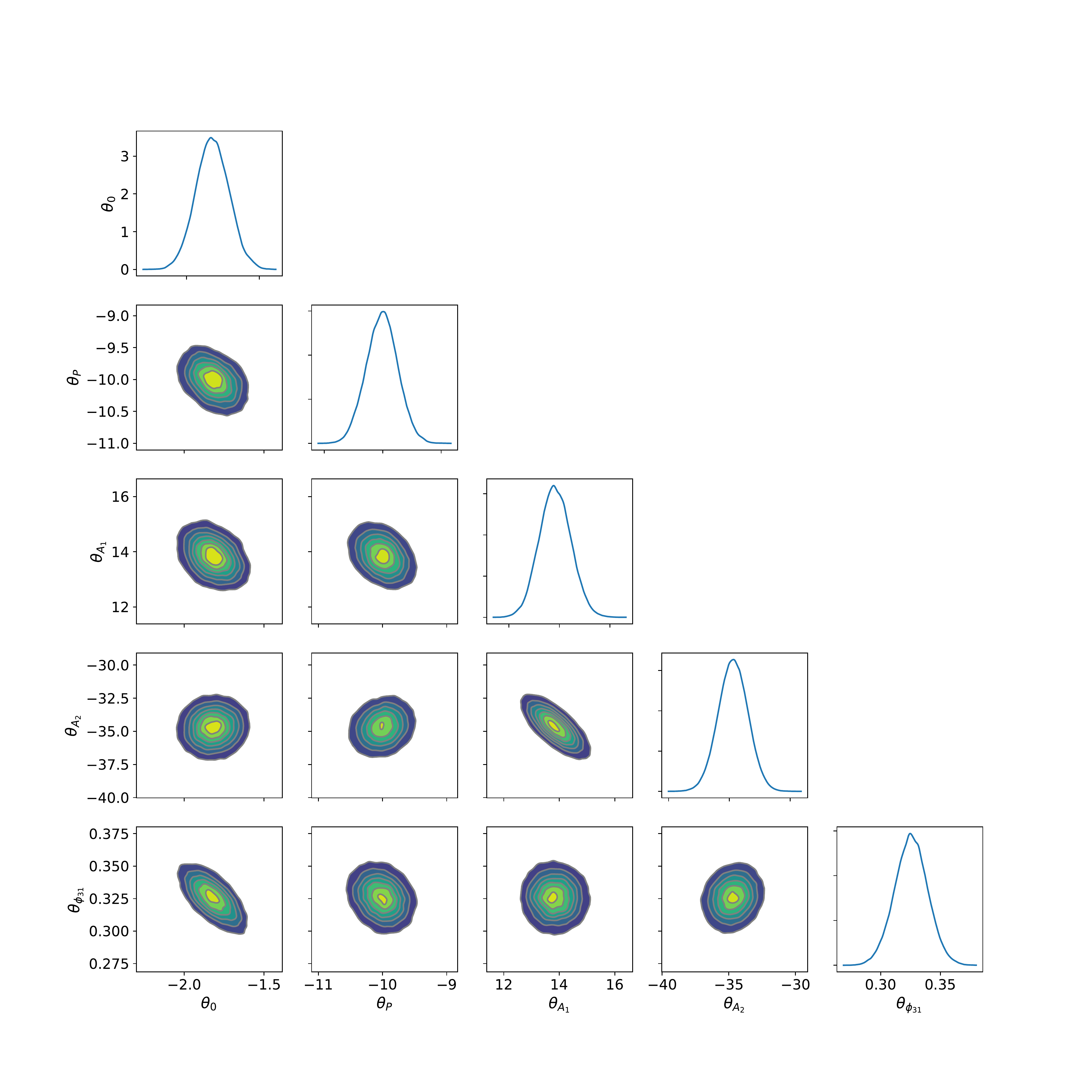}
	\caption{The same as Fig.~\ref{fig:posterior_rrab} but for RRc stars.
		\label{fig:posterior_rrc}}
\end{figure*}

Figures~\ref{fig:posterior_rrab} and \ref{fig:posterior_rrc} show marginal and paired kernel density estimates of the resulting  posterior distributions for the components of $\bm{\theta}$ for RRab and RRc stars, respectively. Credible intervals were determined by computing highest density intervals (HDIs) of the marginal posteriors for $90\%$ and $95\%$ probabilities. Table~\ref{tab:posterior_stat} summarizes the statistics of the posterior distributions of our [Fe/H] regression models' parameters.

\begin{deluxetable*}{l|lrrrrrr}
	%\tablenum{1}
	%\movetabledown=1in
	\tablecaption{Posterior distribution statistics of the [Fe/H] regression models' parameters 
		\label{tab:posterior_stat}}
	%	\tablewidth{0pt}
	%\tablewidth{700pt}
	\tablehead{
		\colhead{Type} \vline & \colhead{Parameter} & \colhead{mean} & \colhead{$\sigma$\tablenotemark{a}} 
		& \colhead{HDI $2.5\%$} & \colhead{HDI $5\%$} & \colhead{HDI $95\%$} & \colhead{HDI $97.5\%$}
	}
	%\decimalcolnumbers
	\startdata
	\multirow{4}{*}{RRab}   & $\theta_0$           &  $ -5.819$ & $0.149$ & $-6.124$ & $-6.079$ & $-5.582$ & $-5.524$  \\
	                        & $\theta_P$           &  $ -6.350$ & $0.067$ & $-6.484$ & $-6.461$ & $-6.240$ & $-6.220$  \\
	                        & $\theta_{A_2}$       &  $  5.785$ & $0.320$ & $ 5.158$ & $ 5.253$ & $ 6.308$ & $ 6.423$  \\
	                        & $\theta_{\phi_{31}}$ &  $  1.248$ & $0.020$ & $ 1.211$ & $ 1.216$ & $ 1.282$ & $ 1.290$  \\
	\hline
	\multirow{5}{*}{RRc}    & $\theta_0$           &  $ -1.821$ & $0.113$ & $ -2.0494$ & $ -2.008$ & $ -1.636$ & $ -1.604$ \\
	                        & $\theta_P$           &  $-10.014$ & $0.269$ & $-10.5464$ & $-10.469$ & $ -9.578$ & $ -9.493$ \\
	                        & $\theta_{A_1}$       &  $ 13.835$ & $0.623$ & $ 12.6454$ & $ 12.799$ & $ 14.838$ & $ 15.071$ \\
	                        & $\theta_{A_2}$       &  $-34.704$ & $1.211$ & $-37.0904$ & $-36.704$ & $-32.721$ & $-32.340$ \\
	                        & $\theta_{\phi_{31}}$ &  $  0.325$ & $0.014$ & $  0.2984$ & $  0.302$ & $  0.348$ & $  0.353$ \\
	\enddata
	\tablenotetext{a}{Standard deviation of the marginal posterior.}
	%\tablecomments{}
\end{deluxetable*}

\begin{deluxetable*}{l|r@{.}lr@{.}lr@{.}lr@{.}l|r@{.}lr@{.}lr@{.}lr@{.}l}
	\tablecaption{Covariance and Pearson correlation matrices of the RRab [Fe/H] prediction model's parameters computed from their joint posterior distribution.}
	\label{tab:matr_rrab} 
	\tablehead{
		& \multicolumn{8}{c}{Covariance Matrix} \vline& \multicolumn{8}{c}{Pearson Correlation Matrix} \\
		\hline
		& \multicolumn{2}{c}{$\theta_0$} & \multicolumn{2}{c}{$\theta_P$} & \multicolumn{2}{c}{$\theta_{A_2}$} & \multicolumn{2}{c}{$\theta_{\phi_{31}}$} \vline
		& \multicolumn{2}{c}{$\theta_0$} & \multicolumn{2}{c}{$\theta_P$} & \multicolumn{2}{c}{$\theta_{A_2}$} & \multicolumn{2}{c}{$\theta_{\phi_{31}}$}
	}
	\startdata
    $\theta_0$             &$ 0$&$0222$ & $-0$&$0042$ & $-0$&$0394$ & $-0$&$0027$ &    $ 1$&$0000$ & $-0$&$4227$ & $-0$&$8240$ & $-0$&$9287$ \\
    $\theta_P$             &$-0$&$0042$ & $ 0$&$0045$ & $ 0$&$0105$ & $ 0$&$0001$ &    $-0$&$4227$ & $ 1$&$0000$ & $ 0$&$4858$ & $ 0$&$0952$ \\
    $\theta_{A_2}$         &$-0$&$0394$ & $ 0$&$0105$ & $ 0$&$1026$ & $ 0$&$0039$ &    $-0$&$8240$ & $ 0$&$4858$ & $ 1$&$0000$ & $ 0$&$6246$ \\
    $\theta_{\phi_{31}}$   &$-0$&$0027$ & $ 0$&$0001$ & $ 0$&$0039$ & $ 0$&$0004$ &    $-0$&$9287$ & $ 0$&$0952$ & $ 0$&$6246$ & $ 1$&$0000$ \\
	\enddata
	%\tablenotetext{}{}
	%\tablecomments{}
\end{deluxetable*}

\begin{deluxetable*}{l|r@{.}lr@{.}lr@{.}lr@{.}lr@{.}l|r@{.}lr@{.}lr@{.}lr@{.}lr@{.}l}
	\tablecaption{The same as Table ~\ref{tab:matr_rrab}, but for RRc stars.}
	\label{tab:matr_rrc} 
	\tablehead{
		& \multicolumn{10}{c}{Covariance Matrix} \vline& \multicolumn{10}{c}{Pearson Correlation Matrix} \\
		\hline
		& \multicolumn{2}{c}{$\theta_0$} & \multicolumn{2}{c}{$\theta_P$} & \multicolumn{2}{c}{$\theta_{A_1}$} & \multicolumn{2}{c}{$\theta_{A_2}$} & \multicolumn{2}{c}{$\theta_{\phi_{21}}$} \vline
		& \multicolumn{2}{c}{$\theta_0$} & \multicolumn{2}{c}{$\theta_P$} & \multicolumn{2}{c}{$\theta_{A_1}$} & \multicolumn{2}{c}{$\theta_{A_2}$} & \multicolumn{2}{c}{$\theta_{\phi_{21}}$} 
	}
	\startdata
	$\theta_0$           &$ 0$&$0129$ & $-0$&$0109$ & $-0$&$0221$ & $ 0$&$0063$ & $-0$&$0010$ &    $ 1$&$0000$ & $-0$&$3556$ & $-0$&$3123$ & $ 0$&$0461$ & $-0$&$6558$  \\
	$\theta_P$           &$-0$&$0109$ & $ 0$&$0725$ & $-0$&$0551$ & $ 0$&$0574$ & $-0$&$0007$ &    $-0$&$3556$ & $ 1$&$0000$ & $-0$&$3281$ & $ 0$&$1759$ & $-0$&$1989$  \\
	$\theta_{A_1}$       &$-0$&$0221$ & $-0$&$0551$ & $ 0$&$3881$ & $-0$&$5421$ & $-0$&$0003$ &    $-0$&$3123$ & $-0$&$3281$ & $ 1$&$0000$ & $-0$&$7186$ & $-0$&$0355$  \\
	$\theta_{A_2}$       &$ 0$&$0063$ & $ 0$&$0574$ & $-0$&$5421$ & $ 1$&$4663$ & $ 0$&$0027$ &    $ 0$&$0461$ & $ 0$&$1759$ & $-0$&$7186$ & $ 1$&$0000$ & $ 0$&$1588$  \\
	$\theta_{\phi_{31}}$ &$-0$&$0010$ & $-0$&$0007$ & $-0$&$0003$ & $ 0$&$0027$ & $ 0$&$0002$ &    $-0$&$6558$ & $-0$&$1989$ & $-0$&$0355$ & $ 0$&$1588$ & $ 1$&$0000$  \\
	\enddata
	%\tablenotetext{}{}
	%\tablecomments{}
\end{deluxetable*}

\begin{figure*}
	\gridline{
		\fig{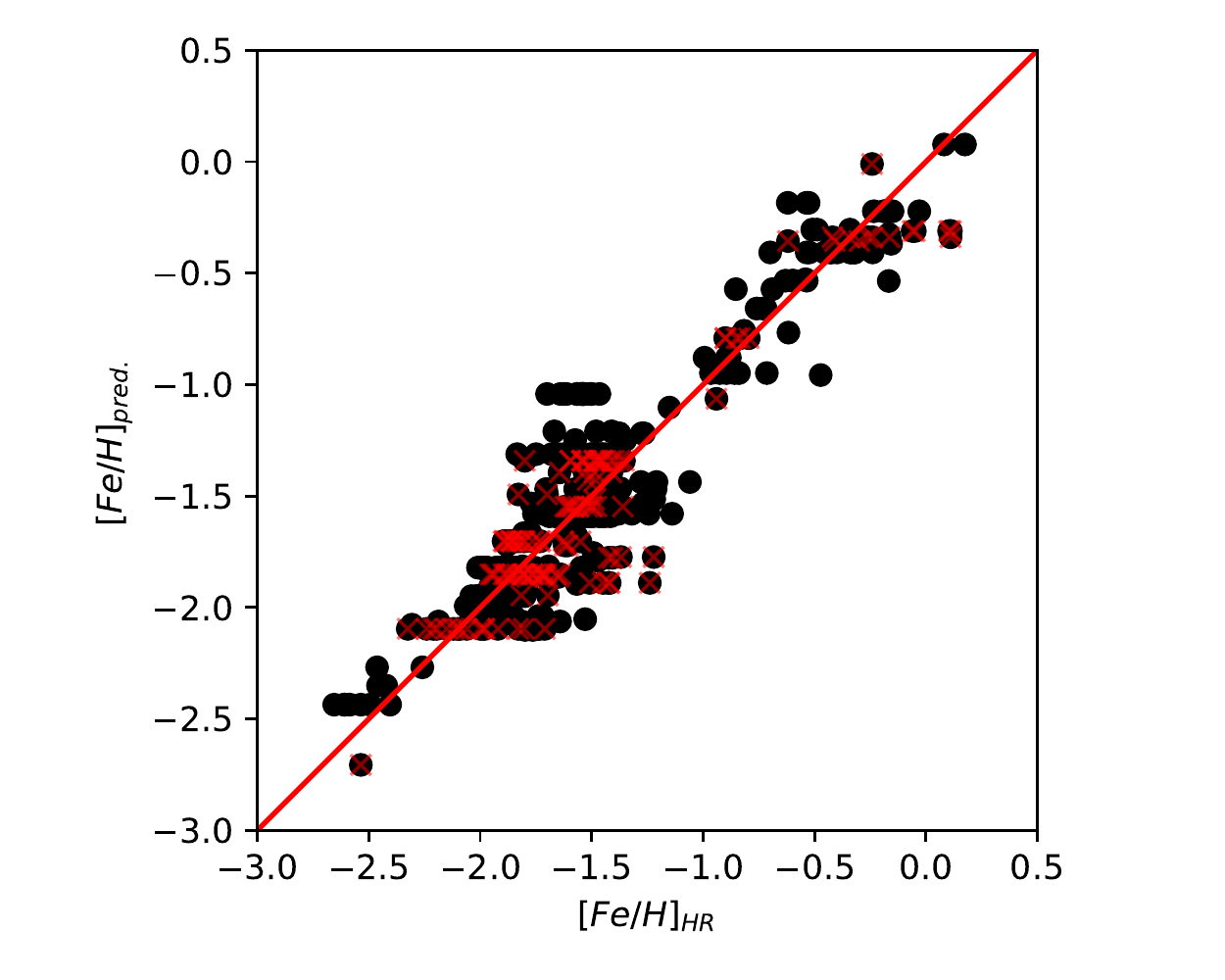}{0.35\textwidth}{}
		\hskip-0.8cm
		\fig{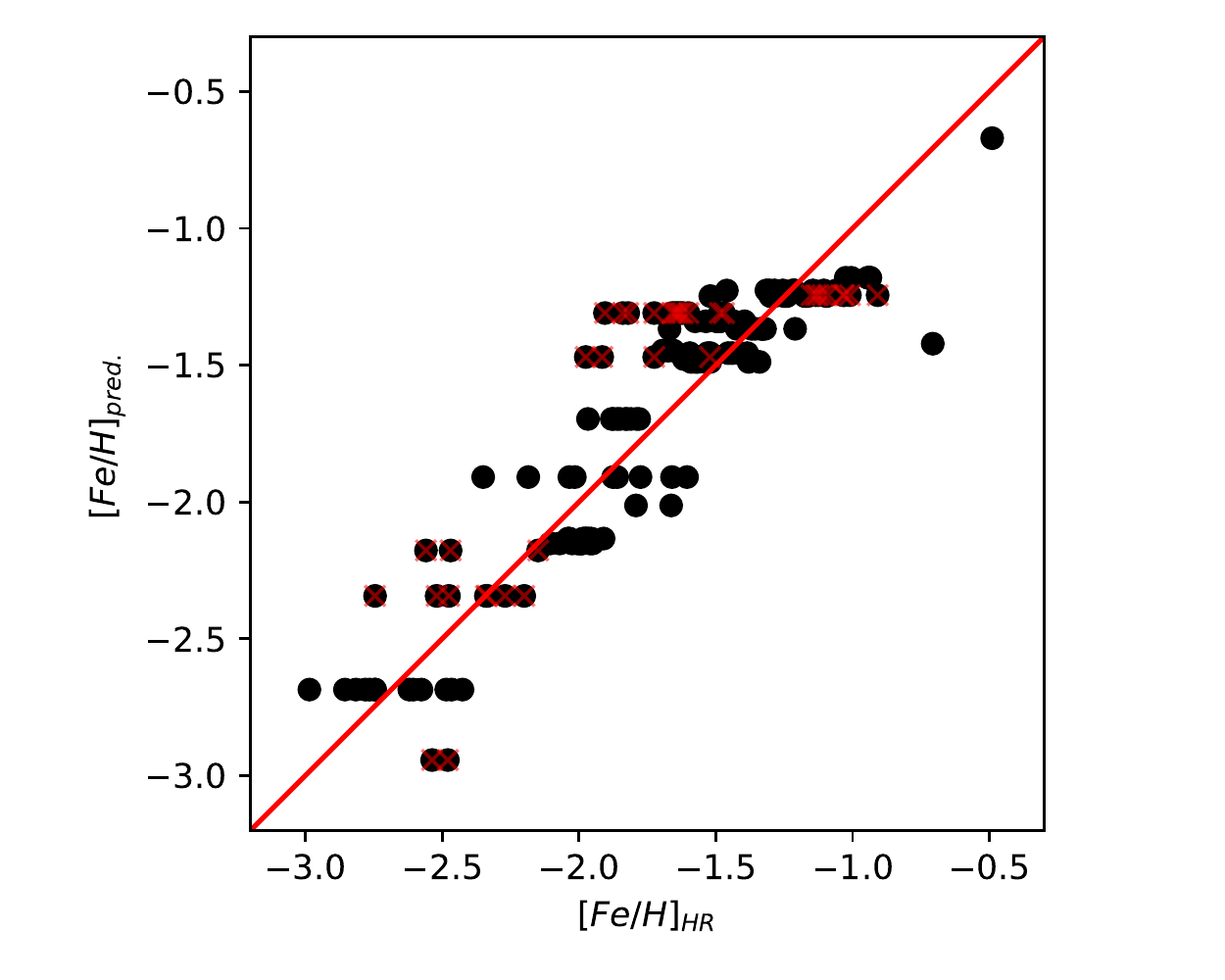}{0.35\textwidth}{}
		\hskip-0.8cm
		\fig{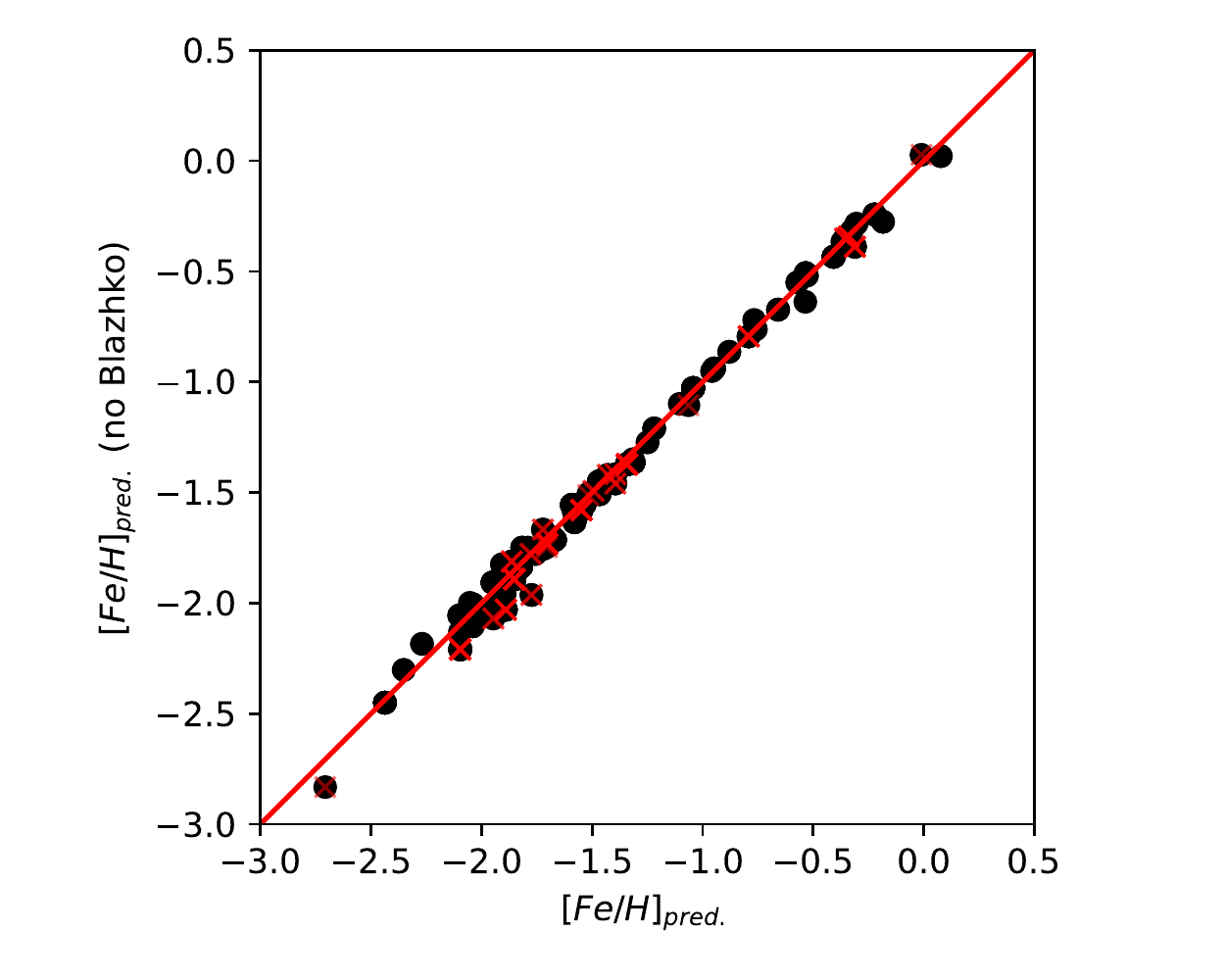}{0.35\textwidth}{}
	}
	\caption{
		Predicted {\em vs} true metallicities for the RRab (left) and RRc (right) stars in the development set. Black points denote the entire development set, red crosses mark data of only those stars that show the Blazhko effect. Note the different scales of the figures.
		Right: comparison of RRab metallicity predictions by our Bayesian regression models trained on data sets including/excluding Blazhko-modulated stars.
		\label{fig:residuals}}
\end{figure*}

Figure~\ref{fig:residuals} shows the predicted [Fe/H] values against their HR spectroscopic equivalents for the RR~Lyrae stars in our development set using the parameter distributions' means from Table~\ref{tab:posterior_stat}. The residual standard deviations of the fits are $0.20$~dex for RRab and $0.21$~dex for RRc stars. The scatter of the RRab predictions is slightly higher than that of the S05 formula ($0.14$) but the latter was trained on a tiny data set of 28 objects with single spectroscopic measurements, which alone can result in a downward bias of the dispersion.

%\subsection{On the influence of the Blazhko effect}\label{subsec:blazhko}

We investigated how the inclusion of Blazhko-modulated stars in the development set might influence our predictive models. In Fig.~\ref{fig:residuals}, data corresponding to stars with the Blazhko effect are highlighted with red crosses. No systematic deviation or increased scatter of these data with respect to the rest of the samples can be seen. We also performed the Bayesian regression on the RRab development set after excluding the data of all objects that show the Blazhko effect in their $I$ light curves. In the right panel of Fig.~\ref{fig:residuals}, we compare the [Fe/H] values predicted by the modified model {\em vs} those by the original one for the {\em entire} development set, with the Blazhko stars highlighted. There is no significant difference between the predictions of the two models; thus we conclude that the photometric metallicity prediction of RRab stars is robust against the presence of the Blazhko effect in the light curves, in case the modulation cycle is well sampled. Finally, we see no point in performing a separate test for RRc stars since the excessively reduced sample size of only 18 non-modulated objects would have a large negative effect on the resulting model's performance.
	
\begin{figure*}
	\gridline{
		\fig{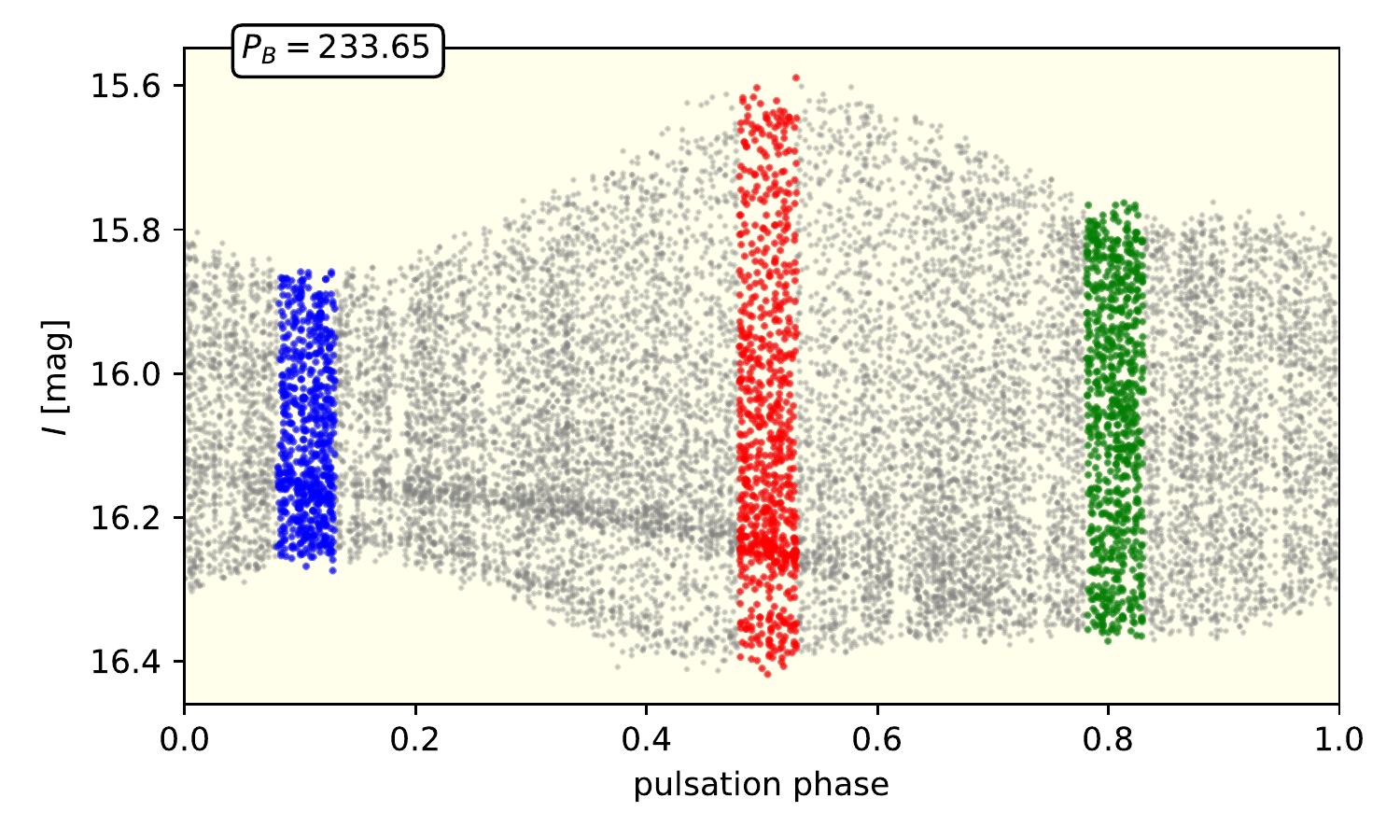}{0.45\textwidth}{}
		\hskip-0.8cm
		\fig{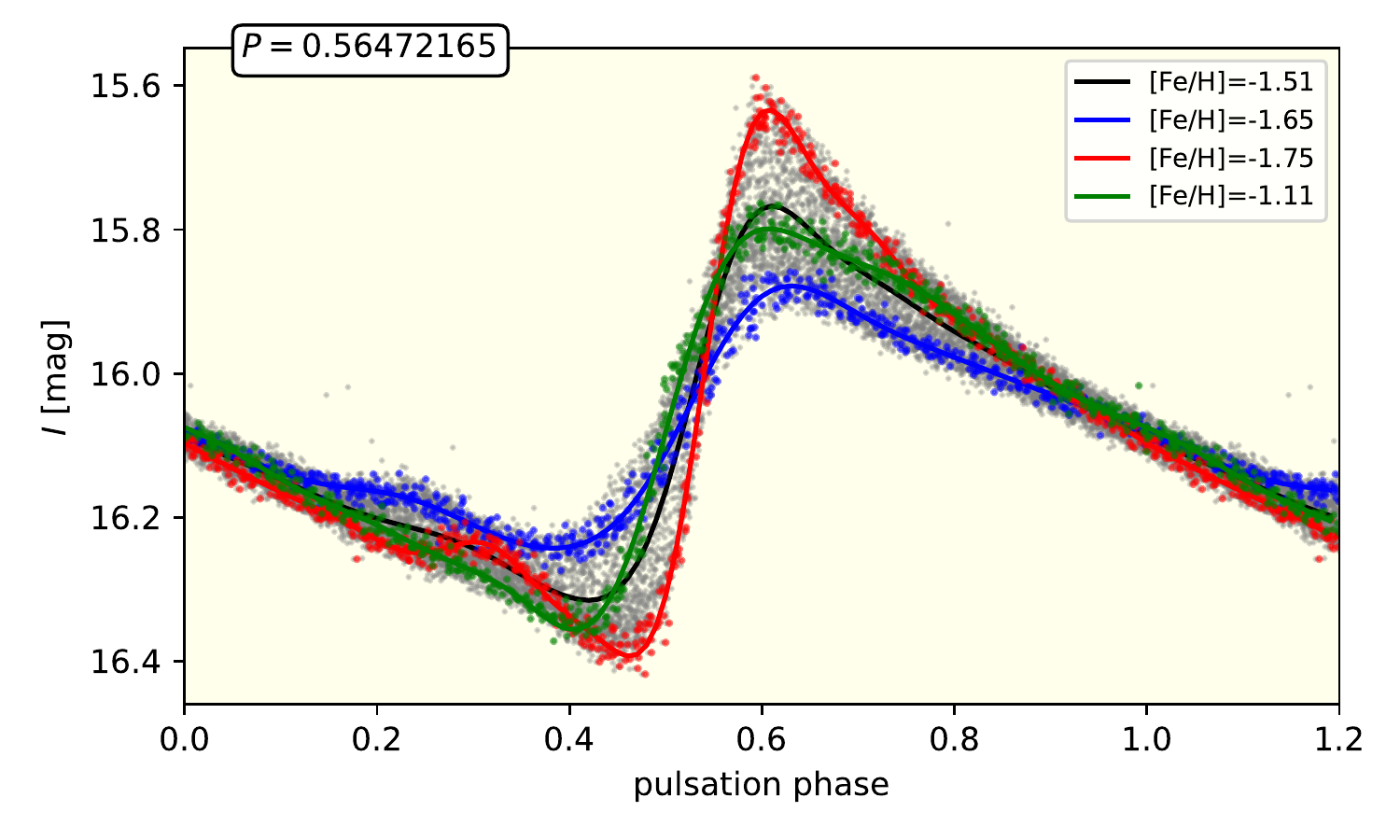}{0.45\textwidth}{}
	}
	\caption{
		OGLE-IV $I$-band photometric time-series of the bulge RRab star OGLE-BLG-RRLYR-06473 as a function of the modulation phase (left) and the pulsation phase (right). Gray points denote all data; the blue, red, and green points show different segments of the data in equally long modulation phase intervals of $0.05$. The black and colored curves show GPR fits to the entire light curve and its various segments plotted with the corresponding colors, respectively. The inset of the right panel shows [Fe/H] values derived from the Fourier representations of the various GPR fits using our Bayesian predictive model. The left and right figure headers show the modulation period \citep{2020MNRAS.494.1237S} and the pulsation period in days, respectively.
		\label{fig:blazhko}}
\end{figure*}

Concerning the predictive model's application on Blazhko RR~Lyrae stars, we emphasize that in case of large phase and/or amplitude modulation, a partial photometric coverage of the Blazhko cycle can obviously lead to rather biased light-curve parameters with respect to those of the mean light variation of the same star, and may consequently result in a biased [Fe/H] prediction. 
% Unfortunately, our development set lacks objects with strong modulation and sufficiently densely sampled Blazhko cycles that could be used to study such biases with respect to spectroscopic measurements.
To demonstrate the above effect, Fig.~\ref{fig:blazhko} shows the light curve of a very strongly modulated bulge RRab star of a rather rare morphological type with a very long modulation cycle of $233.65$\,d \citep[see][]{2020MNRAS.494.1237S}. Observations with a limited time baseline would severely undersample the light curves of similar objects, which we illustrate with three $\sim$11.5-day long light-curve segments at different modulation phases, marked with different colors in Fig.~\ref{fig:blazhko}. The light-curve solutions of these various segments yield biased [Fe/H] predictions with respect to the value derived from the mean light curve model based on all data, with errors of up to 0.4~dex. This caveat should be considered when photometrically estimating the [Fe/H] for individual objects or small samples.

\subsection{Comparison with earlier [Fe/H]-estimators}\label{subsec:comparison}

\begin{figure*}
		\gridline{
		\fig{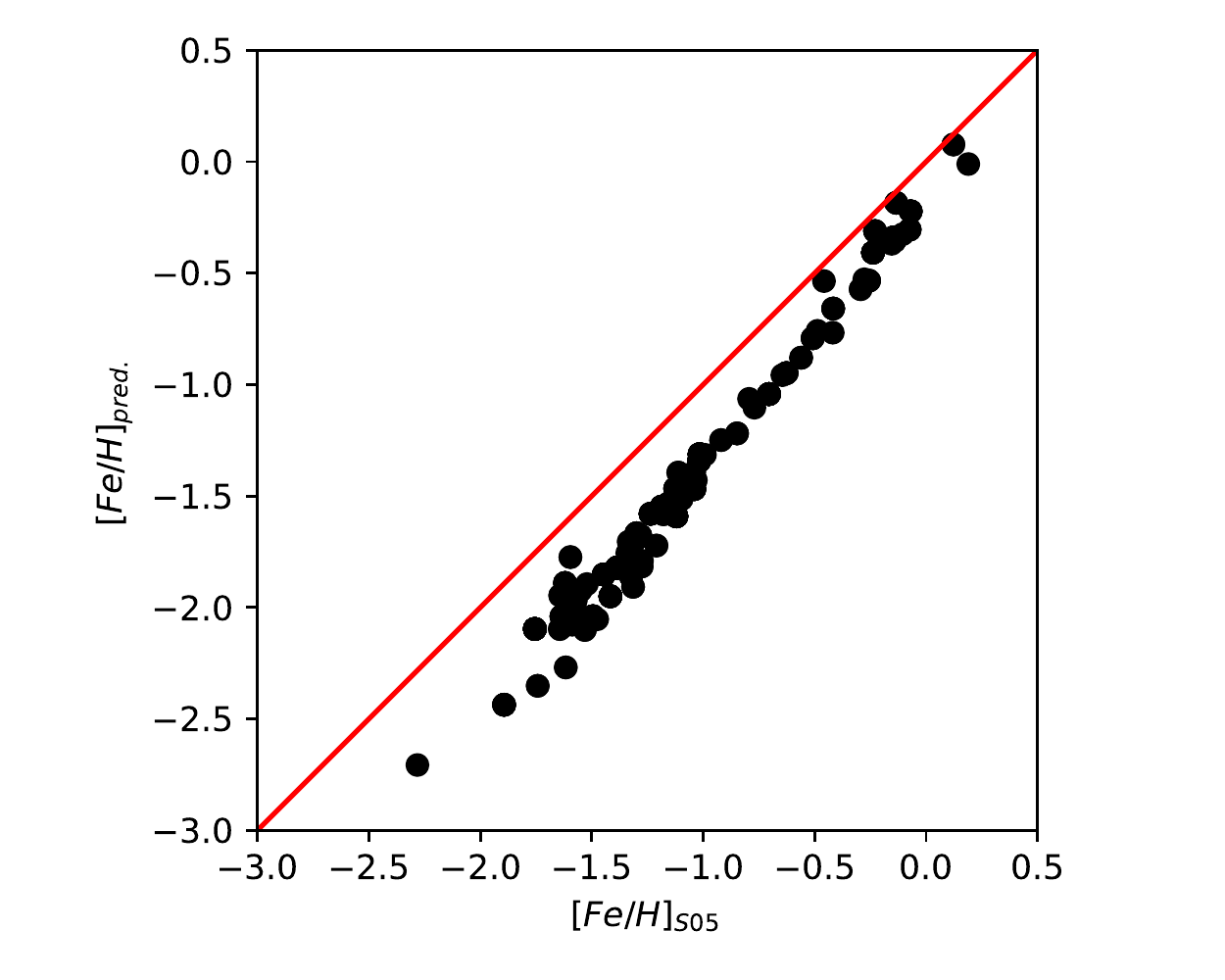}{0.35\textwidth}{}
		\hskip-0.8cm
		\fig{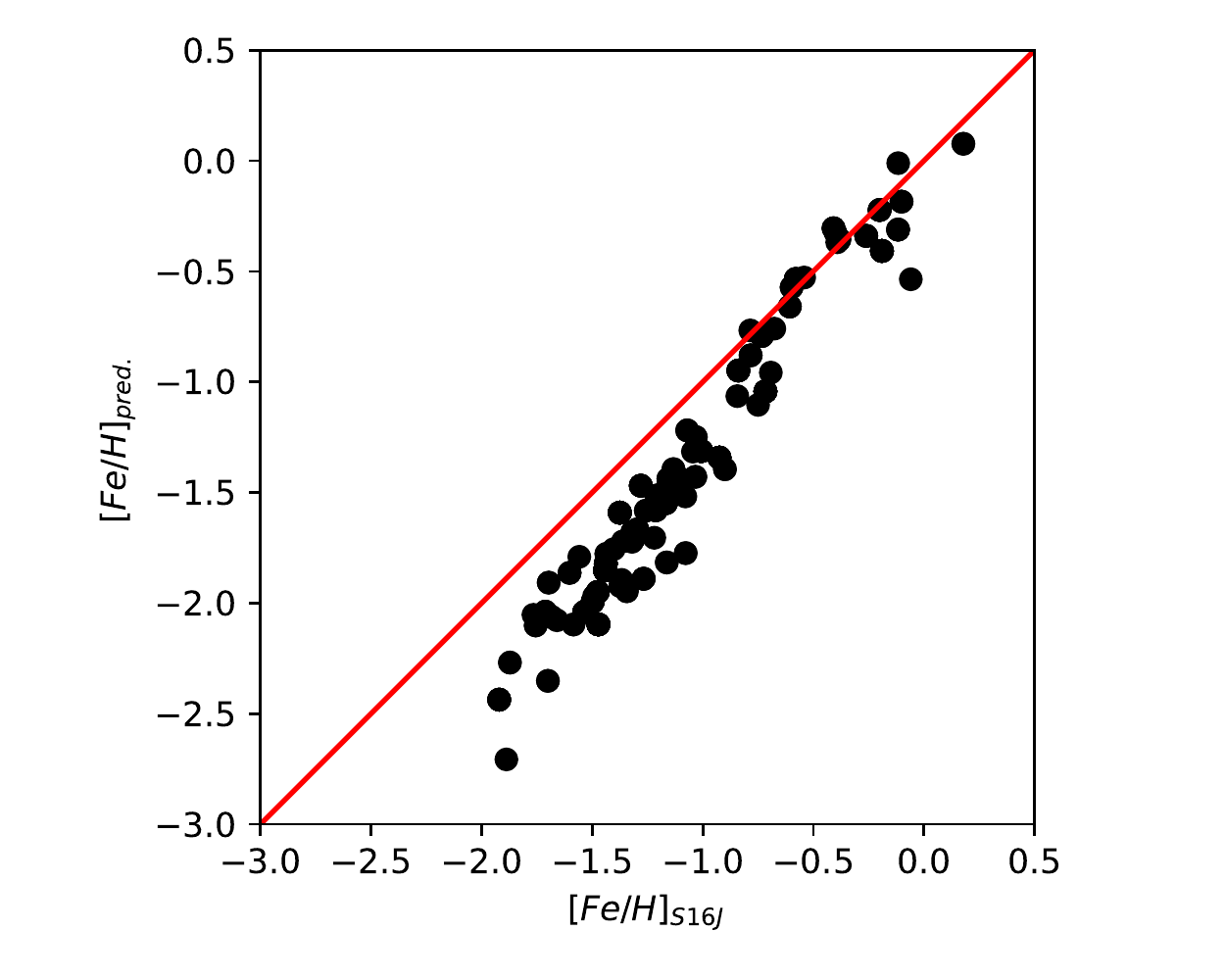}{0.35\textwidth}{}
		\hskip-0.8cm
		\fig{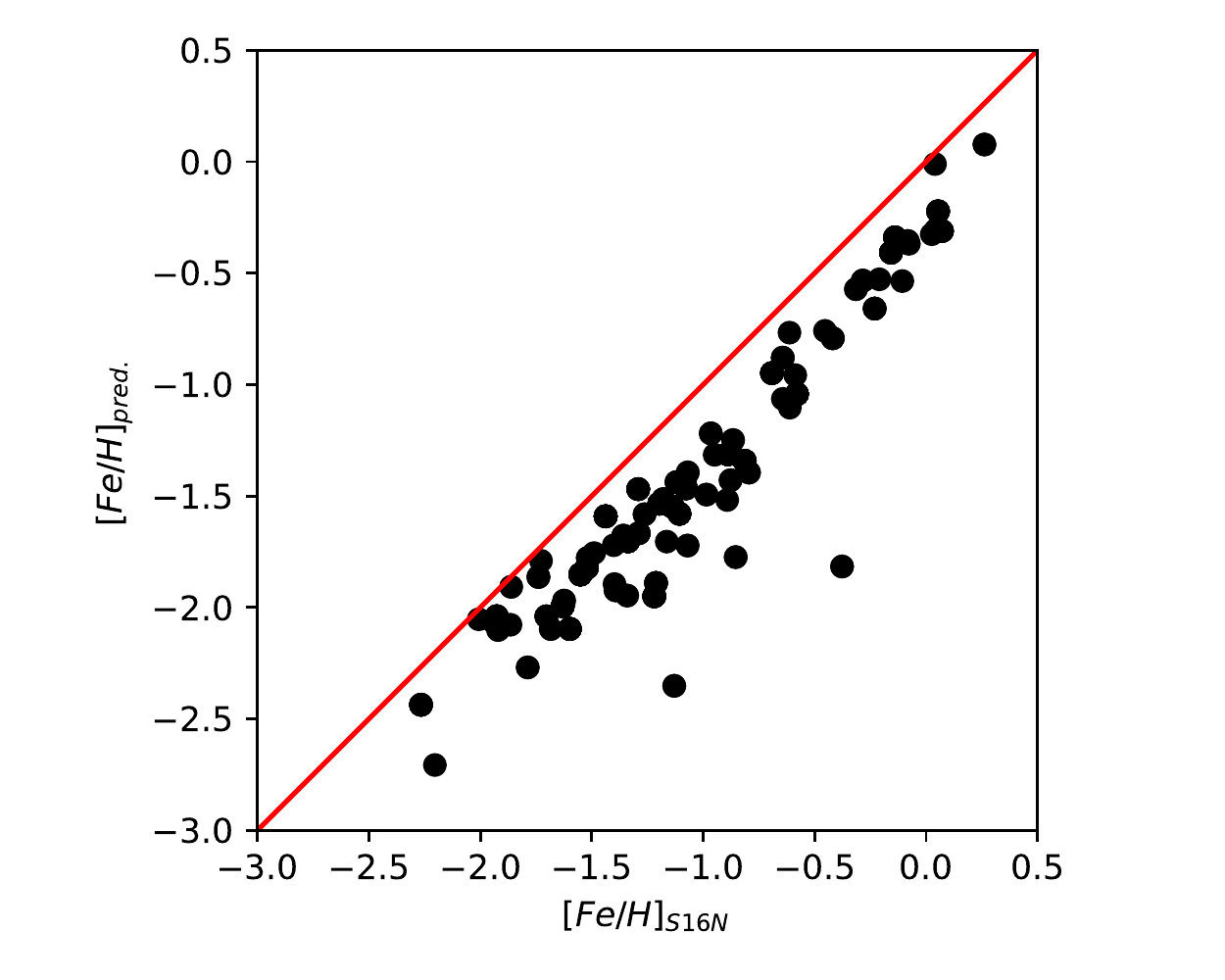}{0.35\textwidth}{}
	}
	%\plotthree{feh_rrab_new_vs_s05.pdf}{feh_rrab_new_vs_s16j.pdf}{feh_rrab_new_vs_s16n.pdf}
	\caption{
		Metallicities of the RRab stars in the development set predicted by our regression model, in comparison with their equivalents obtained by the S05 three-parameter formula (left), the JK96 formula (middle), and the N13 formula (right). The latter two were transformed into the $I$ band following the method of \citet{skowron_ogle-ing_2016}.
		\label{fig:comp_jk96}}
\end{figure*}

To visualize the differences between our Bayesian predictive model and other $I$-band photometric [Fe/H] formulae from the literature, we confronted the metallicity predictions from these with our development data set of RRab stars in Fig.~\ref{fig:comp_jk96}. From left to right, we plotted our predictions against those from the S05 formula, as well as the JK96 and N13 formulae transformed into the $I$-band, following the method of \citet{skowron_ogle-ing_2016}. We refer to these formulae as S16J and S16N, respectively. The predictions from the S05 and S16J formulae are on the \citet{jurcsik_determination_1996} metallicity scale, which was established on the basis of $\Delta S$ measurements of field RR~Lyrae stars \citep[e.g.,][]{layden_metallicities_1994,suntzeff_summary_1994}, while the S16N formula was directly tied to spectroscopic [Fe/H] measurements of 26 RRab stars. We can observe a significant upward bias of the [Fe/H] values computed from the S05 and JK96 formulae with respect to our predictions. The deviations from our model's output show a dependence on the [Fe/H], with a systematically increasing positive bias toward lower metallicities, amounting to as much as $\sim$0.5~dex for $[{\rm Fe/H}]\lesssim-2$. While the J16N predictions are in a better agreement at low metallicities, they too show a positive bias with respect to our values over the entire [Fe/H] range, in addition to a large scatter.

\begin{figure}
	\plotone{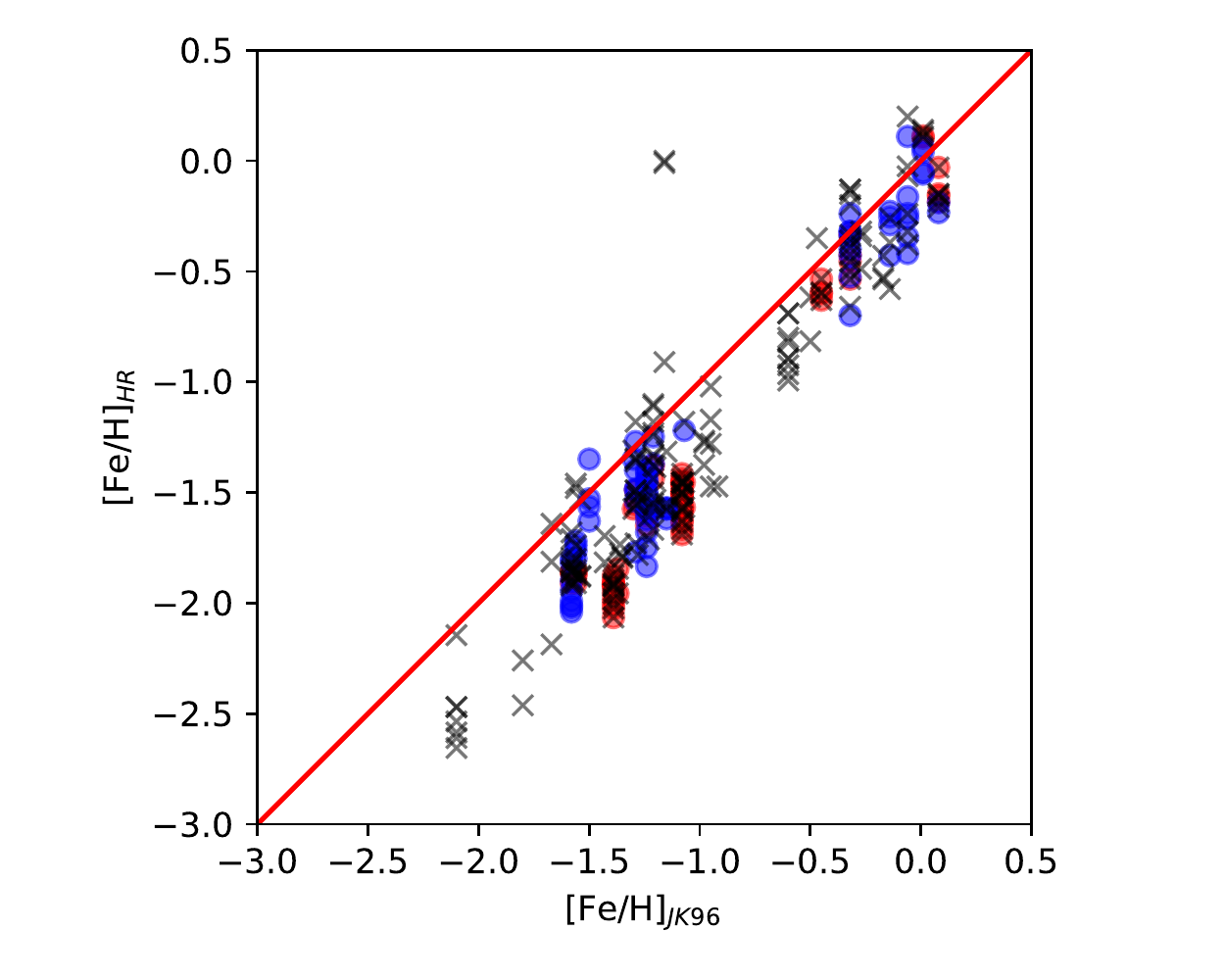}
		\caption{Individual [Fe/H] measurements from our HR spectroscopic data set plotted against their counterparts on the JK96 scale obtained from low-resolution spectra. Circles: objects used for calibrating the $I$-band S05 formula, which have HR spectroscopic [Fe/H] measurements either from the CFCS data set (red) or from other sources (blue, after being shifted to the CFCS scale). Black crosses: objects in the ``basic data set'' of JK96.}
		\label{fig:hr_vs_jk}
\end{figure}

Similar discrepancies were found by \citet{mullen_metallicity_2021} when comparing [Fe/H] predictions from their $V$-band regression model (trained on CFCS metallicities) to those obtained from the original JK96 and N13 formulae. To directly investigate the source of the deviating trend between our predictions and the S05 and S16J ones, we cross-matched our HR spectroscopic sample to their calibration data sets. Figure~\ref{fig:hr_vs_jk} shows individual HR spectroscopic metallicity measurements for the 19 stars in common with the S05 calibration data set plotted against their low-resolution counterparts on the JK96 scale. We can observe a similar positive bias of the S05 calibration data with respect to HR spectroscopy. The same [Fe/H] measurements (with only one exception) were also part of the data set used by JK96 for the calibration of their $V$-band [Fe/H] prediction formula. In Fig.~\ref{fig:comp_jk96} we also compare the [Fe/H] values from their sample with more recent HR spectroscopic measurements (all of these are from the CFCS data set). The same trend between the two is apparent, leading us to the conclusion that the systematically lower photometric metallicity estimates from our work with respect to the S05 and S16J formulae stem mainly from a linear bias in the JK96 scale. The latter may arise from the very modest amount of HR spectroscopic RR~Lyrae [Fe/H] determinations at that time to which they tied the $\Delta S$ measurements. Apart from this, a further cause of the discrepancy between the photometric metallicities may be biased regression parameters due to the small underlying data sets deficient of very metal-poor stars particularly in case of the N13 and S05 regressions. The former includes non-linear terms that apparently do not generalize well, the latter is constrained to $[{\rm Fe/H}]\gtrsim-1.6$.

\section{Photometric metallicity distributions} \label{sec:MDFs}

The systematic, metallicity-dependent biases in S05 and S16 estimators have various implications. Perhaps the most important one is their effect on the photometric MDFs derived for various stellar populations in and around the Milky Way, while capitalizing on the vast amount of accurate $I$-band time-series data from the OGLE surveys of the bulge, southern disk, and the Magellanic Clouds. Most recently, \citet{pietrukowicz_properties_2020} exploited the entire OGLE RR~Lyrae catalog to study the MDFs of the old stellar populations of the Galaxy, the LMC, SMC, and the Sagittarius dwarf galaxy, and drew inferences on their formation histories therefrom. To assess the effect of the systematic difference between the various earlier formulae and our new regression model on our view of the MDFs of these populations, we deployed all of them on the $I$-band light curves of RR~Lyrae stars in the OCVS and computed the MDFs of the same cosmic environments. We used the same simple positional and brightness criteria in selecting the pertinent stellar populations as the ones that were applied by \citet{pietrukowicz_properties_2020}, which are sufficient to keep sample contamination on the level of a few percent, and aid direct comparability and reproducibility. With the additional quality criteria of $\{C>0.9$ and $S/N>100$ and $N_{\rm ep.}>100\}$, we obtained RR~Lyrae samples for the inner and outer bulge, bulge-disk transition area, southern disk, Sagittarius dwarf spheroidal galaxy, LMC, SMC, and Galactic halo.
%, are summarized in Table~\ref{tab:mdfs_sel_crit}.

\begin{figure*}
	\gridline{
		\fig{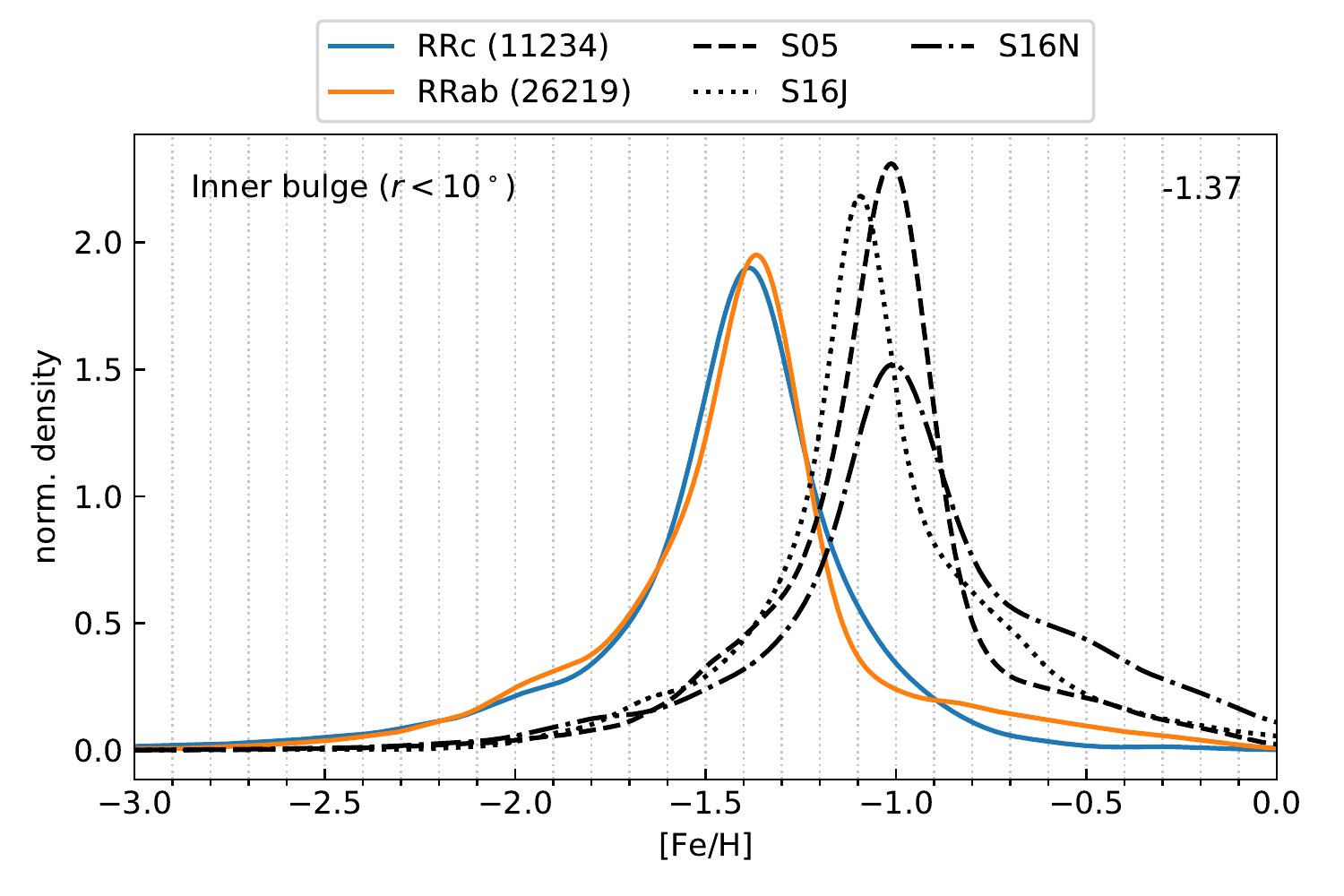}{0.4\textwidth}{}
		\fig{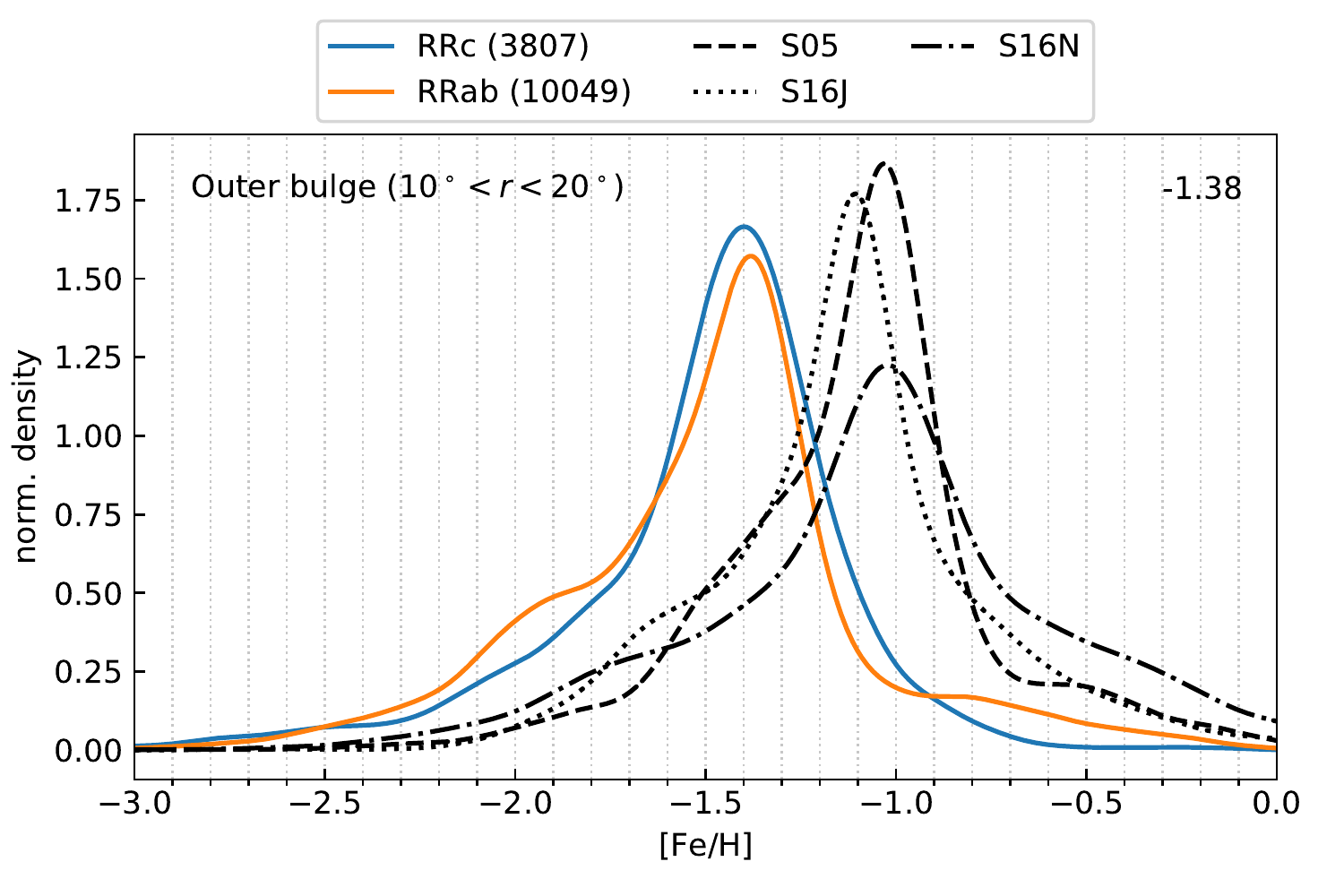}{0.4\textwidth}{}
	}
	\vskip-0.8cm
		\gridline{
		\fig{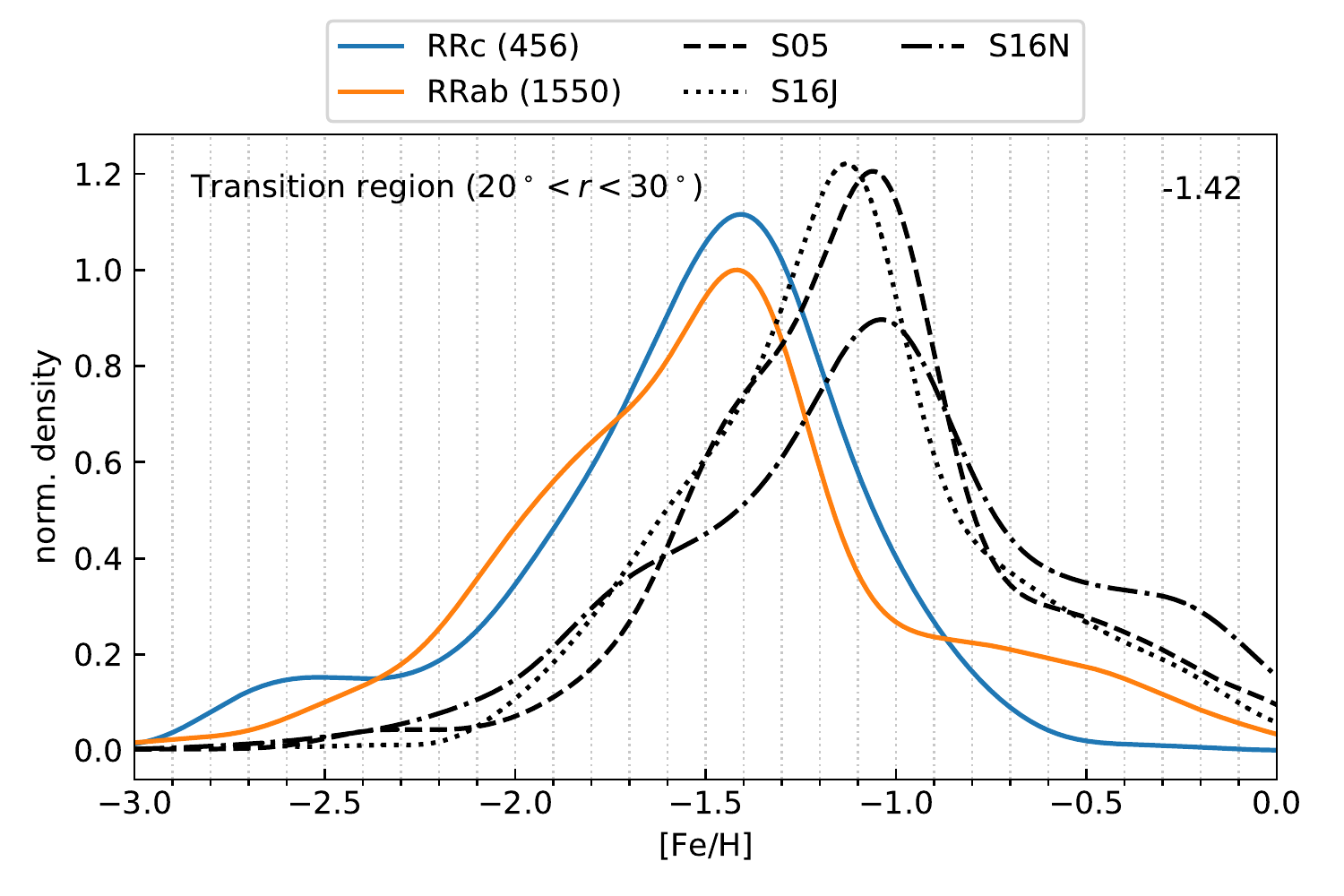}{0.4\textwidth}{}
		\fig{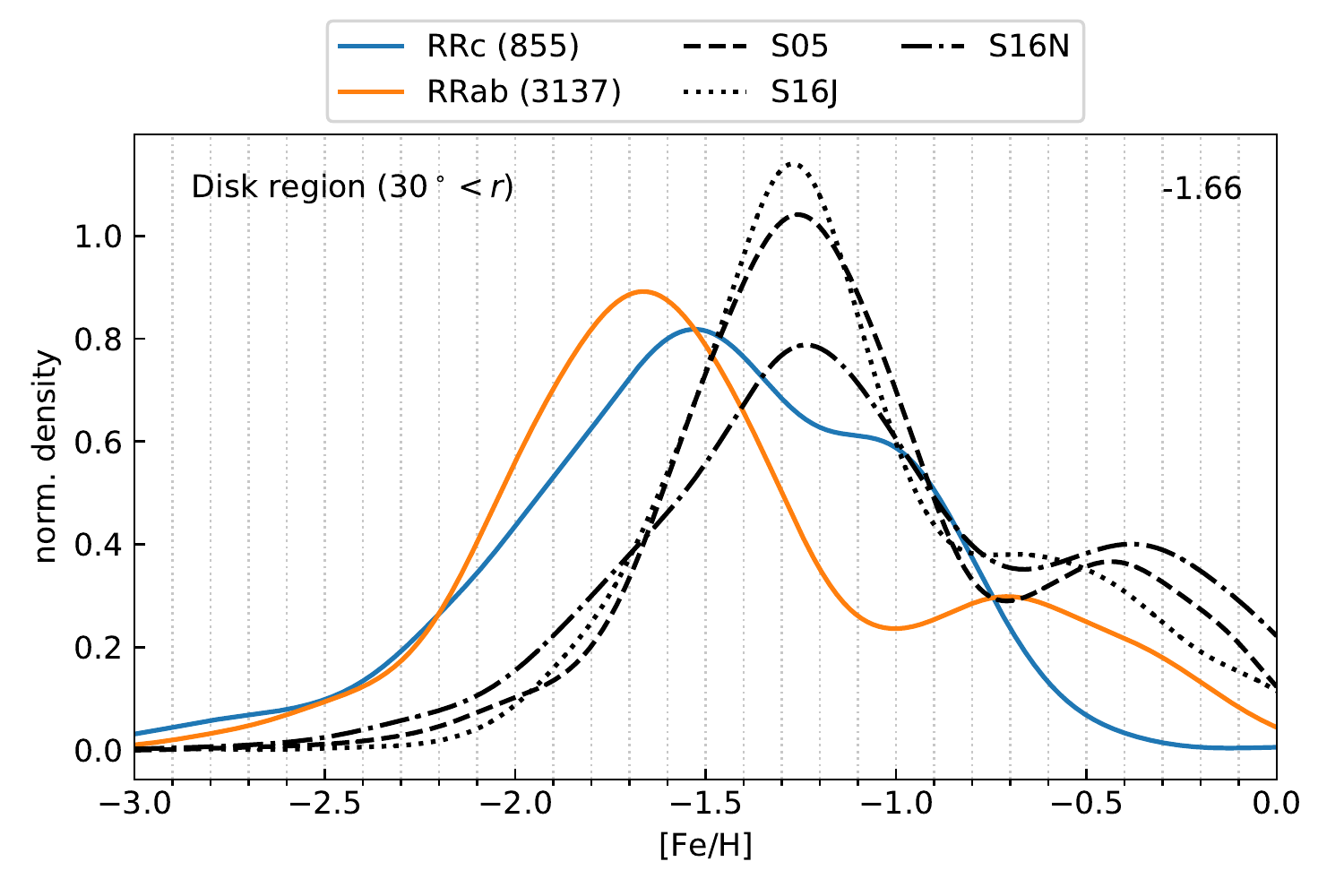}{0.4\textwidth}{}
	}
	\vskip-0.8cm
		\gridline{
		\fig{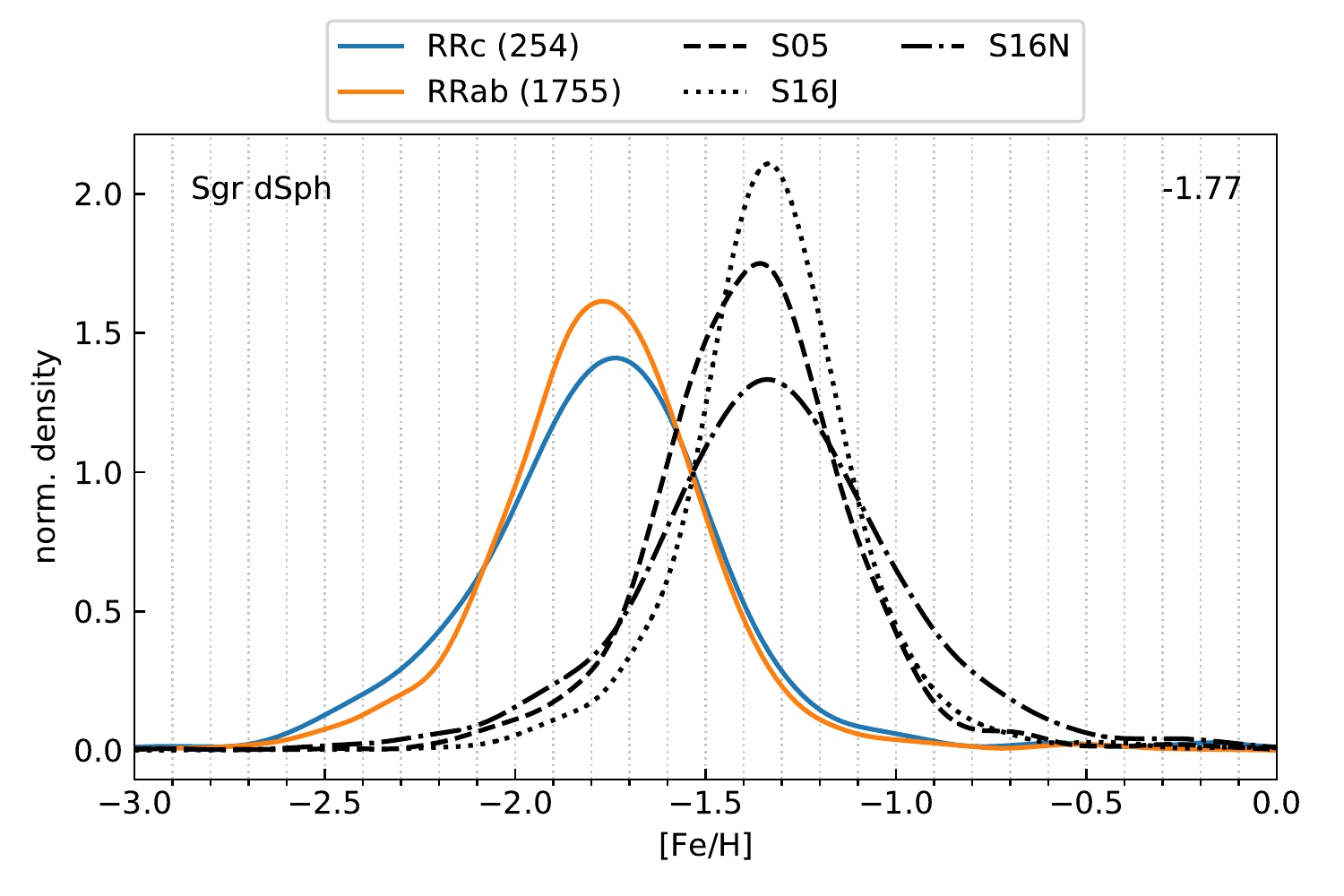}{0.4\textwidth}{}
		\fig{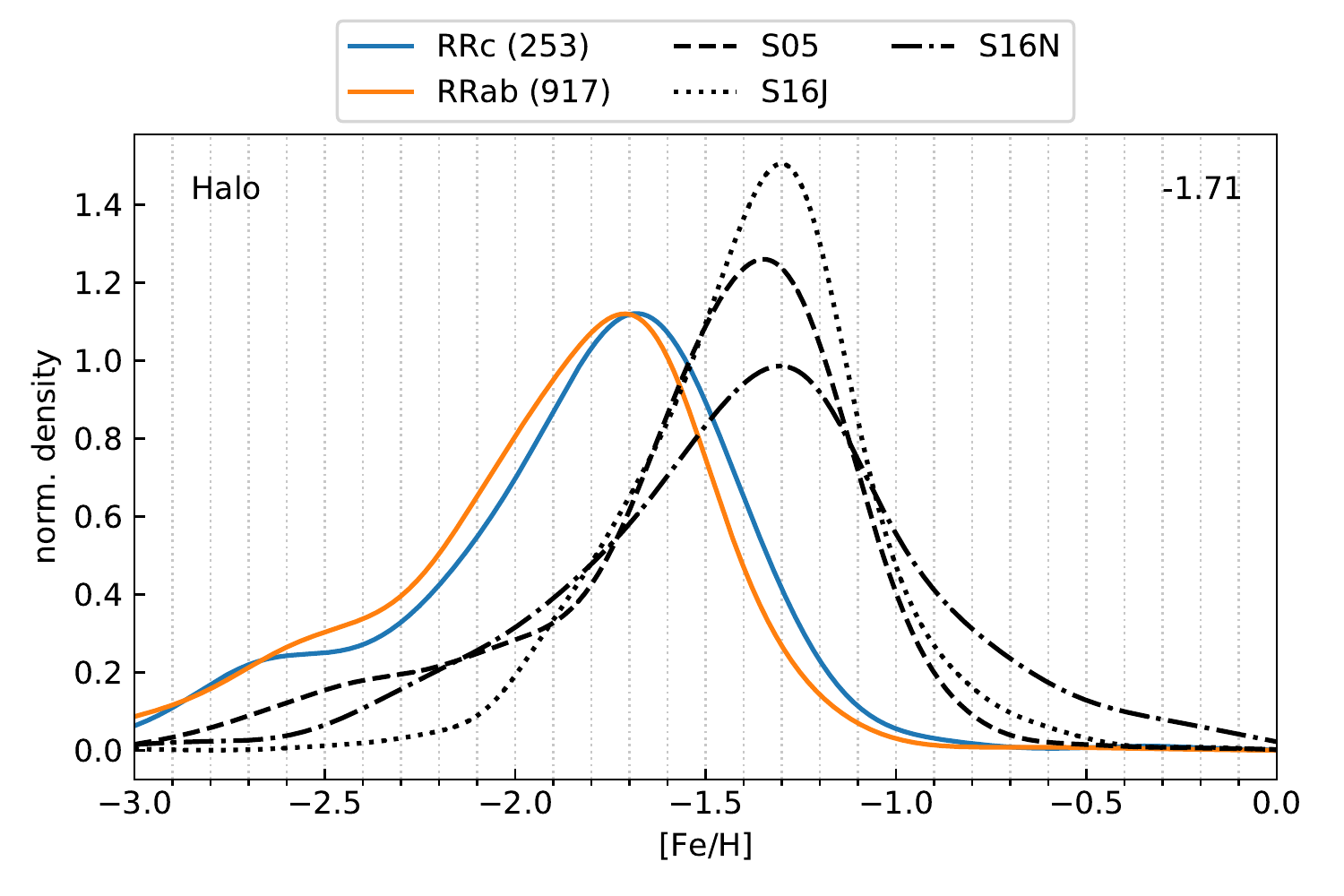}{0.4\textwidth}{}
	}
	\vskip-0.8cm
		\gridline{
		\fig{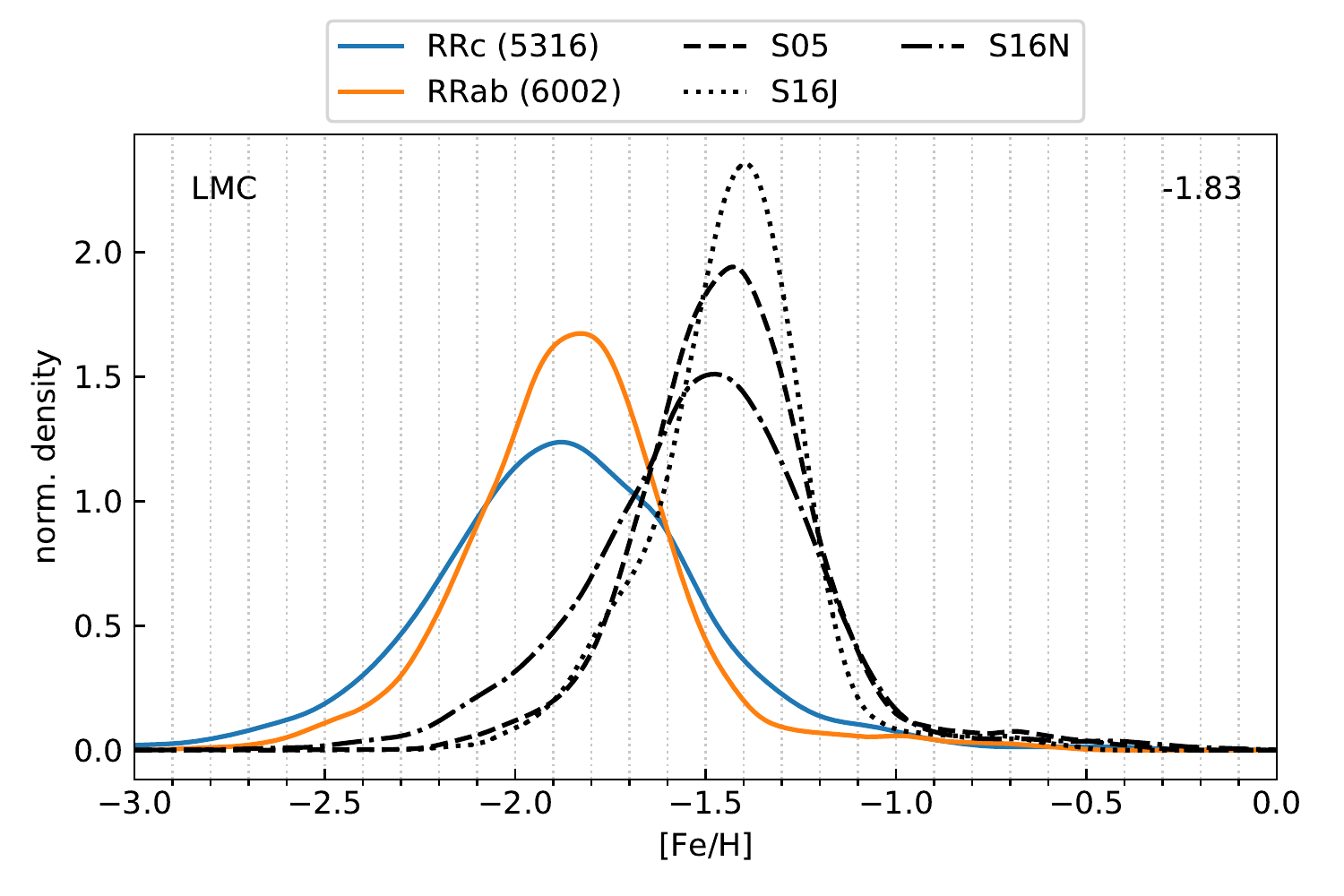}{0.4\textwidth}{}
		\fig{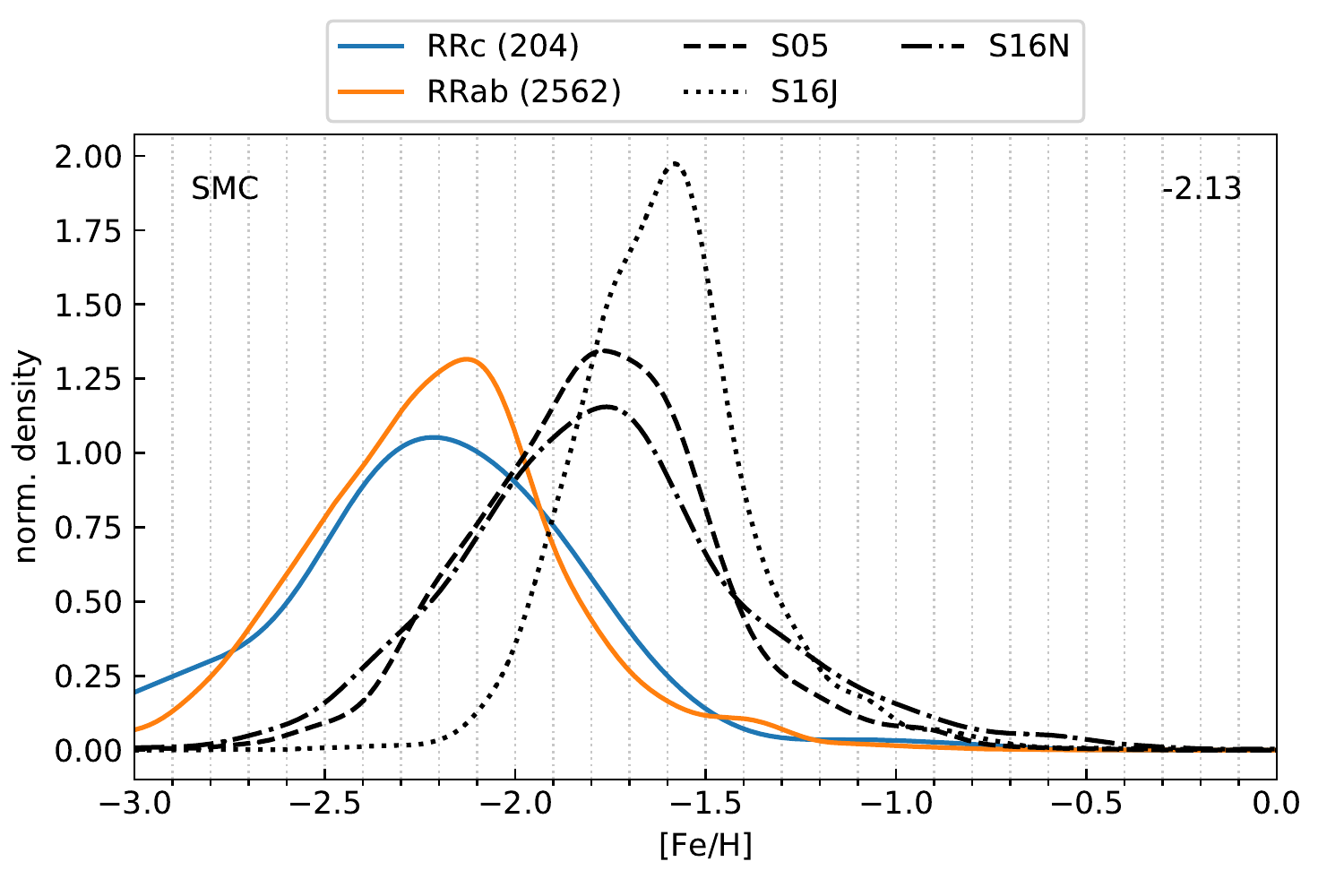}{0.4\textwidth}{}
	}
	\caption{Photometric MDFs of RR~Lyrae stars from the OCVS in different Local Group environments shown in the upper left corners of each panel. The curves denote kernel density estimates using Gaussian kernels and bandwidths according to Scott's rule. The modes estimated from the RRab (orange) density curves obtained from our predictive model are shown in the upper right corners of the individual panels. Orange and blue curves show the results computed using the [Fe/H] predictive models of this work for RRab and RRc stars, respectively. The dashed, dotted, and dash-dotted black curves show the MDFs obtained by using the S05 three-parameter formula, the S16J, and the S16N formulae, respectively. The number of stars used for computing the density estimates are shown in parentheses in each panel's legend. The angular Galactocentric distance ($r$) ranges of the relevant samples are also shown.
		\label{fig:mdfs}}
\end{figure*}

We computed the light-curve features of all OCVS RR~Lyrae stars in the above samples by applying the same light-curve fitting procedures as to the development data set, as discussed in \ref{subsec:lightcurves}, and used them as input of the predictive models. The resulting photometric MDFs of the various stellar populations are shown in Fig.~\ref{fig:mdfs}. The most obvious difference between the RRab MDFs from the earlier formulae and this work is the significant offset of their modes by 0.3---0.4~dex. As opposed to earlier estimates of $\sim -1$~dex for the bulge MDF's mode, in fact it is located at $-1.37$~dex and $-1.38$~dex for the inner and outer bulge samples, respectively. These latter values show a remarkable  agreement with the results of \citet{savino_age_2020}, who determined a median of $-1.39$~dex for their bulge RR~Lyrae MDF obtained from medium-resolution BRAVA-RR spectra \citep{kunder_before_2016}. The shape of their bulge MDF also shows phenomenological consistency with our results ({\em cf.} their Fig.~3). The bulge MDF's mode seems to shift slightly toward lower metallicity with increasing Galactocentric distance, reaching $-1.42$ for the transition region sample, albeit this is based on an order of magnitude smaller sample than the bulge values.

As for the shape of the MDFs, the ones derived by the S05 formula show the closest resemblance to our new solutions. The inner bulge distribution shows weak excesses around the main mode, similarly to the S05 version of the MDF, while the S16 formulae predict a much stronger excess at the metal-rich side. Their relative sizes of the distribution's tails increase toward the outer bulge, also apparent (although shifted) in the S05 and S16 MDFs, but they show significantly less spread in the S05 and S16 versions. The latter can be attributed to the metallicity-varying upward bias of the older formulae having a shrinking effect on the MDFs' tails. This effect is counterbalanced in the S16 formulae by the increased noise due to the applied waveband transformation and the non-linear terms in the case of S16N.

The metal-rich excess evolves into a secondary mode of the MDF as we increase the sample's Galactocentric angular distance along the southern disk. At the same time, the main mode becomes even more metal-poor, converging to the mode of the halo MDF at approximately $-1.7$~dex. We note that the halo MDF shown in Fig.~\ref{fig:mdfs} is based on less than a thousand RRab stars toward the Magellanic cloud footprint of the OGLE survey, which were separated from endemic LMC and SMC objects based on their apparent magnitudes following the criteria of \citet{pietrukowicz_properties_2020}. The mode of the halo MDF is in good agreement with the approximately $-1.6$~dex median metallicity of the halo population derived by  \citet{crestani_deltaS} from a sample of 211 RR~Lyrae stars with HR spectroscopy. Furthermore, our photometric halo MDF has a strong tail toward extremely low metallicities, which is seen greatly diminished in the MDFs based on the S05 and S16 estimators.

The Sagittarius dwarf spheroidal galaxy shows a Gaussian-like [Fe/H] distribution, and it closely matches the mode of the halo's MDF. The [Fe/H] distribution of the LMC RRab stars is also close to normal, with its mode at $-1.83$~dex. The SMC is significantly more metal-poor than the LMC with a [Fe/H] mode at $-2.13$~dex, and its MDF is slightly skewed toward low metallicities. The derived modes are significantly more metal-poor than earlier photometric Fourier-decomposition-based RR Lyrae metallicity
estimates \citep{2012AJ....143...48H,pietrukowicz_properties_2020}.

Figure~\ref{fig:mdfs} also shows MDFs derived from the RRc stars using our predictive model discussed in Sect.~\ref{sec:prediction}. This model has serious limitations with its training data being largely constrained to $[{\rm Fe/H}]\lesssim -1$. In addition, the RRc sample sizes are significantly smaller than those for RRab stars in all of the 8 studied environments. In spite of this, the modes of the resulting RRc MDFs show good agreement with those of the RRab stars, with less than $0.1$~dex offsets between the two. However, subtler features of the RRab MDFs, are generally poorly recovered by the RRc stars, especially at higher metallicities, such as the secondary mode of the disk MDF.

\subsection{On the effect of the He-abundance}

Finally, we would like to emphasize certain important limitations of the general applicability of the RR~Lyrae stars' photometric metallicity estimation. While our predictive model captures the dependency of the light-curve parameters on the bulk heavy-element content, there are other physical variables that can significantly influence the light-curve shapes of RR~Lyrae stars, but which remained latent in the current formulation of the [Fe/H] regression problem; and which currently cannot be taken into account due to the lack of data. For example, helium abundance is arguably such a latent physical quantity, which is hard to measure directly but known to vary across different Galactic environments, and can have a significant spread even within monometallic globular clusters \citep[for a review, see][]{2018ARA&A..56...83B}.
	
Our predictive model cannot be expected to generalize well to data outside the (not precisely known, but assumed to be close to canonical) He-abundance range of its training set of field RR~Lyrae stars; and deploying it on such data poses the risk of significantly biased [Fe/H] predictions. We illustrate this on the extreme example of the metal-rich globular cluster NGC\,6441, which hosts RR~Lyrae stars with anomalously long periods \citep{2001AJ....122.2600P}, interpreted as a result of a very large helium abudance of $Y\gtrsim0.35$ \citep{2007A&A...463..949C}. We analyzed the OGLE $I$-band photometry of its known RRab stars \citep[see][]{2001AJ....122.2600P,2006AJ....132.1014C} following the procedure in Sect.~\ref{subsec:lightcurves}, applied the same quality criteria as earlier, and omitted suspected blends, which resulted in a sample of 14 objects with very high-quality light curves. Applying our predictive model on this data set results in a predicted mean metallicity of $\langle[{\rm Fe}/{\rm H}]\rangle\approx-1.8$, with individual estimates ranging from $-1.5$ to $-2.3$~dex. These values are extremely biased with respect to the cluster's mean spectroscopic metallicity of ${\rm [Fe/H]}\approx-0.4$ measured by \citet{2005ApJ...630L.145C}. Although this is probably one of the most exaggerated cases for the effect of He-abundance variations, it underlines our caveat to the applicability of the photometric metallicity estimator.

\section{Discussion and Conclusions} \label{sec:conclusions}

In this work, we have provided new empirical relations for predicting the metallicities of RR~Lyrae stars from Cousins $I$-band light curves. Our results are based on a homogeneous set of [Fe/H] abundances derived from HR spectroscopic measurements compiled from the contemporary literature, and high-quality photometric time series.

We used robust fitting techniques to derive unbiased parametric representations of the light-curves. Using a standard machine-learning approach, we concluded that with the available data, the best trade-off between [Fe/H] predictive performance and model complexity for RRab stars can be obtained by linear models using the pulsation period and the $\phi_{31}$ and $A_2$ parameters, while the optimal results for RRc stars can be achieved with the $\{P,A_1,A_2,\phi_{31}\}$ feature set. We used cross-validation to measure the mean absolute error in the predictions to be $0.16$ and $0.18$~dex for the RRab and RRc stars, respectively, which are on par with the typical error level of low-resolution spectroscopic [Fe/H] measurements \citep[e.g.,][]{layden_metallicities_1994}. The final results were obtained by following a full Bayesian approach, yielding robust estimates of the optimal regression model parameters, as well as their correlations and credible intervals. Being trained on a sizable HR spectroscopic dataset, our predictive model yields accurate photometric metallicity estimates over a wide range (from solar to $[{\rm Fe}/{\rm H}]\approx-2.5$). At the same time, our formula for the [Fe/H] estimation of RRc stars requires the important caveat that it is unconstrained for $[{\rm Fe}/{\rm H}]\gtrsim-1$ due to the limited range and small size of its training data set.

We deployed our predictive models on a large number RR~Lyrae light curves from the OGLE survey to obtain photometric metallicity distributions for various stellar populations in the Local Group. We found that despite its limitations, our predictive model for RRc stars recovers the distributions' modes computed from RRab data with an error of less than 0.1~dex. We would like to emphasize that targeted $I$-band time-series observations of only about a dozen metal-rich field RRc stars with existing HR spectroscopic abundance measurements would already result in an enormous improvement of the [Fe/H] prediction formula.

Comparisons with individual [Fe/H] estimates and MDFs obtained from the previously employed $I$-band [Fe/H] prediction formulae of \citet{smolec_metallicity_2005} and \citet{skowron_ogle-ing_2016} show that the latter have metallicity-dependent positive biases. In case of the S05 and the $I$-band transformed JK96 formulae, we attribute these to a similar bias in the underlying \citet{jurcsik_determination_1996} metallicity scale. We found that all previously used formulae overestimate the modes of the MDFs of the Galactic bulge, disk, and the Magellanic Clouds by 0.35---0.4~dex and cause distortions in their shapes. This prevalent bias in the RR~Lyrae photometric metallicities affects their use as standard candles as well, by making the objects appear closer. Considering the state-of-the-art theoretical period-luminosity-metallicity relations of \citet{marconi_new_2015}, a systematic error of the aforementioned size in the [Fe/H] corresponds to an offset of 0.06 magnitudes in the distance modulus. This translates to a bias of roughly 200~pc at the distance of the bulge, but amounts to an offset of almost 1.4~kpc at the distance of the LMC, and errors of similar size may propagate into the distances of halo objects as well.

Having unbiased photometric metallicity estimates of RR~Lyrae stars in various common photometric passbands is a key ingredient to their employment as distant indicators and old population tracers. By relating their accurate metallicity distribution through population synthesis models \citep[see][]{savino_age_2020} could give stringent constraints on the formation epoch and early evolution of various stellar environments. Furthermore, in tandem with widely used intermediate-age population tracers such as red clump giants, they can provide an important ingredient to the inference of chemical enrichment histories.

\begin{acknowledgments}
The authors are grateful to the anonymous referee for his/her constructive suggestions that helped to improve this study. The authors say thanks to Dorota Skowron and Zden\v{e}k Prudil for the fruitful and enlightening discussions. I.D. and E.K.G. were supported by the Deutsche Forschungsgemeinschaft (DFG, German Research Foundation) -- Project-ID 138713538 -- SFB 881 (``The Milky Way System'', subproject A03).
\end{acknowledgments}

%% To help institutions obtain information on the effectiveness of their 
%% telescopes the AAS Journals has created a group of keywords for telescope 
%% facilities.
%
%% Following the acknowledgments section, use the following syntax and the
%% \facility{} or \facilities{} macros to list the keywords of facilities used 
%% in the research for the paper.  Each keyword is check against the master 
%% list during copy editing.  Individual instruments can be provided in 
%% parentheses, after the keyword, but they are not verified.

\vspace{5mm}
% \facilities{}

%% Similar to \facility{}, there is the optional \software command to allow 
%% authors a place to specify which programs were used during the creation of 
%% the manuscript. Authors should list each code and include either a
%% citation or url to the code inside ()s when available.

\software{
	numpy \citep{harris2020array},
	scikit-learn \citep{scikit-learn},
	pymc3 \citep{salvatier_probabilistic_2016},
	ArviZ \citep{arviz_2019}
}

%% Appendix material should be preceded with a single \appendix command.
%% There should be a \section command for each appendix. Mark appendix
%% subsections with the same markup you use in the main body of the paper.

%% Each Appendix (indicated with \section) will be lettered A, B, C, etc.
%% The equation counter will reset when it encounters the \appendix
%% command and will number appendix equations (A1), (A2), etc. The
%% Figure and Table counter will not reset.

%\appendix

% A handy "cheat sheet" that provides the necessary \latex\ to produce 17 
% different types of tables is available at
% \url{http://journals.aas.org/authors/aastex/aasguide.html#table_cheat_sheet}.

%% For this sample we use BibTeX plus aasjournals.bst to generate the
%% the bibliography. The sample631.bib file was populated from ADS. To
%% get the citations to show in the compiled file do the following:
%%
%% pdflatex sample631.tex
%% bibtext sample631
%% pdflatex sample631.tex
%% pdflatex sample631.tex

\bibliography{references}{}

\begin{thebibliography}{}
\expandafter\ifx\csname natexlab\endcsname\relax\def\natexlab#1{#1}\fi
\providecommand{\url}[1]{\href{#1}{#1}}
\providecommand{\dodoi}[1]{doi:~\href{http://doi.org/#1}{\nolinkurl{#1}}}
\providecommand{\doeprint}[1]{\href{http://ascl.net/#1}{\nolinkurl{http://ascl.net/#1}}}
\providecommand{\doarXiv}[1]{\href{https://arxiv.org/abs/#1}{\nolinkurl{https://arxiv.org/abs/#1}}}

\bibitem[{Andrievsky {et~al.}(2018)Andrievsky, Wallerstein, Korotin, Lyashko,
  Kovtyukh, Tsymbal, Davis, Gomez, Huang, \&
  Farrell}]{andrievsky_relationship_2018}
Andrievsky, S., Wallerstein, G., Korotin, S., {et~al.} 2018, Publications of
  the Astronomical Society of the Pacific, 130, 024201,
  \dodoi{10.1088/1538-3873/aa9783}

\bibitem[{{Bastian} \& {Lardo}(2018)}]{2018ARA&A..56...83B}
{Bastian}, N., \& {Lardo}, C. 2018, Annual Reviews of Astronomy \&
  Astrophysics, 56, 83, \dodoi{10.1146/annurev-astro-081817-051839}

\bibitem[{{Caloi} \& {D'Antona}(2007)}]{2007A&A...463..949C}
{Caloi}, V., \& {D'Antona}, F. 2007, Astronomy \& Astrophysics, 463, 949,
  \dodoi{10.1051/0004-6361:20066074}

\bibitem[{Catelan \& Smith(2015)}]{catelan_pulsating_2015}
Catelan, M., \& Smith, H.~A. 2015, Pulsating Stars (Wiley-VCH), 2015.
\newblock \url{http://adsabs.harvard.edu/abs/2015pust.book.....C}

\bibitem[{Chadid {et~al.}(2017)Chadid, Sneden, \&
  Preston}]{chadid_spectroscopic_2017}
Chadid, M., Sneden, C., \& Preston, G.~W. 2017, The Astrophysical Journal, 835,
  187, \dodoi{10.3847/1538-4357/835/2/187}

\bibitem[{Clementini {et~al.}(1995)Clementini, Carretta, Gratton, Merighi,
  Mould, \& McCarthy}]{clementini_composition_1995}
Clementini, G., Carretta, E., Gratton, R., {et~al.} 1995, The Astronomical
  Journal, 110, 2319, \dodoi{10.1086/117692}

\bibitem[{{Clementini} {et~al.}(2005){Clementini}, {Gratton}, {Bragaglia},
  {Ripepi}, {Martinez Fiorenzano}, {Held}, \& {Carretta}}]{2005ApJ...630L.145C}
{Clementini}, G., {Gratton}, R.~G., {Bragaglia}, A., {et~al.} 2005, \apjl, 630,
  L145, \dodoi{10.1086/491789}

\bibitem[{Clementini {et~al.}(2019)Clementini, Ripepi, Molinaro, Garofalo,
  Muraveva, Rimoldini, Guy, Jevardat~de Fombelle, Nienartowicz, Marchal,
  Audard, Holl, Leccia, Marconi, Musella, Mowlavi, Lecoeur-Taibi, Eyer,
  De~Ridder, Regibo, Sarro, Szabados, Evans, \& Riello}]{clementini_gaia_2019}
Clementini, G., Ripepi, V., Molinaro, R., {et~al.} 2019, Astronomy \&
  Astrophysics, 622, A60, \dodoi{10.1051/0004-6361/201833374}

\bibitem[{{Corwin} {et~al.}(2006){Corwin}, {Sumerel}, {Pritzl}, {Smith},
  {Catelan}, {Sweigart}, \& {Stetson}}]{2006AJ....132.1014C}
{Corwin}, T.~M., {Sumerel}, A.~N., {Pritzl}, B.~J., {et~al.} 2006, \aj, 132,
  1014, \dodoi{10.1086/505745}

\bibitem[{{Crestani} {et~al.}(2021){Crestani}, {Fabrizio}, {Braga}, {Sneden},
  {Preston}, {Ferraro}, {Iannicola}, {Bono}, {Alves-Brito}, {Nonino},
  {D'Orazi}, {Inno}, {Monelli}, {Storm}, {Altavilla}, {Chaboyer}, {Dall'Ora},
  {Fiorentino}, {Gilligan}, {Grebel}, {Lala}, {Lemasle}, {Marengo}, {Marinoni},
  {Marrese}, {Mart{\'\i}nez-V{\'a}zquez}, {Matsunaga}, {Mullen}, {Neeley},
  {Prudil}, {da Silva}, {Stetson}, {Th{\'e}venin}, {Valenti}, {Walker}, \&
  {Zoccali}}]{crestani_deltaS}
{Crestani}, J., {Fabrizio}, M., {Braga}, V.~F., {et~al.} 2021, \apj, 908, 20,
  \dodoi{10.3847/1538-4357/abd183}

\bibitem[{Deng {et~al.}(2012)Deng, Newberg, Liu, Carlin, Beers, Chen, Chen,
  Christlieb, Grillmair, Guhathakurta, Han, Hou, Lee, Lépine, Li, Liu, Pan,
  Sellwood, Wang, Wang, Yang, Yanny, Zhang, Zhang, Zheng, \&
  Zhu}]{deng_lamost_2012}
Deng, L.-C., Newberg, H.~J., Liu, C., {et~al.} 2012, Research in Astronomy and
  Astrophysics, 12, 735, \dodoi{10.1088/1674-4527/12/7/003}

\bibitem[{Dékány \& Grebel(2020)}]{dekany_near-infrared_2020}
Dékány, I., \& Grebel, E.~K. 2020, The Astrophysical Journal, 898, 46,
  \dodoi{10.3847/1538-4357/ab9d87}

\bibitem[{Dékány {et~al.}(2019)Dékány, Hajdu, Grebel, \&
  Catelan}]{dekany_into_2019}
Dékány, I., Hajdu, G., Grebel, E.~K., \& Catelan, M. 2019, The Astrophysical
  Journal, 883, 58, \dodoi{10.3847/1538-4357/ab3b60}

\bibitem[{Dékány {et~al.}(2018)Dékány, Hajdu, Grebel, Catelan, Elorrieta,
  Eyheramendy, Majaess, \& Jordán}]{dekany_near-infrared_2018}
Dékány, I., Hajdu, G., Grebel, E.~K., {et~al.} 2018, The Astrophysical
  Journal, 857, 54, \dodoi{10.3847/1538-4357/aab4fa}

\bibitem[{Fernley \& Barnes(1996)}]{fernley_metal_1996}
Fernley, J., \& Barnes, T.~G. 1996, Astronomy \& Astrophysics, 312, 957.
\newblock \url{http://adsabs.harvard.edu/abs/1996A%26A...312..957F}

\bibitem[{For {et~al.}(2011)For, Sneden, \& Preston}]{for_chemical_2011}
For, B.-Q., Sneden, C., \& Preston, G.~W. 2011, {\textbackslash}apjs, 197, 29,
  \dodoi{10.1088/0067-0049/197/2/29}

\bibitem[{Govea {et~al.}(2014)Govea, Gomez, Preston, \&
  Sneden}]{govea_chemical_2014}
Govea, J., Gomez, T., Preston, G.~W., \& Sneden, C. 2014, The Astrophysical
  Journal, 782, 59, \dodoi{10.1088/0004-637X/782/2/59}

\bibitem[{Hajdu {et~al.}(2018)Hajdu, Dékány, Catelan, Grebel, \&
  Jurcsik}]{hajdu_data-driven_2018}
Hajdu, G., Dékány, I., Catelan, M., Grebel, E.~K., \& Jurcsik, J. 2018, The
  Astrophysical Journal, 857, 55, \dodoi{10.3847/1538-4357/aab4fd}

\bibitem[{Harris {et~al.}(2020)Harris, Millman, van~der Walt, Gommers,
  Virtanen, Cournapeau, Wieser, Taylor, Berg, Smith, Kern, Picus, Hoyer, van
  Kerkwijk, Brett, Haldane, del R{'{\i}}o, Wiebe, Peterson,
  G{'{e}}rard-Marchant, Sheppard, Reddy, Weckesser, Abbasi, Gohlke, \&
  Oliphant}]{harris2020array}
Harris, C.~R., Millman, K.~J., van~der Walt, S.~J., {et~al.} 2020, Nature, 585,
  357, \dodoi{10.1038/s41586-020-2649-2}

\bibitem[{{Haschke} {et~al.}(2012){Haschke}, {Grebel}, {Duffau}, \&
  {Jin}}]{2012AJ....143...48H}
{Haschke}, R., {Grebel}, E.~K., {Duffau}, S., \& {Jin}, S. 2012, \aj, 143, 48,
  \dodoi{10.1088/0004-6256/143/2/48}

\bibitem[{Hoﬀman \& Gelman(2014)}]{hoffmann_nuts}
Hoﬀman, M.~D., \& Gelman, A. 2014, Journal of Machine Learning Research, 31

\bibitem[{Iorio \& Belokurov(2020)}]{iorio_chemo-kinematics_2020}
Iorio, G., \& Belokurov, V. 2020, arXiv e-prints, 2008, arXiv:2008.02280.
\newblock \url{http://adsabs.harvard.edu/abs/2020arXiv200802280I}

\bibitem[{Jacyszyn-Dobrzeniecka {et~al.}(2017)Jacyszyn-Dobrzeniecka, Skowron,
  Mróz, Soszyński, Udalski, Pietrukowicz, Skowron, Poleski, Kozłowski,
  Wyrzykowski, Pawlak, Szymański, \&
  Ulaczyk}]{jacyszyn-dobrzeniecka_ogle-ing_2017}
Jacyszyn-Dobrzeniecka, A.~M., Skowron, D.~M., Mróz, P., {et~al.} 2017, Acta
  Astronomica, 67, 1, \dodoi{10.32023/0001-5237/67.1.1}

\bibitem[{Jurcsik \& Kov\'acs(1996)}]{jurcsik_determination_1996}
Jurcsik, J., \& Kov\'acs, G. 1996, Astronomy \& Astrophysics, 312, 111

\bibitem[{Jurcsik {et~al.}(2005)Jurcsik, S\'odor, V\'aradi, Szeidl, Washuettl,
  Weber, D\'ek\'any, Hurta, Lakatos, Posztob\'anyi, Szing, \&
  Vida}]{jurcsik_blazhko_2005}
Jurcsik, J., S\'odor, A., V\'aradi, M., {et~al.} 2005, Astronomy \&
  Astrophysics, 430, 1049, \dodoi{10.1051/0004-6361:20041784}

\bibitem[{Kumar {et~al.}(2019)Kumar, Carroll, Hartikainen, \&
  Martin}]{arviz_2019}
Kumar, R., Carroll, C., Hartikainen, A., \& Martin, O. 2019, Journal of Open
  Source Software, 4, 1143, \dodoi{10.21105/joss.01143}

\bibitem[{Kunder {et~al.}(2016)Kunder, Rich, Koch, Storm, Nataf, De~Propris,
  Walker, Bono, Johnson, Shen, \& Li}]{kunder_before_2016}
Kunder, A., Rich, R.~M., Koch, A., {et~al.} 2016, The Astrophysical Journal
  Letters, 821, L25, \dodoi{10.3847/2041-8205/821/2/L25}

\bibitem[{Lambert {et~al.}(1996)Lambert, Heath, Lemke, \&
  Drake}]{lambert_chemical_1996}
Lambert, D.~L., Heath, J.~E., Lemke, M., \& Drake, J. 1996, The Astrophysical
  Journal Supplement Series, 103, 183, \dodoi{10.1086/192274}

\bibitem[{Layden(1994)}]{layden_metallicities_1994}
Layden, A.~C. 1994, The Astronomical Journal, 108, 1016, \dodoi{10.1086/117132}

\bibitem[{{Le Borgne} {et~al.}(2007){Le Borgne}, {Paschke}, {Vandenbroere},
  {Poretti}, {Klotz}, {Bo{\"e}r}, {Damerdji}, {Martignoni}, \&
  {Acerbi}}]{2007A&A...476..307L}
{Le Borgne}, J.~F., {Paschke}, A., {Vandenbroere}, J., {et~al.} 2007, Astronomy
  \& Astrophysics, 476, 307, \dodoi{10.1051/0004-6361:20077957}

\bibitem[{Liu {et~al.}(2020)Liu, Huang, Zhang, Xiang, Ren, Chen, Yuan, Wang,
  Yang, Tian, Wang, \& Liu}]{liu_probing_2020}
Liu, G.-C., Huang, Y., Zhang, H.-W., {et~al.} 2020, The Astrophysical Journal
  Supplement Series, 247, 68, \dodoi{10.3847/1538-4365/ab72f8}

\bibitem[{Liu {et~al.}(2013)Liu, Zhao, Chen, Takeda, \&
  Honda}]{liu_abundances_2013}
Liu, S., Zhao, G., Chen, Y.-Q., Takeda, Y., \& Honda, S. 2013, Research in
  Astronomy and Astrophysics, 13, 1307, \dodoi{10.1088/1674-4527/13/11/003}

\bibitem[{Marconi {et~al.}(2015)Marconi, Coppola, Bono, Braga, Pietrinferni,
  Buonanno, Castellani, Musella, Ripepi, \& Stellingwerf}]{marconi_new_2015}
Marconi, M., Coppola, G., Bono, G., {et~al.} 2015, The Astrophysical Journal,
  808, 50, \dodoi{10.1088/0004-637X/808/1/50}

\bibitem[{Minniti {et~al.}(2010)Minniti, Lucas, Emerson, Saito, Hempel,
  Pietrukowicz, Ahumada, Alonso, Alonso-Garcia, Arias, Bandyopadhyay, Barbá,
  Barbuy, Bedin, Bica, Borissova, Bronfman, Carraro, Catelan, Clariá, Cross,
  de~Grijs, Dékány, Drew, Fari{\textbackslash}textasciitilde~na, Feinstein,
  Fernández~Lajús, Gamen, Geisler, Gieren, Goldman, Gonzalez, Gunthardt,
  Gurovich, Hambly, Irwin, Ivanov, Jordán, Kerins, Kinemuchi, Kurtev,
  López-Corredoira, Maccarone, Masetti, Merlo, Messineo, Mirabel, Monaco,
  Morelli, Padilla, Palma, Parisi, Pignata, Rejkuba, Roman-Lopes, Sale,
  Schreiber, Schröder, Smith, Sodré, Soto, Tamura, Tappert, Thompson, Toledo,
  Zoccali, \& Pietrzynski}]{minniti_vista_2010}
Minniti, D., Lucas, P.~W., Emerson, J.~P., {et~al.} 2010, {\textbackslash}na,
  15, 433, \dodoi{10.1016/j.newast.2009.12.002}

\bibitem[{Monson {et~al.}(2017)Monson, Beaton, Scowcroft, Freedman, Madore,
  Rich, Seibert, Kollmeier, \& Clementini}]{monson_standard_2017}
Monson, A.~J., Beaton, R.~L., Scowcroft, V., {et~al.} 2017, The Astronomical
  Journal, 153, 96, \dodoi{10.3847/1538-3881/153/3/96}

\bibitem[{Muhie {et~al.}(2021)Muhie, Dambis, Berdnikov, Kniazev, \&
  Grebel}]{muhie_kinematics_2021}
Muhie, T.~D., Dambis, A.~K., Berdnikov, L.~N., Kniazev, A.~Y., \& Grebel, E.~K.
  2021, arXiv e-prints, arXiv:2101.03899

\bibitem[{Mullen {et~al.}(2021)Mullen, Marengo, Martínez-Vázquez, Neeley,
  Bono, Dall'Ora, Chaboyer, Thévenin, Braga, Crestani, Fabrizio, Fiorentino,
  Gilligan, Monelli, \& Stetson}]{mullen_metallicity_2021}
Mullen, J.~P., Marengo, M., Martínez-Vázquez, C.~E., {et~al.} 2021, arXiv
  e-prints, 2103, arXiv:2103.09372.
\newblock \url{http://adsabs.harvard.edu/abs/2021arXiv210309372M}

\bibitem[{Muraveva {et~al.}(2018)Muraveva, Delgado, Clementini, Sarro, \&
  Garofalo}]{muraveva_rr_2018}
Muraveva, T., Delgado, H.~E., Clementini, G., Sarro, L.~M., \& Garofalo, A.
  2018, Monthly Notices of the Royal Astronomical Society, 481, 1195,
  \dodoi{10.1093/mnras/sty2241}

\bibitem[{Nemec {et~al.}(2013)Nemec, Cohen, Ripepi, Derekas, Moskalik, Sesar,
  Chadid, \& Bruntt}]{nemec_metal_2013}
Nemec, J.~M., Cohen, J.~G., Ripepi, V., {et~al.} 2013, The Astrophysical
  Journal, 773, 181, \dodoi{10.1088/0004-637X/773/2/181}

\bibitem[{Ngeow {et~al.}(2016)Ngeow, Yu, Bellm, Yang, Chang, Miller, Laher,
  Surace, \& Ip}]{ngeow_palomar_2016}
Ngeow, C.-C., Yu, P.-C., Bellm, E., {et~al.} 2016, The Astrophysical Journal
  Supplement Series, 227, 30, \dodoi{10.3847/1538-4365/227/2/30}

\bibitem[{Pancino {et~al.}(2015)Pancino, Britavskiy, Romano, Cacciari,
  Mucciarelli, \& Clementini}]{pancino_chemical_2015}
Pancino, E., Britavskiy, N., Romano, D., {et~al.} 2015, Monthly Notices of the
  Royal Astronomical Society, 447, 2404, \dodoi{10.1093/mnras/stu2616}

\bibitem[{Pedregosa {et~al.}(2011)Pedregosa, Varoquaux, Gramfort, Michel,
  Thirion, Grisel, Blondel, Prettenhofer, Weiss, Dubourg, Vanderplas, Passos,
  Cournapeau, Brucher, Perrot, \& Duchesnay}]{scikit-learn}
Pedregosa, F., Varoquaux, G., Gramfort, A., {et~al.} 2011, Journal of Machine
  Learning Research, 12, 2825

\bibitem[{Pietrukowicz {et~al.}(2020)Pietrukowicz, Udalski, Soszyński,
  Skowron, Wrona, Szymański, Poleski, Ulaczyk, Koz{\textbackslash}lowski,
  Skowron, Mróz, Rybicki, Iwanek, \& Gromadzki}]{pietrukowicz_properties_2020}
Pietrukowicz, P., Udalski, A., Soszyński, I., {et~al.} 2020, Acta Astronomica,
  70, 121, \dodoi{10.32023/0001-5237/70.2.3}

\bibitem[{Pojmanski(1997)}]{pojmanski_all_1997}
Pojmanski, G. 1997, Acta Astronomica, 47, 467.
\newblock \url{http://adsabs.harvard.edu/abs/1997AcA....47..467P}

\bibitem[{{Pritzl} {et~al.}(2001){Pritzl}, {Smith}, {Catelan}, \&
  {Sweigart}}]{2001AJ....122.2600P}
{Pritzl}, B.~J., {Smith}, H.~A., {Catelan}, M., \& {Sweigart}, A.~V. 2001, \aj,
  122, 2600, \dodoi{10.1086/323447}

\bibitem[{Salvatier {et~al.}(2016)Salvatier, Wiecki, \&
  Fonnesbeck}]{salvatier_probabilistic_2016}
Salvatier, J., Wiecki, T.~V., \& Fonnesbeck, C. 2016, PeerJ Computer Science,
  2, e55, \dodoi{10.7717/peerj-cs.55}

\bibitem[{Savino {et~al.}(2020)Savino, Koch, Prudil, Kunder, \&
  Smolec}]{savino_age_2020}
Savino, A., Koch, A., Prudil, Z., Kunder, A., \& Smolec, R. 2020, Astronomy \&
  Astrophysics, 641, A96, \dodoi{10.1051/0004-6361/202038305}

\bibitem[{{Skarka} {et~al.}(2020){Skarka}, {Prudil}, \&
  {Jurcsik}}]{2020MNRAS.494.1237S}
{Skarka}, M., {Prudil}, Z., \& {Jurcsik}, J. 2020, \mnras, 494, 1237,
  \dodoi{10.1093/mnras/staa673}

\bibitem[{Skowron {et~al.}(2016)Skowron, Soszyński, Udalski, Szymański,
  Pietrukowicz, Skowron, Poleski, Wyrzykowski, Ulaczyk, Kozłowski, Mróz, \&
  Pawlak}]{skowron_ogle-ing_2016}
Skowron, D.~M., Soszyński, I., Udalski, A., {et~al.} 2016, Acta Astronomica,
  66, 269.
\newblock \url{http://adsabs.harvard.edu/abs/2016AcA....66..269S}

\bibitem[{Smolec(2005)}]{smolec_metallicity_2005}
Smolec, R. 2005, Acta Astronomica, 55, 59

\bibitem[{{Smolec}(2016)}]{2016pas..conf...22S}
{Smolec}, R. 2016, in 37th Meeting of the Polish Astronomical Society, ed.
  A.~{R{\'o}{\.z}a{\'n}ska} \& M.~{Bejger}, Vol.~3, 22--25.
\newblock \doarXiv{1603.01252}

\bibitem[{Sneden {et~al.}(2017)Sneden, Preston, Chadid, \&
  Adamów}]{sneden_rrc_2017}
Sneden, C., Preston, G.~W., Chadid, M., \& Adamów, M. 2017,
  {\textbackslash}apj, 848, 68, \dodoi{10.3847/1538-4357/aa8b10}

\bibitem[{Soszyński {et~al.}(2019)Soszyński, Udalski, Wrona, Szymański,
  Pietrukowicz, Skowron, Skowron, Poleski, Kozłowski, Mróz, Ulaczyk, Rybicki,
  Iwanek, \& Gromadzki}]{soszynski_over_2019}
Soszyński, I., Udalski, A., Wrona, M., {et~al.} 2019, Acta Astronomica, 69,
  321, \dodoi{10.32023/0001-5237/69.4.2}

\bibitem[{Stringer {et~al.}(2020)Stringer, Drlica-Wagner, Macri,
  Martínez-Vázquez, Vivas, Ferguson, Pace, Walker, Neilsen, Tavangar, Wester,
  Abbott, Aguena, Allam, Bacon, Bechtol, Bertin, Brooks, Burke, Carnero~Rosell,
  Carrasco~Kind, Carretero, Costanzi, Crocce, da~Costa, Pereira, De~Vicente,
  Desai, Diehl, Doel, Ferrero, García-Bellido, Gaztanaga, Gerdes, Gruen,
  Gruendl, Gschwend, Gutierrez, Hinton, Hollowood, Honscheid, Hoyle, James,
  Kuehn, Kuropatkin, Li, Maia, Marshall, Menanteau, Miquel, Morgan, Ogando,
  Palmese, Paz-Chinchón, Plazas, Roodman, Sanchez, Schubnell, Serrano,
  Sevilla-Noarbe, Smith, Soares-Santos, Suchyta, Tarle, Thomas, To, Varga,
  Wilkinson, Zhang, \& {the DES Collaboration}}]{stringer_identifying_2020}
Stringer, K.~M., Drlica-Wagner, A., Macri, L., {et~al.} 2020, arXiv e-prints,
  2011, arXiv:2011.13930.
\newblock \url{http://adsabs.harvard.edu/abs/2020arXiv201113930S}

\bibitem[{Suntzeff {et~al.}(1994)Suntzeff, Kraft, \&
  Kinman}]{suntzeff_summary_1994}
Suntzeff, N.~B., Kraft, R.~P., \& Kinman, T.~D. 1994, The Astrophysical Journal
  Supplement Series, 93, 271, \dodoi{10.1086/192055}

\bibitem[{Szczygieł {et~al.}(2009)Szczygieł, Pojmański, \&
  Pilecki}]{szczygiel_galactic_2009}
Szczygieł, D.~M., Pojmański, G., \& Pilecki, B. 2009, Acta Astronomica, 59,
  137.
\newblock \url{http://adsabs.harvard.edu/abs/2009AcA....59..137S}

\bibitem[{Udalski {et~al.}(2015)Udalski, Szymański, \&
  Szymański}]{udalski_ogle-iv_2015}
Udalski, A., Szymański, M.~K., \& Szymański, G. 2015, Acta Astronomica, 65,
  1.
\newblock \url{http://adsabs.harvard.edu/abs/2015AcA....65....1U}

\bibitem[{Yanny {et~al.}(2009)Yanny, Rockosi, Newberg, Knapp, Adelman-McCarthy,
  Alcorn, Allam, Allende~Prieto, An, Anderson, Anderson, Bailer-Jones, Bastian,
  Beers, Bell, Belokurov, Bizyaev, Blythe, Bochanski, Boroski, Brinchmann,
  Brinkmann, Brewington, Carey, Cudworth, Evans, Evans, Gates, Gänsicke,
  Gillespie, Gilmore, Nebot Gomez-Moran, Grebel, Greenwell, Gunn, Jordan,
  Jordan, Harding, Harris, Hendry, Holder, Ivans, Ivezič, Jester, Johnson,
  Kent, Kleinman, Kniazev, Krzesinski, Kron, Kuropatkin, Lebedeva, Lee,
  French~Leger, Lépine, Levine, Lin, Long, Loomis, Lupton, Malanushenko,
  Malanushenko, Margon, Martinez-Delgado, McGehee, Monet, Morrison, Munn,
  Neilsen, Nitta, Norris, Oravetz, Owen, Padmanabhan, Pan, Peterson, Pier,
  Platson, Re~Fiorentin, Richards, Rix, Schlegel, Schneider, Schreiber,
  Schwope, Sibley, Simmons, Snedden, Allyn~Smith, Stark, Stauffer, Steinmetz,
  Stoughton, SubbaRao, Szalay, Szkody, Thakar, Sivarani, Tucker, Uomoto,
  Vanden~Berk, Vidrih, Wadadekar, Watters, Wilhelm, Wyse, Yarger, \&
  Zucker}]{yanny_segue_2009}
Yanny, B., Rockosi, C., Newberg, H.~J., {et~al.} 2009, The Astronomical
  Journal, 137, 4377, \dodoi{10.1088/0004-6256/137/5/4377}

\end{thebibliography}
\bibliographystyle{aasjournal}

%% This command is needed to show the entire author+affiliation list when
%% the collaboration and author truncation commands are used.  It has to
%% go at the end of the manuscript.
%\allauthors

%% Include this line if you are using the \added, \replaced, \deleted
%% commands to see a summary list of all changes at the end of the article.
%\listofchanges

\end{document}